\documentclass[fleqn,usenatbib]{mnras}

\usepackage{framed}
\usepackage{graphicx,epsfig,psfig,multicol}
\usepackage[dvipsnames]{xcolor}
\usepackage[]{inputenc,amssymb}
\usepackage{amsmath}
\usepackage{color}
\usepackage{epsfig}
\usepackage{epstopdf}
\usepackage{textcomp}
\usepackage{xcolor,cancel}
\usepackage{comment}
\usepackage{subcaption}

\usepackage[export]{adjustbox}% this is for telling the positioning of the figure eg. center ..
\usepackage{adjustbox}
\usepackage{placeins}% This is for using the float barrier ...

\usepackage{float}
\usepackage{morefloats}

\usepackage{lipsum}
\usepackage{mwe}
\usepackage{arydshln} %for dashed line

\definecolor{webgreen}{rgb}{0,.5,0}
\definecolor{webbrown}{rgb}{.6,0,0}
\usepackage[toc,page]{appendix}

\def \beq{\begin{equation}}
\def \eeq{\end{equation}}
\def \bea{\begin{eqnarray}}
\def \eea{\end{eqnarray}}
\newcommand{\bfx}{\mbox{${\mathbf{ x}}$}}
\newcommand{\bfr}{\mbox{${\mathbf{ r}}$}}

\newcommand{\bfp}{\mbox{${\mathbf{ p}}$}}
\newcommand{\bfv}{\mbox{${\mathbf{ v}}$}}
\newcommand{\bfI}{\mbox{${\mathbf{ I}}$}}
\newcommand{\bfw}{\mbox{${\mathbf{ w}}$}}
\newcommand{\bfl}{\mbox{$\boldsymbol{ l}$}}
\newcommand{\bfOmg}{\mbox{${\mathbf{\Omega}}$}}
\newcommand{\Ftilda}{\mbox{$\widetilde{F}$}}
\newcommand{\Phitilda}{\mbox{$\widetilde{\Phi}$}}
\newcommand{\rmi}{\mbox{ ${\rm i}$ }}
\newcommand{\p}{\partial}
\newcommand{\rmd}{\mbox{${\rm d}$}}

\newcommand{\im}{\mbox{${\rm Im}$}}

\newcommand{\Msun}{\mbox{$ M_\odot $  }}

\newcommand{\scre}{{\cal E}}

\usepackage[T1]{fontenc}

\DeclareRobustCommand{\VAN}[3]{#2}
\let\VANthebibliography\thebibliography
\def\thebibliography{\DeclareRobustCommand{\VAN}[3]{##3}\VANthebibliography}

\title[Density Wakes]{Density Wakes due to Dynamical Friction in Cored Potentials}

\author[K. Kaur et al.]{
Karamveer Kaur,\thanks{E-mail: karamveer.kaur@mail.huji.ac.il}
Nicholas C. Stone
\\
Racah Institute of Physics, The Hebrew University, 91904, Jerusalem, Israel}

\date{Accepted XXX. Received YYY; in original form ZZZ}

\pubyear{2021}

\begin{document}

\label{firstpage}
\pagerange{\pageref{firstpage}--\pageref{lastpage}}
\maketitle

\begin{abstract}
Dynamical friction is often modeled with reasonable accuracy by the  widely used
Chandrasekhar formula. However, in some circumstances, Chandrasekhar's local and uniform approximations can break down severely.  An astrophysically important example %of this
is the ``core stalling'' phenomenon seen in $N$-body simulations of massive perturber inspiralling into
the near-harmonic potential of a stellar system's constant-density core (and possibly also in direct observations of dwarf galaxies with globular clusters).  In this paper we use the linearized collisionless Boltzmann equation to calculate the global response of a cored galaxy to the presence of a massive perturber. We evaluate the density deformation, or wake, due to the perturber and study its  geometrical structure to better understand the phenomenon of core stalling. We also evaluate the dynamical friction torque acting on perturber from the Lynden-Bell--Kalnajs (LBK) formula. % \citep{TremaineWeinberg84}.
In agreement with past work,  %\citep{KaurSridhar18},
we find that the dynamical friction force arising from corotating resonances is greatly weakened, relative to the Chandrasekhar formula, inside a constant density core.  In contrast to past work, however, we find that a population of previously neglected high-order and non-corotating resonances sustain a minimum level of frictional torque at $\sim 10\%$ of the torque from Chandrasekhar formula.  This suggests that complete core stalling likely requires phenomena beyond the LBK approach; %of linear perturbation theory
 we discuss several possible explanations. Additionally, to study %with the aim of studying
 core stalling for multiple perturbers, we investigate approximate
 secular dynamical interactions (akin to Lidov-Kozai dynamics) between two perturbers orbiting a cored stellar system and derive a criterion for instability arising due to their %  for the appearance of
 close encounters. % between two stalled perturbers.   
\end{abstract}


%\fontsize{12}{15}
%\selectfont

%\onecolumn

\section{Introduction}
Dynamical friction is the statistical force created by many 
gravitational encounters between a massive object and a large population of scatterers.  Over time, dynamical friction (DF) can transfer energy and angular momentum between the massive perturber and the background stellar population\footnote{Throughout this paper, we refer to the constituents of the background mass distribution as ``stars'', though they might be either stellar objects or dark matter particles in a real system.}, with astrophysically important consequences. The concept of DF was first formulated in a kinetic theory describing an infinite and homogeneous stellar system, resulting in a quasi-local, decelerating 
force on the perturber \citep{Chandrasekhar43}.  Although Chandrasekhar's seminal paper correctly predicted that a massive perturber passing through a population of lighter background stars can be efficiently decelerated, it reached this conclusion only for a simplified system of uniform and homogeneous background stars with an isotropic velocity distribution.  This approach is usually applied as an approximate local theory to non-uniform and finite astrophysical stellar systems, which in reality have a more complicated orbital structure.\footnote{The assumption of velocity isotropy is sometimes relaxed in the local theory, as in \citet{Binney77}.}
Eventually it was realized that in the local (hereafter ``Chandrasekhar'') picture of DF, the massive perturber deflects stars into an overdense wake that trails behind it \citep{Marochnik1968,Kalnajs71, Kalnajs72,Mulder1983}. The gravitational field of this wake decelerates the perturber, causing it to slow down.  

Later, \citet{Kalnajs71} showed that in a flattened (2D) stellar disk, a {\it global} calculation (accounting for real orbits of stars in the unperturbed disk) of DF torques can differ dramatically from the local picture, and the effect of DF may even vanish completely for a uniformly rotating disk of stars. Later, the global analysis of angular momentum transfer \citep{LyndenBellKalnajs72} for perturbed stellar disks was extended to a DF torque calculation for a general 3D spherical stellar system by \citet[][henceforth TW]{TremaineWeinberg84}. The perturber follows a circular orbit with the radius technically assumed to be fixed so as to evaluate the torque at an instant; this assumption is the \emph{secular} approximation.  
In the TW approach, DF in realistic, finite spherical systems transfers energy and angular momentum only through the subset of stars in resonance with the perturber.  One implementation of this is visible in \citet{KaurSridhar18}, who derived a torque formula applying the original approach of \citet[][hereafter LBK]{LyndenBellKalnajs72}. They first calculated the linear response (distribution function deformation) of the host galaxy to the perturber, using the linearly perturbed collisionless Boltzmann equation. Then, the LBK torque acting on the perturber can be computed as the inverse of the torque acting on the deformed stellar system. \citet{Weinberg1986} calculated the density deformation (or wake) resulting from the linearly deformed distribution function, yielding useful insights on the properties of wakes and the action of DF.  \citet{Weinberg1989} further devised a formalism to compute torque that takes into account the self-gravity (or polarization) of the wake, which was neglected in the TW theory. Recently, \citet{Banik2021} relaxed the secular assumption to account for the \emph{memory effect} that arises due to the finite time evolution of the perturber's orbit. In contrast to the LBK approach, the memory effect leads to a non-vanishing 
contribution to the torque from non-resonant orbits as well, that can even be sometimes anti-frictional. In their more recent non-perturbative approach, \citet{Banik2021b} also identify those stellar orbital families contributing the most to the DF.

Although the account above shows that the full global picture of DF is non-trivial, the simpler local picture seems to work reasonably well in most 
contexts. With suitable modelling and adjustment of the Coulomb logarithm\footnote{Here $b_{\rm max}$ and $b_{\rm min}$ are the chosen limits of maximum and minimum impact parameters for a field star undergoing a hyperbolic encounter with the perturber.} $\ln{\Lambda} = \ln{(b_{\rm max}/b_{\rm min})}$, many numerical simulations find Chandrasekhar's formula a reasonable fit to the orbital evolution of a perturber sinking in a background mass distribution \citep{LinTremaine1983,Cora1997,vandenBosch1999,JiangBinney2000,Hashimoto2003,Kolchin2008,Jiang2008}. The most notable failures of the Chandrasekhar formula tend to occur in the highly flattened density profiles (``cores'') that exist in the centers of globular clusters \citep{King1966}, and possibly in dwarf galaxies as well \citep{FloresPrimack1994, Burkert1995, AmoriscoEvans2012, Oh2015}. Numerous $N$-body simulations have demonstrated that in constant (or nearly-constant) density cores, the inspiral of massive perturbers slows and then effectively stalls, rather than continuing to the center of system as the local theory of DF would predict \citep{Read2006,Goerdt2006,Inoue+09,Inoue2011,Cole2012,Petts2015,Petts2016}.  Stalling in these numerical simulations is generally complete -- a total cessation of the inspiral -- although in the recent work of \citet{Meadows+20}, the inspiral continues inside the core, albeit at a greatly reduced rate.  These numerical predictions are compatible with observations of dwarf galaxy globular cluster populations whose DF inspiral times (in the Chandrasekhar picture) are much less than a Hubble time \citep{Tremaine1976,Durrell1996,Hernandez1998,Miller1998,Oh2000,Vesperini2000,Mackey2003,Lotz2004,Huang2021}.  Without the possibility of core stalling, the survival of such globular populations is a puzzle sometimes known as the ``timing problem,'' and has even motivated consideration of new physics, such as ultralight dark matter \citep{Hui+18, Bar+21} or modified Newtonian dynamics \citep{AngusDiafero09}.

\citet{KaurSridhar18} investigated the core stalling problem analytically in the TW framework and found that the stalling %can be understood as a result of 
results from the depletion and weakening of corotating resonances in the inner core of the galaxy. The corotating (CR) resonant orbits typically have orbital frequencies $\Omega_w \sim \Omega_p$, $\Omega_p$ being the orbital frequency of perturber at an instant in time. The more recent study of \citet{Banik2021} relates stalling to the balance between DF and buoyancy, the anti-frictional torque (resulting from the memory effect) which is also observed in some numerical simulations \citep{Cole2012,DuttaChowdhury2019}. 

In this paper we build on the earlier work of \citet[][hereafter KS18]{KaurSridhar18} by exploring the physical geometry of wakes due to DF in the global, TW picture, focusing on the cored potentials where DF is at its most non-local. Our work is motivated by the illuminating study of \citet{Weinberg1986} which first identified the structure of density wakes in a spherical potential. The resultant density wakes can be decomposed into resonant and non-resonant portions. Resonant stellar orbits are deformed to give rise to the resonant wake, which is anti-symmetric on the leading and trailing sides of perturber. As a result of this anti-symmetry, the gravitational pull of the resonant wake gives rise to a net torque. On the contrary, the non-resonant wake is a symmetric structure about the perturber arising from deformations of non-resonant stellar orbits and hence can not contribute to the torque. In the present work, we examine the geometry of density wakes in the isochrone core potential studied by KS18, which exhibits stalling of a perturber's inspiral at the filtering radius $r_\star$, well inside the galaxy core. Inside $r_\star$, the CR resonances are highly depleted, which leads to suppression of DF and effective stalling of the perturber near $r_\star$.

We aim to better understand the reasons for core stalling by studying the varied geometry of density wakes as the perturber falls into the inner core region.  In particular, we are interested in the non-CR resonances, which have been neglected in the past literature due to their typical subdominance to CR resonances.  Non-CR resonances may play a greater role in the special environment of a flat galactic core, as their usual competition becomes extremely weak inside $r_\star$. %is absent. 
The semi-analytical approach of this paper is well-suited to this problem, as non-CR resonances are often narrower and more distant, and thus easier to smear out in approximate numerical treatments of $N$-body gravity.

Later in the paper, we also attempt to find preliminary, yet useful, insights on the more complicated problem of multi-perturber interactions and their implications for core stalling. This is astrophysically important given the presence of many GCs in low-mass galaxies, like some dwarfs and ultra diffuse galaxies (UDGs). The Fornax dwarf spheroidal, which has motivated the core stalling problem for decades, has six GCs orbiting its core \citep{Buonanno1998, Strader2003,Wang2019,Pace2021}. Many UDGs have a cored mass distribution, and are usually orbited by tens of GCs \citep{vanDokkum2018,Forbes2018,DuttaChowdhury2019,Forbes2020}. $N$-body simulations studying core stalling in the case of multiple perturbers find that mutual interactions among GCs play an important role in their orbital evolution \citep{Inoue+09,DuttaChowdhury2019}. In this paper, we study in particular the secular dynamics of a globular orbiting on a circular orbit of radius $\sim r_\star$ in isochrone core, under the perturbing influence of an outer more massive infalling perturber. 

The setup of our astrophysical model is presented in \S~\ref{sec_phy_set}, including the mathematical formalism used to compute the linearized deformation to the background galactic density profile.  In \S~\ref{sec_2d_wakes} and \S~\ref{sec_3d_str} we present both 2D and 3D (respectively) results for the linearized density wakes produced by an inspiraling perturber. In \S~\ref{sec_non_CR_torq} we analyze the net LBK torque acting on a perturber,  emphasizing the novel role of torques produced by non-corotating resonances.  In \S~\ref{sec_LK_dynamics} we analyze the Lidov-Kozai-like secular interactions between two inspiraling perturbers. We conclude in \S~\ref{sec_conclusion}.

\section{Physical Set-up and Formalism}
\label{sec_phy_set}

We consider a globular cluster\footnote{We use the terms ``globular cluster" or ``perturber" equivalently throughout the paper.} of mass $M_p$ orbiting on a circular orbit of radius $r_p$ well inside the core of a spherical galaxy of total mass $M$. The galaxy is modelled as an isochrone potential with core radius $b \gg r_p$ and an ergodic distribution function \citep{Henon1959a,Henon1959b,Henon1960,BT08}. The orbital radius $r_p$ of the perturber shrinks due to the action of DF. The DF timescale is assumed to be much longer than the orbital timescale. Hence, we make the secular approximation to evaluate the linear density deformation, and at a given time $t$, we assume that the perturber orbits on a circle\footnote{In numerical simulations of core stalling, massive perturbers on initially eccentric orbits exhibit moderate but incomplete circularization before stalling \citep{Inoue+09}.  However, the far greater complexity of global models for eccentric DF means that we must defer an investigation of this for future work.}, centered on the galactic center, of \emph{fixed} radius $r_p$  with a constant orbital frequency $\Omega_p$. In addition, we do not take into account the self-gravity of the deformation
(see e.g. \citealt{Weinberg1989,Chavanis13} for a discussion).

\emph{Canonical coordinates}: Unperturbed orbits of stars in a spherical galaxy are planar rosettes described by plane polar coordinates $\{r, \psi\}$, where $r$ is the radial distance from galactic center and $\psi$ is the true phase of star measured, in the orbital plane, from  ascending node in the anti-clockwise sense.  The 3D physical space canonical coordinates are $\{ \bfr , \bfp \} \equiv $ $  \{r, \theta, \phi ; p_r , p_{\theta} , p_\phi \}$, where radial velocity $p_r = \dot{r}$, $p_\theta = r^2 \dot{\theta}$ and $p_\phi = r^2 \sin^2{\theta} \, \dot{\phi} = L_z$, the $z$-component of specific angular momentum. The magnitude of specific angular momentum of the star $L = \sqrt{ p_\theta^2 + p_\phi^2 / \sin^2{\theta} }$, and its specific energy $E = \{ p_r^2 + p_\theta^2/r^2 + p_\phi^2/(r^2 \sin^2{\theta}) \} /2 + \Phi_0(r)$, $\Phi_0(r)$ being the galaxy potential.  

It is desirable to pursue linear perturbation theory in action-angle (AA) coordinates $\{\bfI , \bfw\} \equiv \{ I,L,L_z , w,g,h \}$ of the spherical galaxy, which are defined as: 
\beq 
\begin{split}
& I = 2 J_r(E,L) + L  \; ; \qquad w = \frac{ \Omega_r}{2} (t - t_p)  \\
& L   \; ; \qquad \qquad g = \chi + \left( \Omega_{\psi} - \frac{\Omega_r}{2}  \right) (t - t_p)  \\ 
& L_z \; ; \qquad \qquad h  .
\end{split}
\label{AAs}
\eeq 
Here $J_r$ is the radial action, $t_p$ is the time since periapse passage, and $\Omega_r$ and $\Omega_\psi$ are the radial and azimuthal orbital frequencies in the plane of a stellar orbit. %The first action in our AA coordinate system ($I$) is 
The angle $w$ is the mean anomaly (i.e. mean orbital phase measured from periapse), $g$ the argument of periapse (measured from ascending node), and $h$ the longitude of ascending node. The above coordinates are specially chosen for a cored density profile, for which $\Omega_{\psi} \simeq \Omega_r / 2$. This makes $g$ a slowly varying angle, and its corresponding angular frequency $\Omega_g <\!\!< \Omega_w$. Hence a star can be considered as orbiting the unperturbed galaxy core on a centered, nearly-closed ellipse with mean orbital frequency $\Omega_w$.  The ellipse undergoes a slow apsidal precession with frequency $\Omega_g$.

\subsection{Unperturbed Galaxy}

Our spherical unperturbed galaxy is described by the isochrone potential:
\beq
\Phi_0 (r) = - \frac{ G M  }{ b + \sqrt{ b^2 + r^2  }  }
\eeq 
with core radius $b$. The resulting mass distribution can be written explicitly for an isochrone galaxy (see \S~3.1 of KS18),
\beq
M_0(r) = M \bigg[ \frac{r^3}{ ( b + \sqrt{b^2 + r^2}  )^2 \sqrt{b^2 +r^2}   }  \bigg] \, . 
\label{M0}
\eeq 
Here $M_0(r)$ is the mass of unperturbed galaxy enclosed within a sphere of radius $r$.
The specific orbital energy $E(I,L)$ for an unperturbed stellar orbit has the following explicit form in terms of actions:
\beq
E(I,L) = - \frac{2 (G M)^2}{[ I + \sqrt{ I_b^2 + L^2  } ]^2} \, ,
\label{E_exact}
\eeq
where $I_b = \sqrt{4 G M b}$. For the isochrone potential, physical coordinates can be analytically expressed in terms of AAs (see \S~3.5.2 of \citet{BT08}). But, for a region well inside the isochrone core ($r <\!\!< b$ or $I <\!\!< I_b$), the form of these relations is greatly simplified and real space coordinates can be described by following approximate relations (see \S~4.1 of KS18 for derivation):
\beq
\begin{split}
& r^2 \simeq \frac{I}{\Omega_b} \left[ 1 - e \cos{(2 w)} \right]  \\[1ex]
& \psi \simeq  g + \begin{cases} \arctan\left( \sqrt{\frac{1+e}{1-e}} \tan{w}  \right) \qquad \quad \! \mbox{for} \; w \in [0,\pi) \\
\pi + \arctan\left( \sqrt{\frac{1+e}{1-e}} \tan{w}  \right)\quad
\mbox{for} \; w \in [\pi,2\pi) 
\end{cases}
.
\end{split}
\label{core_orbits}
\eeq
Here $e = \sqrt{1 - L^2/I^2}$ is a measure of orbital eccentricity and $\sqrt{I/\Omega_b}$  is the mean-squared orbital radius, giving a measure of an average size of the orbit. See appendix~\ref{app_3d_orb} for expressions for 3D physical coordinates $\bfr$ in terms of AAs $\{ \bfI, \bfw \}$. Here $\Omega_b$ is the central azimuthal orbital frequency of galaxy, given as:
\beq
\Omega_b = \frac{1}{2} \sqrt{\frac{G M}{b^3}} \,. 
\eeq 
Similar to KS18, we define core stars with $I \leq I_{\rm max} = \epsilon I_b$, with $\epsilon = 0.1$. For our fiducial values of galaxy mass $M=1.6 \times 10^9 M_\odot$ and core radius $b=1~{\rm kpc}$ (as in KS18), this corresponds to a maximum mean squared radius of 632 pc for a core star. The unperturbed orbital frequencies have the following simplified expressions for core stars:
\beq
\Omega_w(I) = \frac{\p E}{\p I} \simeq \Omega_b \left( 1 - 3 \frac{I }{I_b }  \right) \; , \; \Omega_g(L) = \frac{\p E}{\p L} \simeq \Omega_b \left( \frac{ L }{ I_b  } \right) \, .
\eeq
These expressions are obtained by approximating $E(I,L)$ of equation~(\ref{E_exact}), in the limit $I,L <\!\!< I_b$, upto the first order in the small parameter $\epsilon$.  For this choice of parameters, the dynamical timescale associated with core stars is $\Omega_w^{-1} \simeq \Omega_b^{-1} \simeq 0.02 $ Gyr, and the long term apse precession timescale $\Omega_g^{-1} > ( \epsilon \Omega_b)^{-1}\simeq0.2$Gyr.   

Similar to KS18, we employ the ergodic phase space distribution function $f_0(\bfr,\bfp) = F_0(E)$ (see equation~(4.54) of \citet{BT08}) for our unperturbed isochrone galaxy, 
\begin{align}
& F_0(E) =  \frac{M}{\sqrt{2}\,(2\pi)^3\,(GMb)^{3/2}}
\frac{\sqrt{\scre}}{\left[\,2\left(1 - \scre\right)\,\right]^4}\; \times 
\nonumber\\[1ex]
& \qquad \bigg[ 27 - 66\scre + 320\scre^2 - 240\scre^3 + 64\scre^4 
\nonumber\\
& \qquad\quad + 3\left(16\scre^2 + 28\scre - 9\right)
\frac{\arcsin{\sqrt{\scre}}}{\sqrt{\scre(1-\scre)}}\,
\bigg]\,,
\label{df-iso}
\end{align}
where $\scre = -Eb/GM \in (0,1/2]$, which leads to an isotropic velocity dispersion.

\subsection{Perturber -- Globular Cluster}

We model the inspiraling globular cluster (GC) of mass $M_p$ as a Plummer sphere with core radius $a$. It follows a circular orbit of radius $r_p$, which adiabatically shrinks with time due to DF.  The perturbing potential felt by a star at position vector $\bfx$ (with respect to the galaxy center) is the tidal potential of the GC:
\beq
\Phi_p = - \frac{ G M_p }{ \sqrt{a^2 + | \bfx - \bfr_p |^2  } } + \frac{ G M_p (\bfx \centerdot \bfr_p)  }{(a^2 + r_p^2)^{3/2}  } \, .
\label{Phi_p}
\eeq 
The second term arises due to the choice of a non-inertial frame of reference, whose origin is set at the galaxy center that itself accelerates due to the GC.\footnote{This is analogous to the indirect term of disturbing function $\Phi_p$ in planetary dynamics \citep{MD99}.} 
The orbital frequency of the GC following its circular orbit is modelled as:
\beq
\Omega_p = \sqrt{  \frac{ G [M_0(r_p) + M_p] }{ r_p^3  } },
\eeq
where $M_0(r)$ is the mass of unperturbed galaxy enclosed within radius $r$; see equation~(\ref{M0}).

\subsection{Linear Perturbation Theory}
\label{sec_per_theo}

Here we evaluate linear deformations of the distribution function of our initially spherical galaxy under the perturbing potential exerted by a GC orbiting the galaxy on a \emph{quasi-stationary} circular orbit of radius $r_p$ (where the near stationarity is owed to long DF timescales). Following the secular approximation, the orbital radius $r_p$ (and hence the GC orbital frequency $\Omega_p(r_p)$) is considered to be constant in this analysis. Our frame of reference, centered at the galactic center, is the non-inertial, corotating rest frame of the GC, uniformly rotating with frequency $\Omega_p$. 
We take the $x,y-$plane to be coincident with the orbital plane of GC. Without loss of generality, we assume the GC to lie on the $x$-axis at $x=r_p$.  

The unperturbed galaxy is described by the ergodic distribution function $f_0(\bfr,\bfp) = F_0(E)$. In the rotating frame, the unperturbed Jacobi Hamiltonian $H_{J0} = E(I,L) - \Omega_p L_z$. In order to avoid transients in the galaxy's response, the perturbing potential $\Phi_1^{\rm ext}$ due to the GC is introduced in the system extremely slowly. At a time $t$, it is described as $\Phi_1^{\rm ext} = \exp{[\gamma t]} \Phi_p(\bfr) $, where $\gamma \gtrsim 0$ is an infinitesimally slow growth rate, so that $\Phi_1^{\rm ext}\rightarrow 0$ in the distant past as $t \rightarrow -\infty$. \footnote{Later, we will take the limit of $\gamma \rightarrow 0^{+}$ to get rid of this free parameter.}  Then, the linear deformation in galaxy distribution function can be written as $f_1 = \exp{[\gamma t]} F_1(\bfr , \bfp) $. 
The linear response of the galaxy is governed by the linearly perturbed collisionless Boltzmann equation (\S~2.1 KS18): 
\beq
\frac{\p f_1}{\p t} + [f_1 , H_{J0}] + [f_0, \Phi_1^{\rm ext}] = 0 \, ;   
\eeq   
here the polarization term from the self-consistent gravitational response of $f_1$ \citep{Chavanis13} is neglected for simplicity, as in most previous studies \citep{TremaineWeinberg84,KaurSridhar18}.  
Dividing out the time-dependent common factor $\exp{[\gamma t]}$, the above partial differential equation (PDE) gets reduced to following time-independent PDE:  
\beq
\gamma F_1 + [F_1, H_{J0}] + [F_0, \Phi_p] = 0 \,. 
\label{TI_PDE}
\eeq
It is straight-forward to solve this PDE in terms of the AAs $\{ \bfI, \bfw \}= \{  I,L,L_z ; w,g,h \}$ given by equation~(\ref{AAs}) in the unperturbed galaxy (also see \S~2.3 of KS18). $\Phi_p$ and $F_1$ can be formally written in terms of $\{\bfI,\bfw\}$ by employing equation~(\ref{gen_coord_expl}), and can then be Fourier expanded in the angles $\bfw$: 
\begin{align}
& F_1 = \sum_{\bfl} \Ftilda_{\bfl}(\bfI) \exp{[ \rmi \bfl \centerdot \bfw ]} \label{F1_expansion} \\[1 ex]
& \Phi_p = \sum_{\bfl} \Phitilda_{\bfl}(\bfI) \exp{[ \rmi \bfl \centerdot \bfw ]} \label{Phi_p_expansion} \,,
\end{align}
where $\bfl \equiv \{ n,\ell,m  \}$ and $\bfl \centerdot \bfw = n w + \ell g + m h$. We study the symmetry properties of the perturbing potential $\Phi_p$ and of its Fourier coefficients $\Phitilda_{\bfl}(\bfI)$ in appendix~\ref{app_Phi_nlm_properties}.   

Employing the above Fourier expansions in the PDE of equation~(\ref{TI_PDE}) leads to a linear algebraic relation in Fourier coefficients which can be solved for $\Ftilda_{\bfl}(\bfI)$: 
\beq
\begin{split}
& \Ftilda_{\bfl}(\bfI) \equiv  \Ftilda_{n \ell m }(I,L,L_z)\\
&= \left( n \frac{ \p F_0  }{\p I } + \ell \frac{\p F_0  }{\p L  }  \right) \frac{ \rmi \Phitilda_{n \ell m}  }{[\gamma + \rmi ( n \, \Omega_w + \ell \, \Omega_g - m \, \Omega_p )  ] } \,. 
\end{split}
\eeq 
Here the unperturbed orbital frequencies of a stellar orbit are $\Omega_w = {\p H_{J0}}/{\p I} = {\p E}/{\p I} $, $\Omega_g = {\p H_{J0}}/{\p L} = {\p E}/{\p L}  $ and $\Omega_h = {\p H_{J0}}/{\p L_z} = -\Omega_p $. Note that the choice of the rotating reference frame introduces a uniform retrograde nodal precession with frequency $\Omega_p$ for a general unperturbed stellar orbit. For an ergodic distribution function, $(n \, \p F_0/\p I + \ell \,  \p F_0/ \p L) = (n \Omega_w + \ell\Omega_g) \rmd F_0/\rmd E$. In the desired limit $\gamma \rightarrow 0^{+}$, the above expression for $\Ftilda_{\bfl}(\bfI)$ can be simplified by employing the Plemelj theorem to give:  
\beq 
\Ftilda_{\bfl}(\bfI)  =  (n \Omega_w + \ell \Omega_g) \frac{ \rmd F_0 }{\rmd E  } \Phitilda_{\bfl} \left[ \frac{1}{\bfl \centerdot \bfOmg} + \rmi \pi \delta{(\bfl \centerdot \bfOmg)}  \right] \,.
\label{F_tilda}
\eeq
Here we use the shorthand notations $\bfl \centerdot \bfOmg = n \Omega_w + \ell \Omega_g - m \Omega_p$ and $\Phitilda_{\bfl}(\bfI) \equiv  \Phitilda_{n \ell m }(I,L,L_z)$. 
Employing the above expression for $\Ftilda_{\bfl}(\bfI)$ in equation~(\ref{F1_expansion}) and using the fact that $\Phitilda_{\bfl}$ is real (from property {\bf P1} of appendix~\ref{app_Phi_nlm_properties}), the distribution function's linear deformation $F_1$ can be expressed as: 
\beq
\begin{split}
F_1(\bfI,\bfw) & = \frac{\rmd F_0}{\rmd E} \sum_{n,\ell,m}\!\! (n \Omega_w + \ell \Omega_g) \frac{\Phitilda_{\bfl}}{\bfl \centerdot \bfOmg } \cos{(\bfl \centerdot \bfw)} \\
& \quad - 2 \pi \frac{\rmd F_0}{\rmd E} \Omega_p(r_p)\!\! \sum_{n,\ell,m>0}\!\! m \Phitilda_{\bfl} \delta(\bfl \centerdot \bfOmg) \sin(\bfl \centerdot \bfw) \, .
\end{split}
\label{fin_F1}
\eeq
The second term has a Dirac delta function $\delta$ in the resonant condition $\bfl \centerdot \bfOmg = 0 $ (or $n \Omega_w + \ell \Omega_g = m \Omega_p$); so only stellar orbits which are in resonance with the perturber can contribute to the resonant distribution function deformation $F_{1, \rm res}$, i.e. the second term in the above expression. Note that resonances with $m=0$ do not contribute to $F_{1, \rm res}$. We refer to the first term, with a $\cos(\bfl \centerdot \bfw)$, as the non-resonant distribution function deformation $F_{1, \rm nr}$.

The geometry of the linear density deformations can be understood at a broad level from simple symmetry arguments.  The resonant term $F_{1,\rm res}$ is anti-symmetric under the angular transformation {\bf T0} (see appendix~\ref{app_Phi_nlm_properties} for definitions of the angular transformations used here), while the non-resonant part $F_{1,\rm nr}$ is symmetric under the same. Also, both $F_{1, \rm res}$ and $F_{1, \rm nr}$ are symmetric under transformations {\bf T1}, {\bf T2} and {\bf T3}. Thus, for given actions $\bfI$, $F_{1,\rm res}$ ($F_{1,\rm nr}$) will be anti-symmetric (symmetric) under the combined transformation, {\bf T0} and {\bf T2} (or {\bf T3}), which is equivalent to $\{ x,y,z  \} \rightarrow \{x,-y,z  \}$. Recalling that we have defined our coordinate system to place the perturber on the $x$-axis, resonant (non-resonant) orbits are thus deformed in an anti-symmetric (symmetric) manner on the leading and trailing sides of perturber. As a result, the gravitational pull from the mass contained in the resonant deformation $F_{1, \rm res}$ can exert a DF torque on perturber, but the gravitational pull from the non-resonant deformation cannot.        

Both $F_{1,\rm res}$ and $F_{1,\rm nr}$ are symmetric under the angular transformations {\bf T2} and {\bf T3}, which correspond to coordinate transformations $\{ x,y,z  \} \rightarrow \{x,y,-z  \}$ for fixed $\bfI$. Hence $F_{1,\rm res}$ and $F_{1,\rm nr}$ are symmetric above and below the orbital plane ($x,y$-plane) of perturber.

\subsection{Density deformation of galaxy}
\label{sec_density_deform}

To go beyond the aforementioned symmetry arguments, we can quantify the linear density deformation $\rho_1(\bfr)$ by integrating the distribution function deformation $F_1$ over velocity space $\bfv = \{ \dot{r} , r \, \dot{\theta} , r \sin{\theta} \, \dot{\phi} \} $ $= \{ p_r ,\, p_\theta / r , \, p_\phi / (r \, \sin{\theta}) \}$.  
The density deformation at $\bfr'$ can be expressed as:
\beq
\rho_1(\bfr') = \int \!\rmd^3 \bfv \, F_1(\bfr',\bfp) = \frac{1}{r'^2 \sin{\theta'}}\! \int\! \rmd^3\bfp \, \rmd^3\bfr \, F_1(\bfr , \bfp)\, \delta^3(\bfr - \bfr') \,.  
\eeq 
Integration variables can be conveniently transformed to AAs, owing to invariance of the phase space volume element under canonical transformations, i.e. $\rmd^3 \bfp \, \rmd^3 \bfr = \rmd^3 \bfI \, \rmd^3 \bfw $. We employ the Fourier expansion of $F_1$ of equation~(\ref{F1_expansion}) in the above expression, which gives, 
\beq 
\rho_1(\bfr') = \frac{1}{r'^2 \sin{\theta'}} \sum_{n \ell m} \int \rmd^3 \bfI \; \Ftilda_{n \ell m} (\bfI) \; \mathcal{I}_{\bfl}(\bfI,\bfr') \, ,
\label{rho1_in}
\eeq
where the integral $\mathcal{I}_{\bfl}$ is:
\beq
\mathcal{I}_{\bfl}(\bfI,\bfr') = \int \rmd^3 \bfw \; \exp{[\!\!\rmi \!\! (\bfl \centerdot \bfw ) ]} \, \delta^3(\bfr(\bfI,\bfw) - \bfr') \, .
\eeq
It is easier to evaluate the integral $\mathcal{I}_{\bfl}$ by transforming the integration variables to \bfr. The corresponding volume elements can be related as: 
\beq
\rmd^3 \bfw = \left| \frac{ \p \bfw }{ \p \bfr  }  \right| \rmd^3 \bfr \, . 
\eeq
 For the simple analytic form of equations~(\ref{core_orbits}) and (\ref{polar_ang}) in the isochrone core, we have $ \left| { \p \bfr  }/{ \p \bfw  }  \right| = \left| { \p r  }/{ \p w  } \; { \p \theta  }/{\p g  } \; {\p \phi  }/{\p h  }  \right| $ and the Jacobian simplifies to 
\beq
\left| \frac{ \p \bfw }{ \p \bfr  }  \right| =  \frac{ \Omega_b }{ I e  } \frac{ r \sin \theta }{ \sin{ i} \,| \cos{ \psi} \, \sin{(2 w)} | } \, .  
\eeq
 Hence the integral $\mathcal{I}_{\bfl}$ further simplifies to 
\beq
 \mathcal{I}_{\bfl}(\bfI,\bfr') =  \frac{ \Omega_b r' \sin{\theta'} }{ I e \sin{i}  } \int \rmd^3 \bfr \; \frac{ \exp{[\rmi \bfl \centerdot \bfw(\bfI,\bfr)]} }{ |\cos{\psi} \sin{(2 w)}|  } \delta^3(\bfr - \bfr') 
 \label{Il_integral_initial}
\eeq 
where the angles $\bfw(\bfI,\bfr)$ (and also the mean anomaly $w \equiv w(I,L,r)$, and the true phase in stellar orbital plane $\psi \equiv \psi(L,L_z,\theta)$ in the denominator) are expressed as functions of $\bfr = \{ r,\theta , \phi \}$ for a given $\bfI$ using transformation equations~(\ref{core_orbits}) and (\ref{polar_ang}). We design a scheme to evaluate functions $\bfw' \equiv \bfw(\bfI,\bfr')$ in appendix~\ref{app_8_cases} and find that there are four distinct combinations, {\bf A1} to {\bf A4}, for $\bfw'$ given in table~\ref{tbl:wgh_cases} (the first four rows). Hence, the integral $\mathcal{I}_{\bfl}$ becomes,     
\beq 
\begin{split}
 \mathcal{I}_{\bfl}(\bfI,\bfr') &=  \frac{ \Omega_b r' \sin{\theta'} }{ I e \sin{i}  } 
\; {\sum}' \frac{ \exp{[\rmi \{    n w' + \ell g' + m h' \}]} }{ |\cos{\psi'} \sin{(2 w')} | } \\
&= \, \frac{ \Omega_b r' \sin{\theta'} \sum' \exp{[\rmi \{ n w' +\ell g' + m h'  \}]}  }{ I e \sin{i} |\cos{\psi_1'} \sin{(2 w_1')} |    }  
\end{split}
\eeq
where summation $\,{\sum}' \, $ indicates the summation over 4 combinations of multivalued functions $(w', g' , h')$ as detailed in appendix~\ref{app_8_cases}. Here we use the fact that $|\cos{\psi'} \sin{(2 w')} |$ is equal for all 4 combinations and hence we can display it for {\bf A1} (say).

Using the above expression in equation~(\ref{rho1_in}), we have:
\beq
\begin{split}
\rho_1(\bfr') &=  \frac{\Omega_b }{2 r' } \sum_{n, \ell ,m } \int \rmd^3 \bfI \;  \frac{1}{ I e \sin{i} \, | \cos{\psi_1'} \sin{(2 w_1')}| } \\
& {\sum}' \{ \Ftilda_{\bfl}(\bfI) \exp{[\rmi \bfl \centerdot \bfw' ]} + \Ftilda_{-\bfl}(\bfI) \exp{[-\rmi \bfl \centerdot \bfw']}   \}
\end{split}
\eeq
The expression in parenthesis $``\{ \;  \}"$ can be simplified by employing equation~(\ref{F_tilda}) and using the fact that $\Phitilda_{\bfl}$ is real which implies $\Phitilda_{\bfl} = \Phitilda_{-\bfl}$ (see the property {\bf P1} of appendix~\ref{app_Phi_nlm_properties}),  
\beq
\begin{split}
& \{ \Ftilda_{\bfl}(\bfI) \exp{[\rmi \bfl \centerdot \bfw' ]} + \Ftilda_{-\bfl}(\bfI) \exp{[-\rmi \bfl \centerdot \bfw']}   \} = \\
& \quad 2 \frac{ \rmd F_0  }{ \rmd E } \, ( n \Omega_w + \ell \Omega_g ) \Phitilda_{\bfl} 
 \, \bigg[  \frac{   \cos{(\bfl \centerdot \bfw')}     }{ \bfl \centerdot \bfOmg  } 
 - \pi \, \delta(\bfl \centerdot \bfOmg) \,   \sin{( \bfl \centerdot \bfw' )} 
 \bigg] \,. 
\end{split}
\eeq
Using the above expression, we have the following final form of density deformation
\beq
\begin{split}
& \rho_1(\bfr') =  \frac{ \Omega_b }{r' }\! \sum_{\bfl} \int \!\! \rmd^3 \bfI \;  \,\frac{ ({ \rmd F_0  }/{ \rmd E }) ( n \Omega_w + \ell \Omega_g ) \Phitilda_{\bfl} }{I e \sin{i} \, | \cos{\psi_1'} \sin{(2 w_1')}| } \\
& \qquad \left[ \frac{\displaystyle{{\sum}'}\!\! \cos{(\bfl \centerdot \bfw')} }{ \bfl \centerdot \bfOmg } -\pi \delta(\bfl \centerdot \bfOmg){{\sum}'}\!\! \sin{(\bfl \centerdot \bfw')}  \right]
\end{split}
\label{rho1_fin}
\eeq
where the summation over $\bfl \equiv \{ n, \ell , m   \}$ is restricted to even-only and odd-only combinations %for 
due to 
property {\bf P2} of appendix~\ref{app_Phi_nlm_properties}. The second term in the density deformation (with the resonance condition inside the $\delta$-function) gets contributions only from stellar orbits in resonance with the perturber, and will be referred to as the ``resonant density deformation'' $\rho_{1,\rm res}$. The first term corresponds to the non-resonant part $\rho_{1,\rm nr}$. {The limits of the $\bfI$ integral correspond to $ I \geq \Omega_b {r'}^2/2 $, $ | 1- {r'}^2 \Omega_b /I | \leq e \leq 1$, $|\cos{i}| \leq \sin{\theta'}$, so that the unperturbed stellar orbits with given $\bfI$ can access the physical point $\bfr'$. We further restrict $I \leq I_{\rm max} = \epsilon I_b$ to account for only core stars. }

\begin{figure*}
\centering
\includegraphics[width=0.9\textwidth]{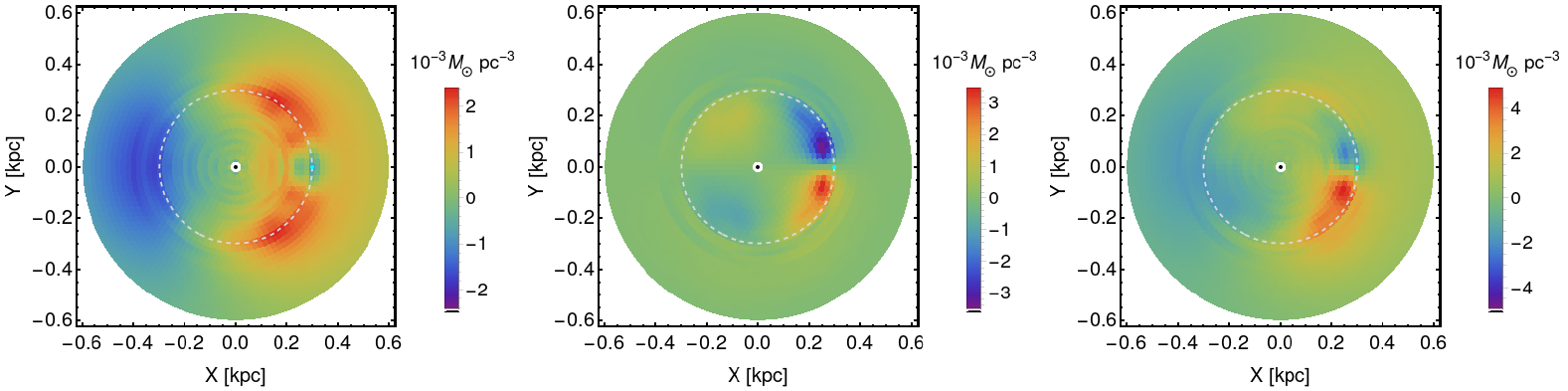}%,trim={5cm} 	
\caption{   Non-resonant (left), resonant (middle), and overall (right) density deformations (in units of $10^{-3} \Msun {\rm pc}^{-3}$) for a perturber located at $r_p = 0.3$kpc, in the perturber's rotating rest frame. These calculations include the full set of Fourier modes $\bfl$, with magnitudes of individual integers taken up to $\{ n, \ell, m\} \leq 10$. The circular orbit of the perturber (shown as a cyan dot) is represented as a dashed white circle.  This 2D slice shows density deformations within the perturber's orbital plane, and illustrates the basic geometry of dynamical friction wakes.  The non-resonant wake is symmetric about the $x$-axis and therefore does not produce net torque, while the resonant wake is antisymmetric about the $x$-axis and thus produces the total retarding torque.  At large perturber radii $r_p$, as is shown here, the net wake resembles the classical predictions of \citet{Chandrasekhar43}.}
\label{fig_wakes}
\end{figure*}

\subsubsection{Properties of $\mathbf{\rho_1}$}
Here we summarize the high-level properties of the linear density deformation $\rho_1$.

\begin{itemize}

\item[\bf S1] The resonant part $\rho_{1,\rm res}$ is anti-symmetric with respect to the $x$-axis ($\phi' \rightarrow - \phi'$), i.e. on the leading and trailing sides of perturber. The non-resonant part $\rho_{1,\rm nr}$ is symmetric under this transformation. 

As is evident from equation~(\ref{rho1_fin}), $\rho_1$ depends upon $\phi'$ only through the function $h'$; see equation~(\ref{g'_h'}). $\sum'$ is summed over the first four angular combinations A1, A2, A3 and A4 of table~\ref{tbl:wgh_cases}. As $\phi' \rightarrow - \phi'$, $\bfl \centerdot \bfw'_i \rightarrow \bfl \centerdot \bfw''_{i}$, with $i = 1,2,3,4$, satisfying these relations:
\begin{equation*}
    \begin{split}
& \bfl \centerdot \bfw''_1 = (n-m)\pi - \bfl \centerdot \bfw'_4 \; , \hspace{0.5cm}  \bfl \centerdot \bfw''_2  = (n-m)\pi -   \bfl \centerdot \bfw'_3  \\
& \bfl \centerdot \bfw''_3  = (n-m)\pi - \bfl \centerdot \bfw'_2 \; , \hspace{0.5cm} \bfl \centerdot \bfw''_4  = (n-m)\pi - \bfl \centerdot \bfw'_1  \,.
\end{split}
\end{equation*}
Since $(n-m)$ is an even integer (owing to the property {\bf P2} of appendix~\ref{app_Phi_nlm_properties}), $\rho_{1,\rm res}$ -- which contains $\sum' \sin{(\bfl \centerdot \bfw')}$ -- changes its sign under the transformation. On the contrary, $\rho_{1,\rm nr}$, which contains $\sum' \cos{(\bfl \centerdot \bfw')}$, remains invariant. 
\\

\item[\bf S2] For even (odd) integers $\bfl$, the non-resonant part of the density deformation (contributed by Fourier mode $\bfl$) is symmetric (anti-symmetric) with respect to the $y$-axis ($\phi' \rightarrow \pi - \phi'$), and vice versa for resonant part.

As $\phi' \rightarrow \pi - \phi'$, $\bfl \centerdot \bfw'_i \rightarrow \bfl \centerdot \bfw''_{i}$, with $i = 1,2,3,4$, satisfying these relations:
\begin{equation*}
    \begin{split}
& \bfl \centerdot \bfw''_1 = n\pi - \bfl \centerdot \bfw'_4 \hspace{2cm}  \bfl \centerdot \bfw''_2  = n\pi -   \bfl \centerdot \bfw'_3  \\
& \bfl \centerdot \bfw''_3  = n\pi - \bfl \centerdot \bfw'_2  \hspace{2cm} \bfl \centerdot \bfw''_4  = n\pi - \bfl \centerdot \bfw'_1  \,.
\end{split}
\end{equation*}  
Hence, the non-resonant part, containing $\sum' \cos{(\bfl \centerdot \bfw')}$, changes its sign by a factor $(-1)^n$; while the resonant part, containing $\sum' \sin{(\bfl \centerdot \bfw')}$, changes its sign by a factor $(-1)^{n+1}$.
\\

\item[\bf S3] Both $\rho_{1,\rm res}$ and $\rho_{1,\rm nr}$ are symmetric above and below the $x,y$-plane ($\theta' \rightarrow \pi - \theta'$).  

{$\rho_1$ depends upon $\theta'$ through functions $\psi_1'$, $g'$ and $h'$ (see equations~\ref{psi1'} and \ref{g'_h'}). Additionally, the dependence in integration limits appears only through $\sin{\theta'}$. As $\theta' \rightarrow \pi - \theta'$, \footnote{As $\theta' \rightarrow \pi - \theta'$, $\psi_1' \rightarrow - \psi_1'$ and $\zeta_1' \rightarrow - \zeta_1'$. } $\bfl \centerdot \bfw'_i \rightarrow \bfl \centerdot \bfw''_{i}$, with $i = 1,2,3,4$, satisfying these relations:
\begin{equation*}
    \begin{split}
& \bfl \centerdot \bfw''_1 = (m-\ell)\pi + \bfl \centerdot \bfw'_2 \; , \hspace{0.5cm}  \bfl \centerdot \bfw''_2  = -(m-\ell)\pi +   \bfl \centerdot \bfw'_1  \\
& \bfl \centerdot \bfw''_3  = (m-\ell)\pi + \bfl \centerdot \bfw'_4 \; , \hspace{0.5cm} \bfl \centerdot \bfw''_4  = -(m-\ell)\pi - \bfl \centerdot \bfw'_3  \,.
\end{split}
\end{equation*}
As $(m-\ell)$ is always an even integer, both $\rho_{1,\rm nr}$ and $\rho_{1,\rm res}$ are symmetric under this transformation. }

\end{itemize}

\section{2D Structure of Density wakes}
\label{sec_2d_wakes}

\begin{table}

\centering

\footnotesize

\begin{tabular}{l  l  l }
  \hline
  
 Quantity & Symbol & Value   \\
  
  \hline
  
Galaxy Mass &  $M$ & $1.6 \times 10^{9} \Msun$  \\ 

Galaxy core radius & $b$ & 1 kpc \\ 

Perturber Mass & $M_p$ & $2 \times 10^5 \Msun$ \\ 
 
Perturber softening & $(a/r_p)^2$ & $10^{-3}$  \\

Perturber orbital radius & $r_p$ & $\{0.225,0.26,0.3\}$kpc \\%[1ex]

 \hline
\end{tabular}
 
\caption{Parameter values chosen for numerical computation of wakes. } 
\label{tbl:parameters}

\end{table}

Using the formalism of \S \ref{sec_density_deform}, we now compute the geometric structure of linear deformations to the background density profile of the isochrone galaxy. The choice of parameters for numerical computation corresponds to KS18 
and are mentioned in table~\ref{tbl:parameters}. The numerical methods are described in appendix~\ref{sec_num_methods}. We present separately the non-resonant part $\rho_{1, \rm nr}$, resonant part $\rho_{1, \rm res}$, and the total density deformation $\rho_1 = \rho_{1,\rm nr} + \rho_{1,\rm res}$ in Figure~\ref{fig_wakes}.  These 2D results are shown in the orbital plane of the perturber in its rest frame, where it is stationary on the $x$-axis at $x = 0.3$kpc. The maximum magnitude of $\rho_1$ is smaller by roughly a factor $\sim  10^{-1}$ compared to the average unperturbed density inside the galaxy core ($\sim 0.04\Msun {\rm pc}^{-3}$). The non-resonant wake, as seen mathematically in previous sections, is symmetric about the $x$-axis with two over-densities sandwiching the perturber, on the leading and trailing side of its orbit. On the contrary, the resonant wake is anti-symmetric with respect to the $x$-axis, with a compact mass overdensity trailing behind the perturber, and a corresponding underdensity leading it. This simple decomposition illustrates geometrically why the antisymmetric resonant wake is the one that leads to a net DF torque on the perturber, and is thus responsible for its slow inspiral. In the remaining part of this section, we investigate the properties of the wakes more quantitatively, and present their dependence on perturber orbital radius $r_p$.      

\begin{figure*}
	\centering
	\includegraphics[width=1 \textwidth]{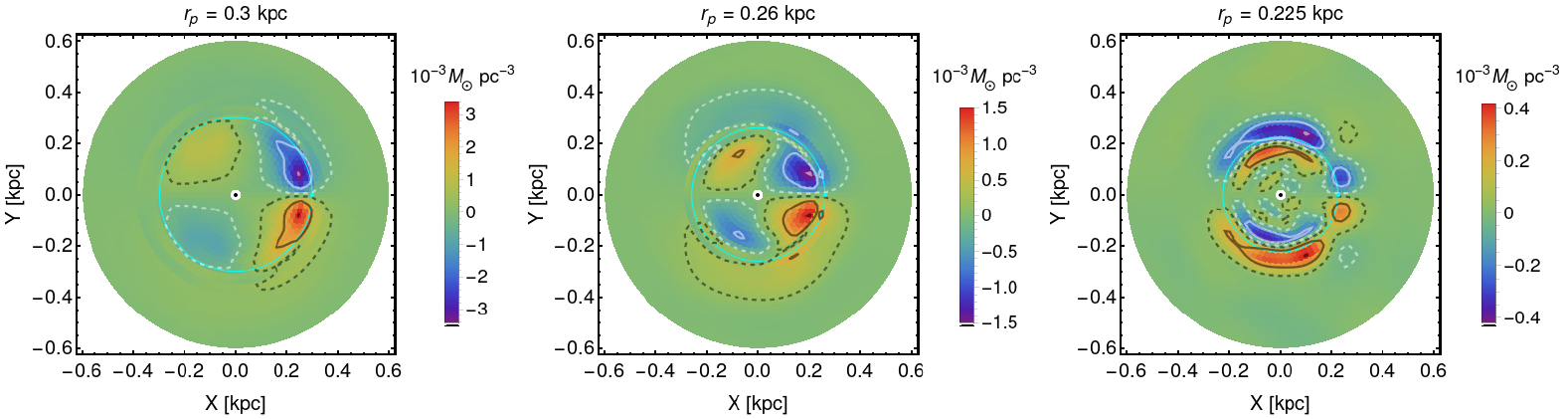}
\caption{ The resonant part $\rho_{1,\rm res}$ of the linear density deformation (in units of $10^{-3}\Msun {\rm pc}^{-3}$) for $r_p = 0.3$kpc ({\it left panel}), 0.26kpc ({\it middle panel}) and 0.225kpc ({\it right panel}).  Color-coded density perturbations are plotted in the rotating frame of the perturber, which is shown as a cyan dot (with its orbit as a cyan colored circle). Overdensities stand out in red, 
while underdensities are in blue. 
The point of maximum (minimum) deformation is shown as a black (white) dot. Solid black (white) contours corresponds to half the maximum (minimum) value of the resonant density deformation, while dashed contours refer to 1/10th of the maximum (minimum) strength. With decreasing $r_p$, the structure of the wake changes from a compact and strong overdensity trailing the perturber (analogous to that in the Chandrasekhar picture) to a more extended/global overdensity, with a more intricate structure and a weaker amplitude. The peak strength of the overdensity falls by roughly an order of magnitude as $r_p$ decays, which explains the suppression of DF torque near the filtering radius $r_{\star} = 0.22$kpc.   }
\label{fig_res_wake2d_rps}
\end{figure*}

\begin{figure*}
\centering
\includegraphics[width=1\textwidth]{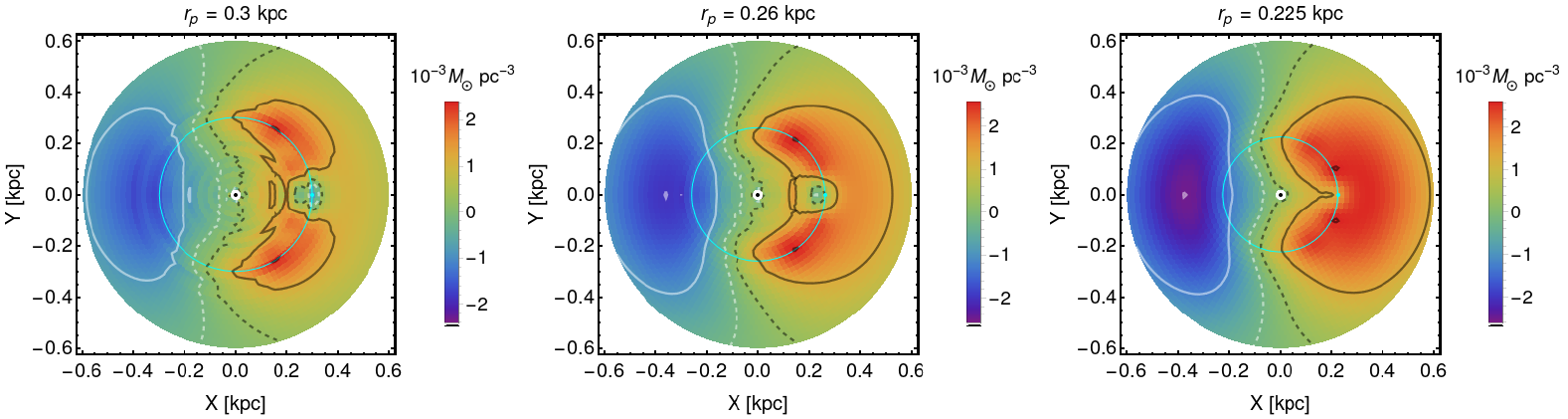}
\caption{ The non-Resonant part $\rho_{1,\rm nr}$ of the density deformation (in units of $10^{-3}\Msun {\rm pc}^{-3}$) for different values of $r_p$ in the perturber's rest frame; the figure format is the same as in figure~\ref{fig_res_wake2d_rps}. As $r_p$ decreases (from left to right), the non-resonant overdensities sandwiching the perturber at $r_p = 0.3$kpc merge to form a single radially extended overdensity at $r_p=0.225$kpc. The peak strength of the overdensity increases, but only a little, at small $r_p$.    }
\label{fig_nr_wake2d_rps}
\end{figure*}

\begin{figure*}
\centering
\includegraphics[width=1\textwidth]{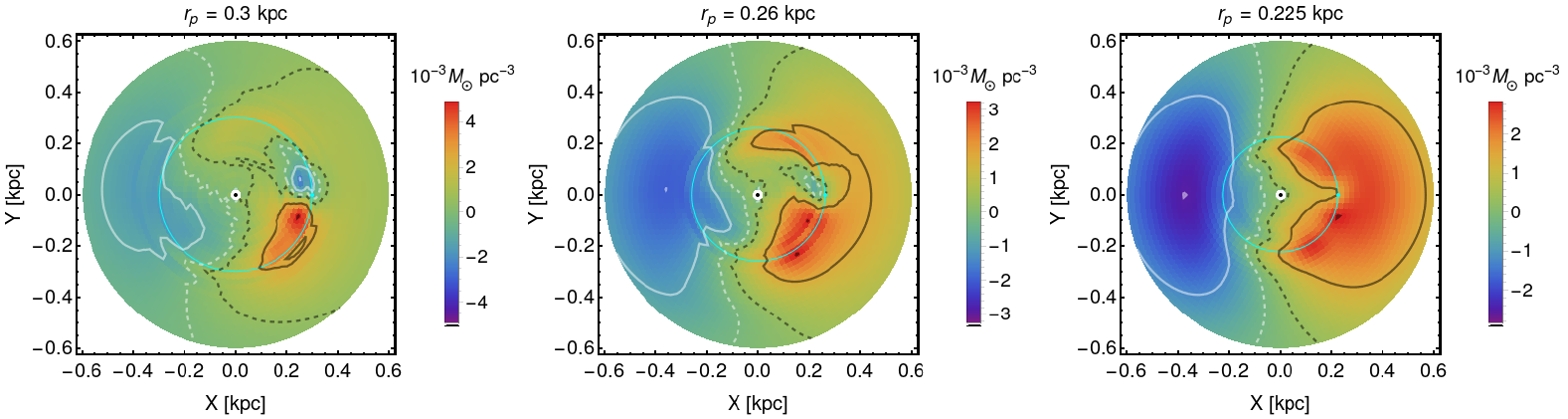}
\caption{ The overall density deformation $\rho_1 = \rho_{1,\rm res} + \rho_{1,\rm nr}$ (in units of $10^{-3}\Msun {\rm pc}^{-3}$) 
for different values of $r_p$ in the perturber's rest frame; the figure format is the same as in figure~\ref{fig_res_wake2d_rps}. As $r_p$ decreases, density wakes composed of a trailing overdensity at $r_p = 0.3$kpc assumes a roughly symmetric structure at $r_p=0.225$kpc, representing a smooth transition from a regime where the resonant wake dominates to one where the non-resonant wake dominates (as the perturber approaches filtering radius $r_{\star}$).   }
\label{fig_wake2d_rps}
\end{figure*}

\begin{figure*}
\centering
\includegraphics[width=1\textwidth]{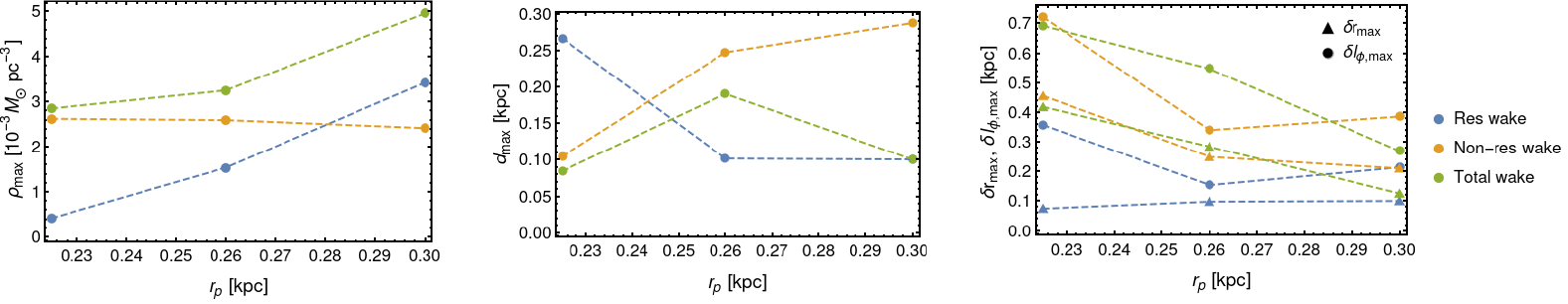}
\caption{Properties defining the 2D structure of wake overdensities (resonant $\rho_{1,\rm res}$ in \emph{blue}, non-resonant $\rho_{1,\rm nr}$ in \emph{yellow} and total $\rho_1$ in \emph{green}) are presented as function of orbital radius $r_p$ of the perturber, for the three representative values $r_p = 0.225, \, 0.26, \, 0.3$kpc studied here. \emph{Left panel}: the peak overdensity $\rho_{1\rm max}$ (in units of $10^{-3}\Msun {\rm pc}^{-3}$) decays by roughly an order of magnitude for $\rho_{1,\rm res}$, while changing only a little for $\rho_{1,\rm nr}$. \emph{Middle panel}: as $r_p$ shrinks, the distance $d_{\rm max}$ of the peak overdensity from the perturber (in kpc) increases for the resonant wake, which assumes a more extended structure for smaller $r_p$.  Conversely, it decreases for the non-resonant wake. \emph{Right panel}: the spatial extents (in kpc) of wakes around their respective maxima in both the radial direction, $\delta r_{\rm max}$ (filled triangles), and in the azimuthal direction $\delta l_{\phi,{\rm max}}$ (filled circles). With decreasing $r_p$, the resonant wake becomes more azimuthally extended (increasing $\delta l_{\phi,{\rm max}}$) while maintaining a roughly fixed $\delta r_{\rm max}$. On the contrary, the non-resonant wake extends in both directions with decreasing $r_p$.}
\label{fig_wake2dover_properties}
\end{figure*}

\begin{figure*}
\centering
\includegraphics[width=1\textwidth]{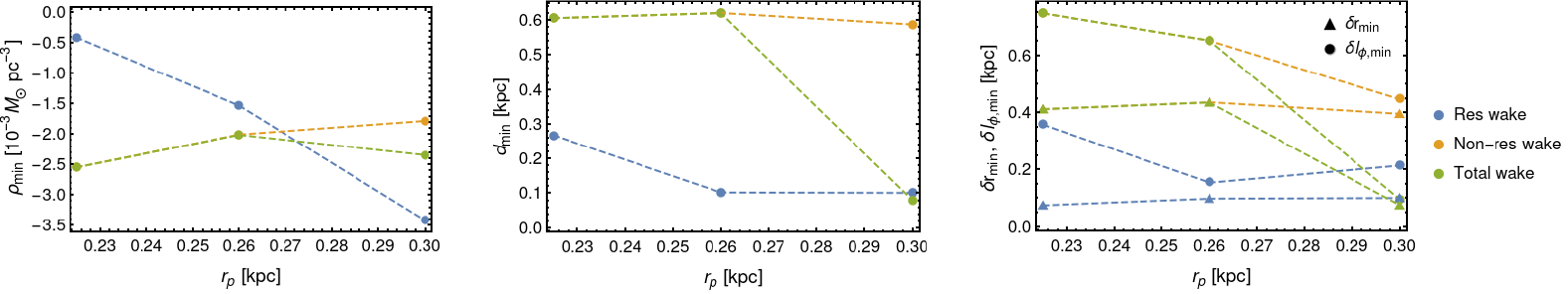}
\caption{The same as figure \ref{fig_wake2dover_properties}, but now showing properties defining the 2D structure of wake {\it underdensities}.  As before, the resonant part $\rho_{1,\rm res}$ is in \emph{blue}, the non-resonant $\rho_{1,\rm nr}$ in \emph{yellow}, and the total wake $\rho_1$ in \emph{green}. 
\emph{Left panel}: the strength of the minimum underdensity $\rho_{1 \rm min}$ (in $10^{-3}\Msun {\rm pc}^{-3}$) decreases significantly for $\rho_{1,\rm res}$ and increases only slightly for $\rho_{1,\rm nr}$ as $r_p$ decreases. \emph{Middle panel}: the distance $d_{\rm min}$ of the minimum point from the perturber (in kpc) increases with decreasing $r_p$ for $\rho_{1,\rm res}$, without a significant change for $\rho_{1,\rm nr}$. \emph{Right panel}: the radial $\delta r_{\rm min}$ (triangles) and azimuthal $\delta l_{\phi,{\rm min}}$ (circles) extents of wake underdensities, both in kpc, around the minimum point. Both the $\rho_{1,\rm res}$ and $\rho_{1,\rm nr}$ underdensities become more azimuthally extended without much change in their radial properties as $r_p$ decreases. }
\label{fig_wake2dunder_properties}
\end{figure*}

\subsection{Variation of wake structure with $\pmb{r_p}$}

As the perturber inspirals closer to the center due to DF, its perturbing effect on the orbits of background stars becomes more and more significant. We therefore investigate how the basic structure of wakes discussed above depends on the orbital radius $r_p$ of the perturber.  Throughout this and the next section on wake structure, we consider three fiducial values of $r_p = \{0.3,0.26,0.225\}$ kpc, lying  in a relatively narrow range of radii near the filtering radius $r_\star = 0.22$ kpc (for the current choice of parameters), as it is in this range where one observes the qualitative change in wake morphology that produces core stalling. The general defining equation of $r_\star$ is $\Omega_p(r_{\star}) = \Omega_b$ (KS18), i.e. at $r_{\star}$, orbital frequency of perturber $\Omega_p$ equates the core orbital frequency $\Omega_b$ of isochrone potential (which is the maximum orbital frequency attainable by a background star). 

\smallskip

\emph{Resonant Wakes}: For the larger orbital radii $r_p = 0.3$kpc and 0.26kpc, the resonant wake is composed of a dominant and compact trailing overdensity following behind the perturber, in addition to a weak overdensity far from the perturber on the leading side; see %left and middle panels of
figure~\ref{fig_res_wake2d_rps}. There are corresponding underdensities owing to the anti-symmetry of $\rho_{1,\rm res}$. The resonant wake is relatively compact, with spatial extents in the radial and azimuthal directions of roughly $\delta r \sim $ 0.1kpc, and $\delta l_{\phi} \sim$ 0.2kpc, respectively. 
We measure the \emph{size} of a wake structure by treating the density deformation contours, corresponding to half of the extremum value, as its boundaries. The wake's radial extent $\delta r$ is the length of the line segment joining the intersection points of a radial vector through the extremum with these contours.
Similarly, the azimuthal extent $\delta l_{\phi}$ about an extremum is the arc length joining the points of intersection of the circle of constant radius passing through the extremum point, and the boundary contours. For large values of $r_p$, the density extrema lie relatively close ($\sim$0.1kpc) to the perturber. Also, the dominant part of the resonant wake primarily lies inside the perturber's orbit. The peak strength of the overdensity falls roughly by half as $r_p$ decreases from 0.3kpc to 0.26kpc; this corresponds to diminishing DF torques for smaller $r_p$.  

An intriguing structural change emerges in resonant wakes as the perturber spirals in further to smaller radii.
For $r_p = 0.225$kpc, the resonant wake structure becomes more intricate (figure~\ref{fig_res_wake2d_rps}), with a dominant trailing overdensity lying just outside the perturber's circular orbit and a comparatively smaller leading overdensity just inside its orbit. These overdensities have corresponding underdensities, to respect the anti-symmetry of the resonant wake. The exterior wake is quite azimuthally extended, with $\delta l_{\phi}\sim $0.3-0.4kpc, but is also localized outside the orbital radius $r_p$ within a small radial extent, about $\delta r \sim$0.1kpc. Continuing the aforementioned trend, the peak overdensity has a magnitude smaller than its counterparts at larger $r_p$; it has decreased roughly by an order of magnitude when compared with $r_p=0.3$kpc. A more quantitative comparison of the structural properties of resonant wakes at these $r_p$ values is shown in figures \ref{fig_wake2dover_properties} and \ref{fig_wake2dunder_properties}.

As the perturber's orbital radius decays, with $r_p$ approaching $r_{\star}$, the resonant wake, apart from becoming weak, also assumes a more extended or global morphology that  
plays an important role in suppressing DF torque near $r_{\star}$. In addition, the compact, dominant wake that lies inside the perturber's orbit at large $r_p$ shifts to lie outside the orbit as the perturber reaches radii close to $r_{\star}$. This can be understood in light of the different types of resonant orbits that torque the perturber effectively at different $r_p$, as we discuss in more detail in appendix~\ref{sec_lbk_tor_comp}. At larger $r_p$ values, corotating resonant (CR) orbits with smaller size contribute dominantly to DF torque; while at small $r_p$s, non-CR orbits of larger size and eccentricity have a dominant relative contribution to the torque; see figures~\ref{fig_res_lines}-\ref{fig_torq_st_ecc}.

%\smallskip

\emph{Non-resonant Wakes}: For an orbital radius $r_p=0.3$kpc, the non-resonant wake is composed of two identical overdensities sandwiching the perturber along its orbit; see figure~\ref{fig_nr_wake2d_rps}. Each of these structures is quite extended both radially ($\sim$0.2kpc) and azimuthally($\sim$0.4kpc). As $r_p$ decreases to 0.26kpc, these overdensities merge together to form a single symmetric horseshoe-shaped structure enclosing a small underdense hole around the perturber. As the perturber's orbit decays further to $r_p =0.225$kpc, this underdense hole shrinks, giving a cashew-nut shaped appearance to the overdensity. To summarize: as $r_p$ shrinks from 0.3kpc to 0.225kpc, the non-resonant wake overdensity assumes a more spatially extended structure, owing to the greater perturbing influence of the massive perturber on the background stars with an enclosed galactic mass $M_0(r_p)$ that has decreased. A dominant part of this structure lies outside the orbit of perturber, especially for small $r_p$. The strength of the overdense peak increases only slightly with decreasing $r_p$. Figures~\ref{fig_wake2dover_properties} and \ref{fig_wake2dunder_properties} compare, respectively, the properties of overdensities and underdensities associated with non-resonant wakes at these different $r_p$ values.\footnote{The dominant underdensity enclosing the point of the minimum non-resonant density deformation is chosen for $r_p=$0.3, 0.26kpc rather than small underdensity close to the perturber. For $r_p=0.3$kpc, the minimum point actually occurs outside the perturber's orbital plane, as will be discussed in \S~\ref{sec_3d_str}. } In general, non-resonant wakes are much more spatially extended than resonant wakes. \footnote{For azimuthal extent $\delta l_{\phi,\rm max}$ of the horse-shoe overdensity at $r_p =0.26$kpc, the arclength covered in the underdensity hole is already subtracted. For $\delta l_{\phi,\rm max}$ of the non-resonant wake at $r_p=0.3$kpc, only one of the overdensities is considered (figures~\ref{fig_wake2dover_properties}).}

%\smallskip

\emph{Net density wakes}: For the larger orbital radius of $r_p = 0.3$kpc, the net wake, which is the sum of the resonant and non-resonant wakes, can be interpreted as a trailing overdensity following behind the perturber (similar to its resonant part); see figure~\ref{fig_wake2d_rps}. This structure is relatively compact, with a radial extent close to 0.1kpc and an  azimuthal extent 0.2-0.3kpc. As the perturber inspirals to an orbital radius $r_p=0.26$kpc, the overdensity becomes more extended and diffuse, and morphs into an asymmetric horse-shoe structure (similar to its non-resonant part, which is a symmetric horse-shoe) with the trailing arm dominant. For the smaller $r_p=0.225$kpc, the overdensity assumes a roughly symmetric cashew-nut structure (similar to its non-resonant component, which is an exactly symmetric cashew-nut), with a slightly dominant trailing arm. In summary, we see a shift from a regime dominated by the resonant wake (at larger $r_p$) to one dominated by the non-resonant wake (at smaller $r_p$); this shift begins slightly outside of $r_{\star}$. 
The relatively rapid change in the structure of the net wake from a predominantly anti-symmetric trailing wake (for larger $r_p$) to a more-or-less symmetric structure (for smaller $r_p$), happens over a narrow range of perturber radii, {\it and this increase in reflection symmetry is the cause of the diminished DF torque at small $r_p$}. The strength of the peak overdensity decreases roughly by half as $r_p$ decreases throughout this radial range of $r_p$ considered, so the loss of DF torque is not due to a change in the overall amplitude of the net density perturbation. Figures~\ref{fig_wake2dover_properties} and \ref{fig_wake2dunder_properties} compare the properties of the net overdensities\footnote{For $r_p=0.26$kpc, there exist two points of maxima for $\rho_1$ as shown in figure~\ref{fig_wake2d_rps}. Hence, in order to evaluate various quantities associated with the wake overdensity shown in the figure~\ref{fig_wake2dover_properties}, we choose the mid-point of the line joining these points of maxima. Also at $r_p=0.26$kpc, the azimuthal stretch of overdensity $\delta l_{\phi,{\rm max}}$ excludes the small underdensity close to the perturber.  
} and underdensities\footnote{For $r_p=0.3$kpc, we consider the smaller underdensity, enclosing the point of minimum net density deformation, which is also close to the perturber. However, the dominant underdensity at $r_p=0.26$kpc is the large structure enclosing the minimum lying on the $x$-axis, on the side of the galaxy opposite to the perturber.}, respectively.

These density deformation computations take into account Fourier modes of integer set $\bfl$, with magnitudes of each integer $\leq 10$. We investigate the effect of higher order terms in appendix~\ref{sec_hi_res}. There we find that the higher order contributions intensify the wake strength close to the perturber without changing its overall qualitative structure significantly; see figures~\ref{fig_wake_2d_hi} and \ref{fig_wakes_fourier_comp} for more details.

\section{3D Structure of Density Wakes}
\label{sec_3d_str}

\begin{figure*}
\begin{subfigure}{1.\textwidth}
\centering
\includegraphics[width=0.9 \textwidth]{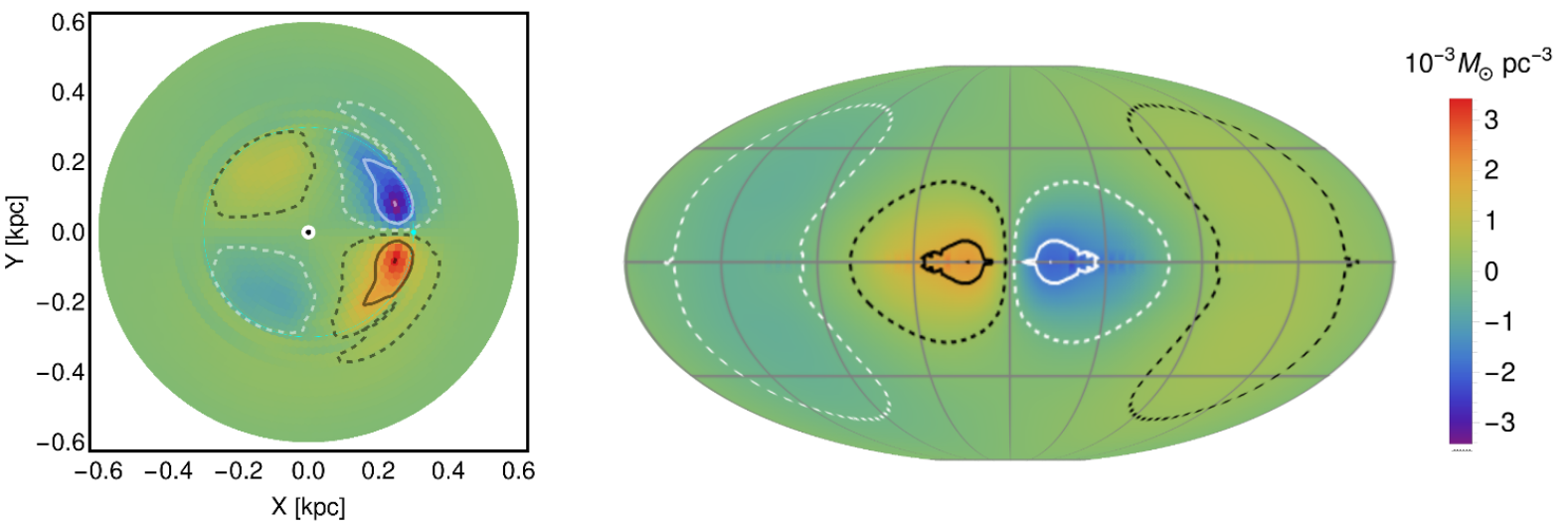}
\subcaption{$r_p = 0.3$kpc}

\end{subfigure}
\begin{subfigure}{1.\textwidth}
\centering
\includegraphics[width=0.9 \textwidth]{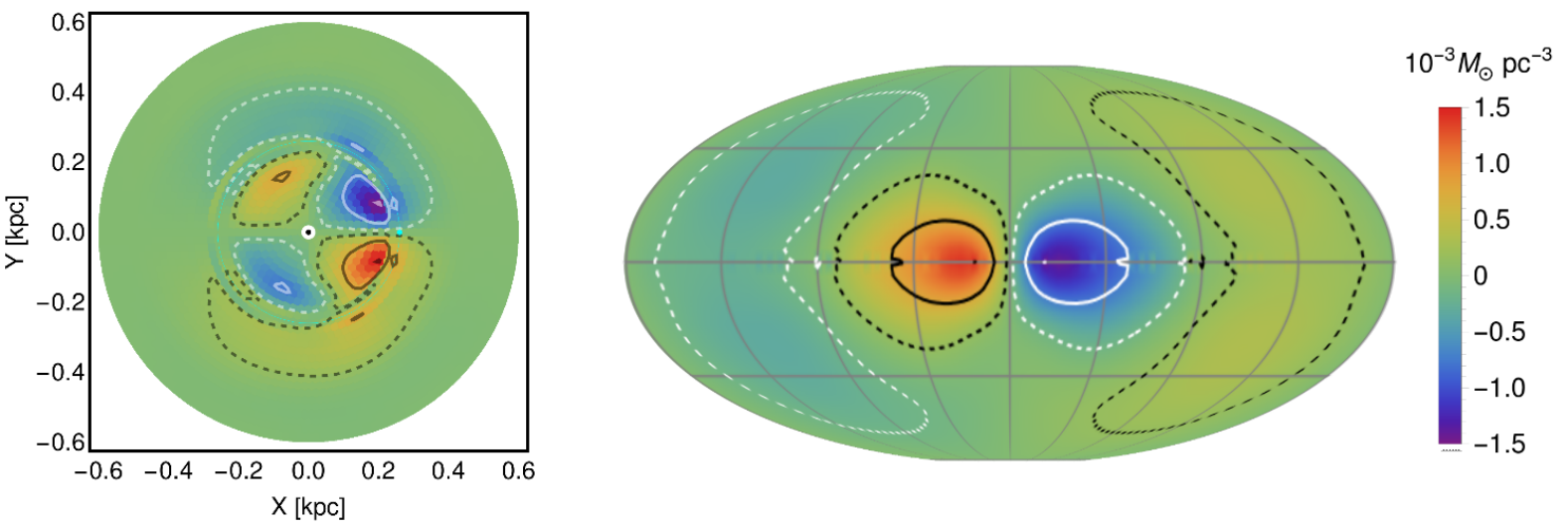}%,trim={5cm 9.5cm 1cm 6cm}
\subcaption{$r_p = 0.26$kpc}
%\label{ilr}
\end{subfigure}
\begin{subfigure}{1. \textwidth}
\centering
\includegraphics[width=0.9 \textwidth]{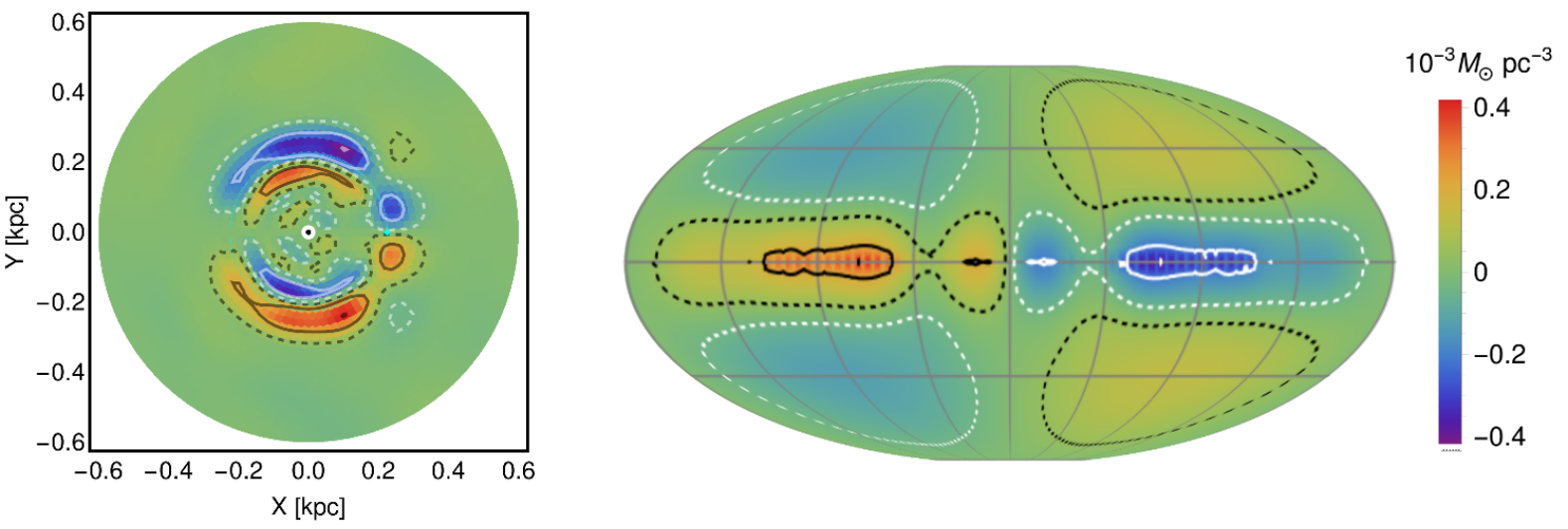}%,trim={5cm 9.5cm 1cm 6cm}
\subcaption{$r_p = 0.225$kpc}
%\label{ilr}
\end{subfigure}
\caption{ 3D Structure of resonant wakes for various $r_p$ values (shown in different rows). \emph{Left panel}: $\rho_{1,\rm res}$ in the orbital plane of perturber in its rest frame (same as what is shown in figure~\ref{fig_res_wake2d_rps}). \emph{Fight panel}: Mollweide projections (with the observer lying on positive $x$-axis such that the central longitude corresponds to the $z$-axis) of $\rho_{1,\rm res}$ on the sphere of  radius $r_{\rm res, max}$, on which the point of the maximum (and also the minimum) $\rho_{1,\rm res}$ lies. Note that $r_{\rm res, max} = 0.26$, 0.22, 0.26kpc for $r_p=$ 0.3, 0.26, 0.225kpc respectively. The color scheme of the contours is the same as in figure~\ref{fig_res_wake2d_rps}. At small $r_p$ ($0.225$kpc), the resonant wake overdensity assumes an azimuthally elongated and global trailing structure significantly weaker in strength, in contrast to a compact and strong overdensity trailing just behind the perturber at larger $r_p$ ($0.3,0.26$kpc). The resonant overdensity maintains its compactness in the radial and vertical directions without significant change throughout this range of $r_p$.           }
\label{fig_res_moll}
\end{figure*}

\begin{figure*}
\begin{framed}
\begin{subfigure}{0.9\textwidth}
\centering
\includegraphics[width=0.8 \textwidth]{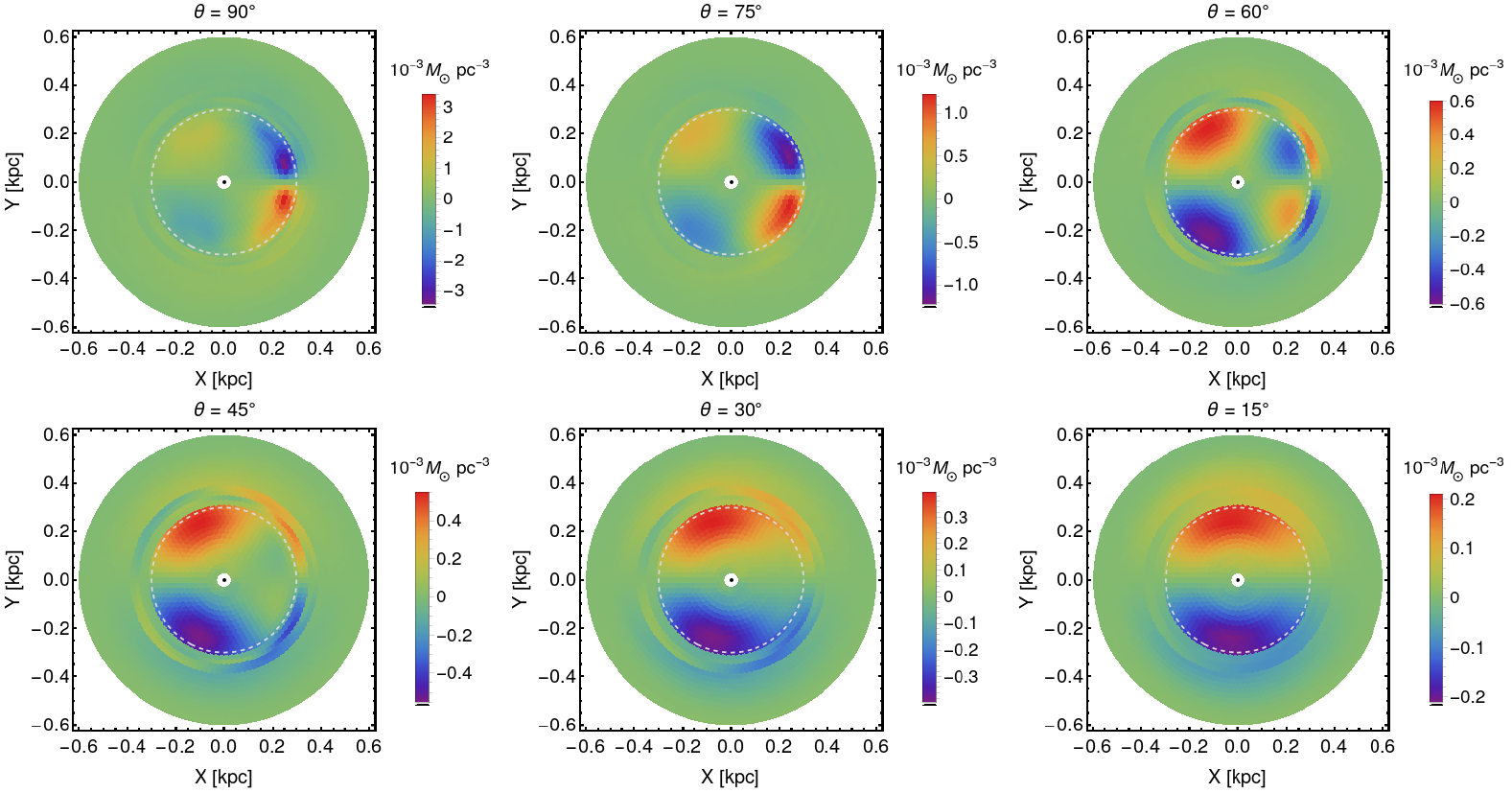}%,trim={5cm 9.5cm 1cm 6cm}
\subcaption{$r_p = 0.3$kpc}
%\label{ilr}
\end{subfigure}

%\hline
\begin{subfigure}{0.9\textwidth}
\centering
\includegraphics[width=0.8 \textwidth]{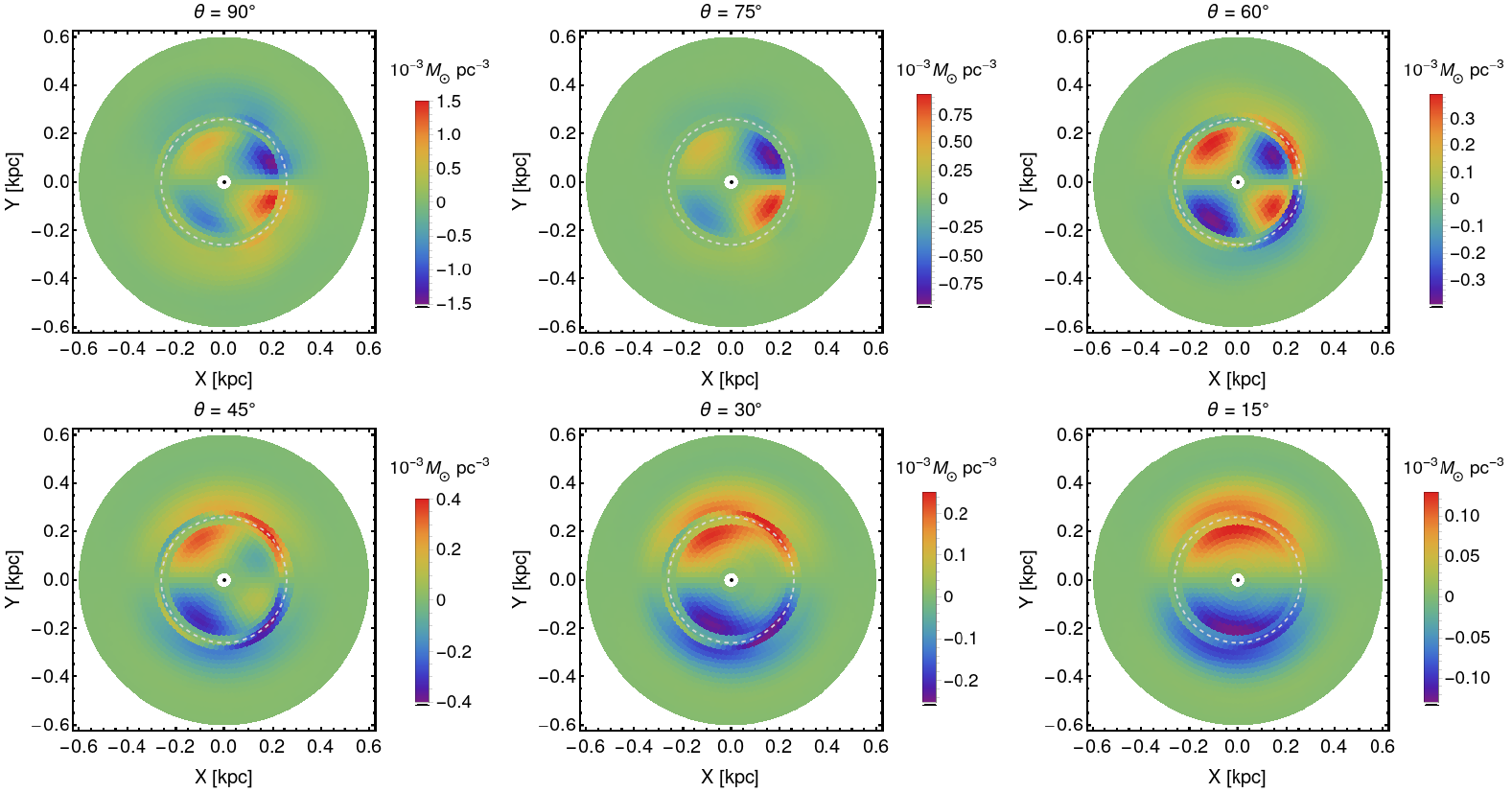}%,trim={5cm 9.5cm 1cm 6cm}
\subcaption{$r_p = 0.26$kpc}
%\label{ilr}
\end{subfigure}
%\hline

\begin{subfigure}{0.9\textwidth}
\centering
\includegraphics[width=0.8 \textwidth]{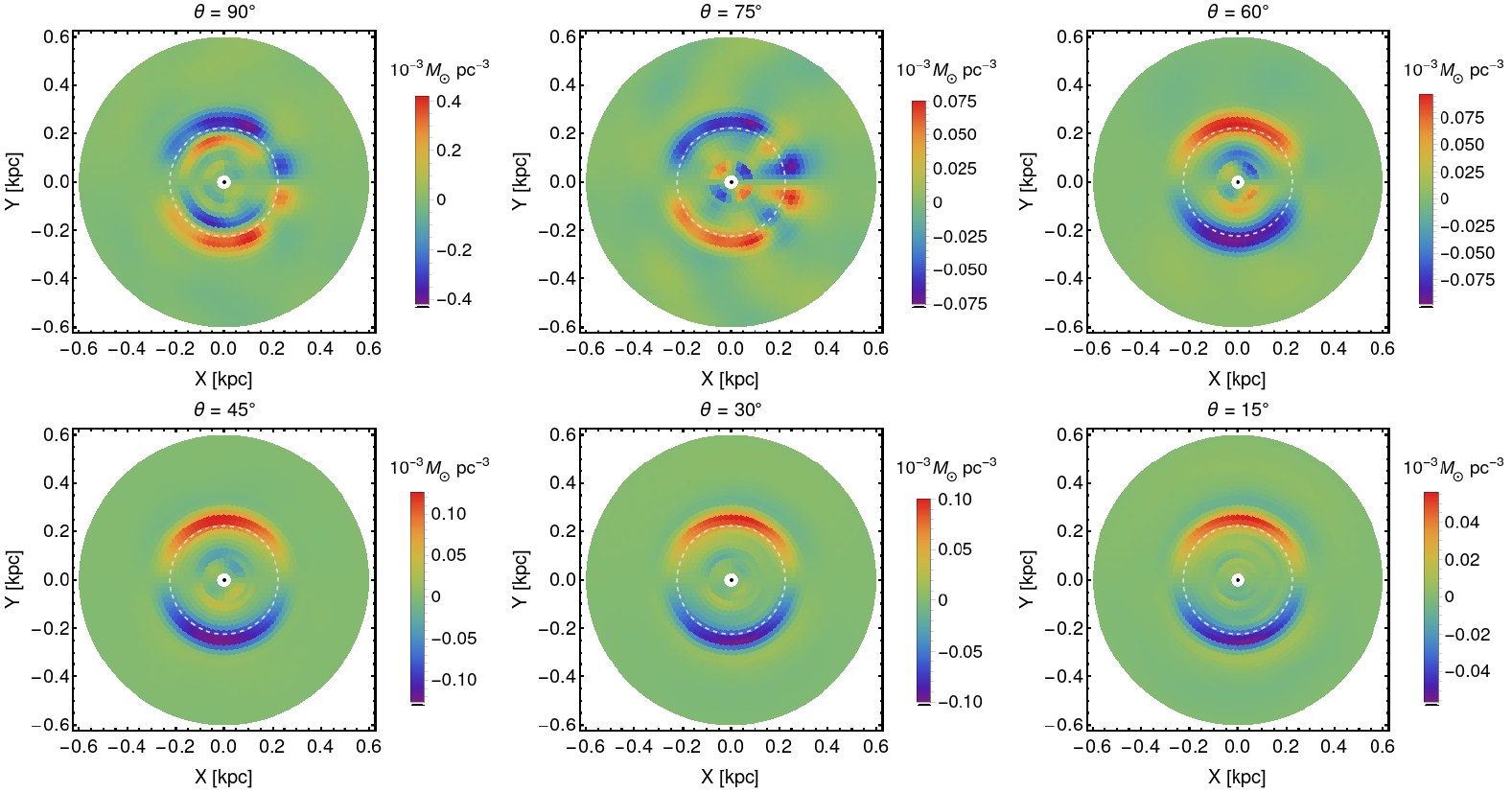}%,trim={5cm 9.5cm 1cm 6cm}
\subcaption{$r_p = 0.225$kpc}
%\label{ilr}
\end{subfigure}
\end{framed}
\caption{
Resonant wakes ($10^{-3}\Msun {\rm pc}^{-3}$) in the perturber's rest frame for three $r_p$ values are shown in three sets. The first (top left) panel of each set shows the wake in the plane of perturber's orbit (i.e. for a colatitude $\theta = 90^{\circ}$), as in figure \ref{fig_res_wake2d_rps}. All other panels showcase unfolded cones with constant $\theta < 90^\circ$, such that the radial distance $r$ and azimuthal angle $\phi$ shown in 2D figures match with the true 3D values. At all $r_p$, there is a relatively strong trailing overdensity (and leading underdensity) for larger $\theta$ near perturber's orbital plane. This trend reverses at smaller $\theta$, away from the perturber's orbital plane.  At these high latitudes, a weak leading overdensity (and trailing underdensity) emerges.
}
\label{fig_res_cones}
\end{figure*}

\begin{figure*}
\begin{subfigure}{1.0\textwidth}
\centering
\includegraphics[width=0.9 \textwidth]{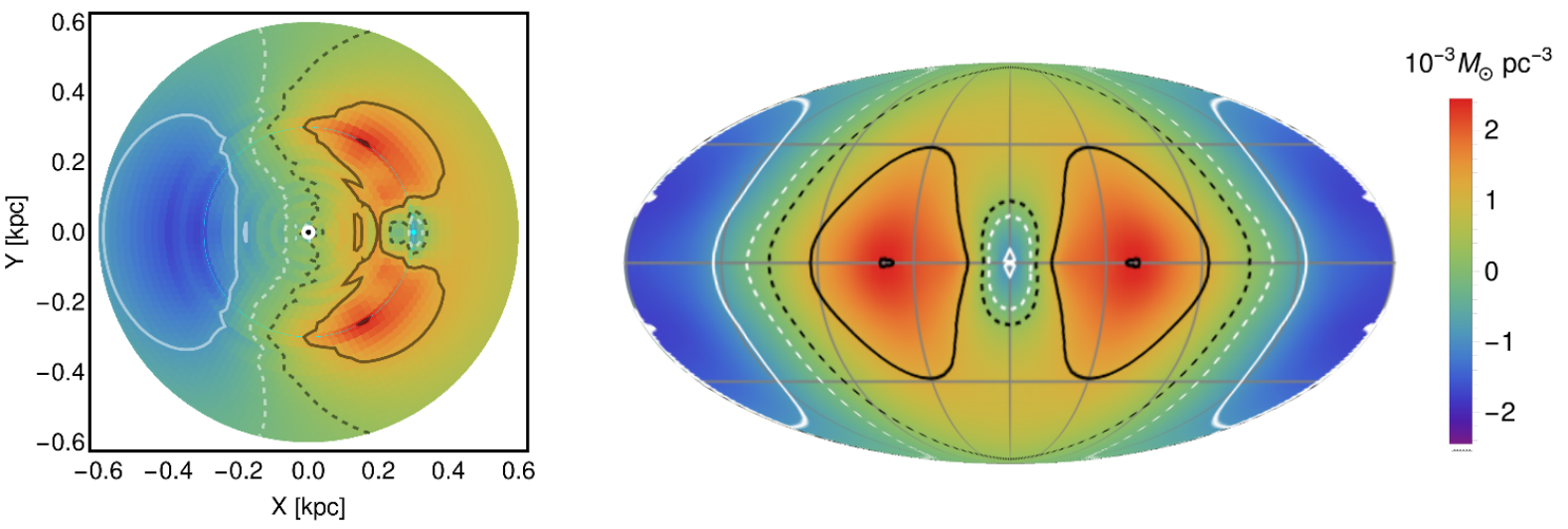}%,trim={5cm 9.5cm 1cm 6cm}
\subcaption{$r_p = 0.3$kpc}
%\label{ilr}
\end{subfigure}
\vspace{1em}
\begin{subfigure}{1.0\textwidth}
\centering
\includegraphics[width=0.9 \textwidth]{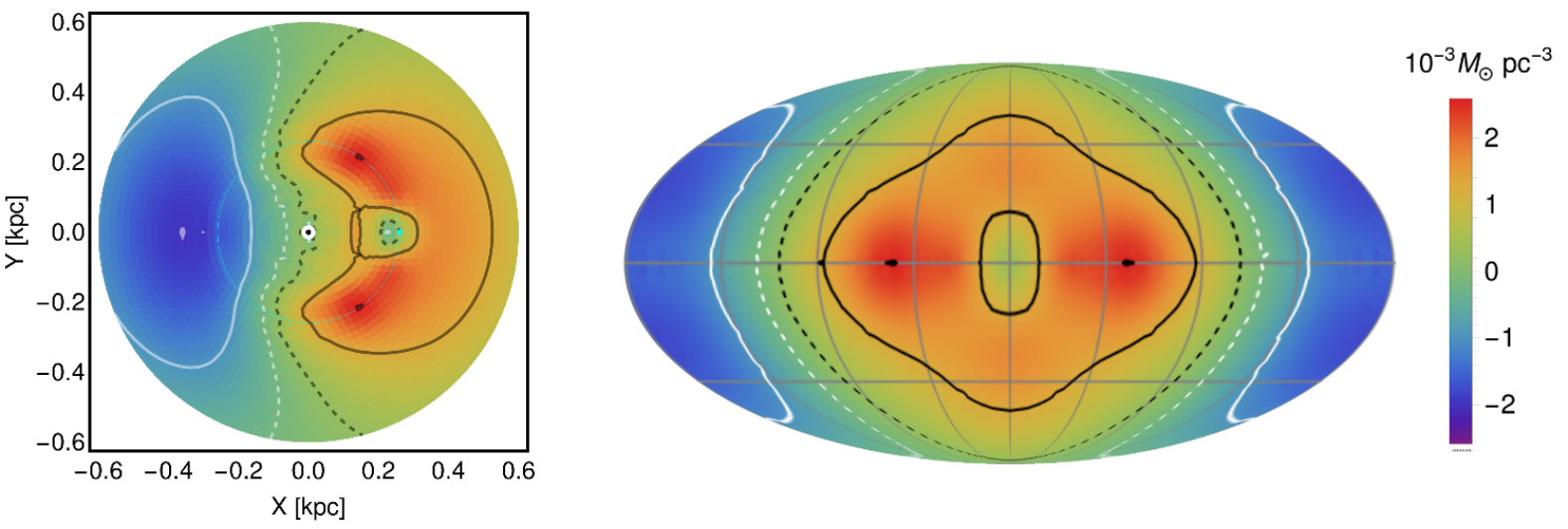}%,trim={5cm 9.5cm 1cm 6cm}
\subcaption{$r_p = 0.26$kpc}
%\label{ilr}
\end{subfigure}
\\
\begin{subfigure}{1.0 \textwidth}
\centering
\includegraphics[width=0.9 \textwidth]{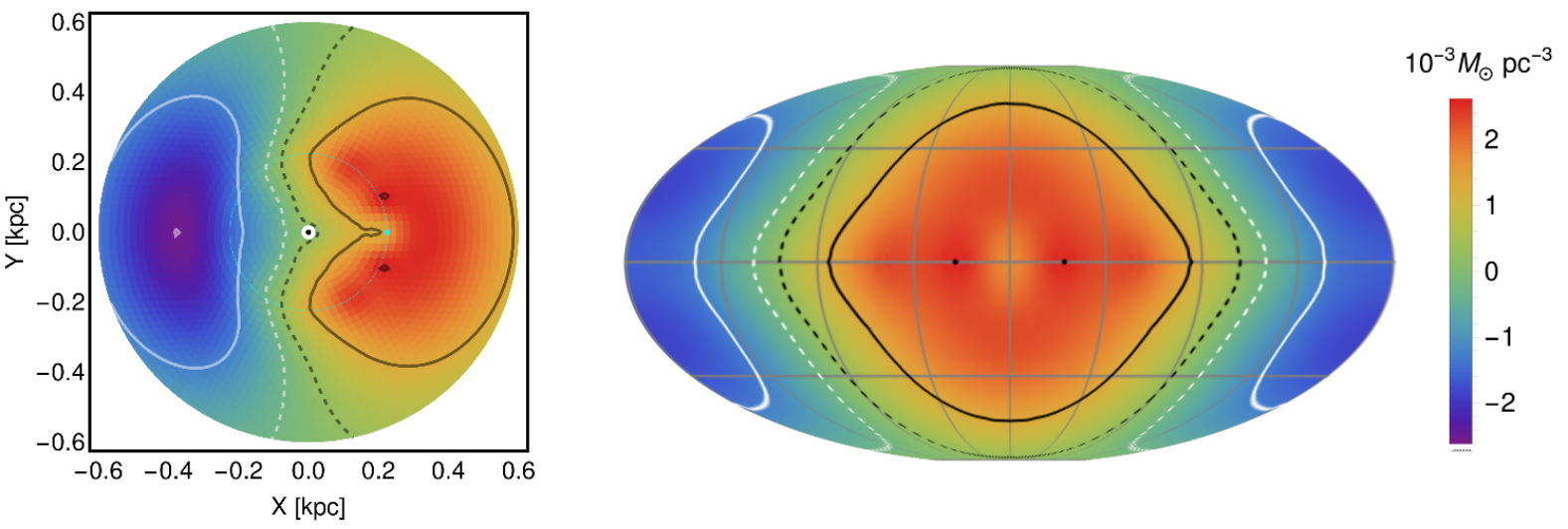}%,trim={5cm 9.5cm 1cm 6cm}
\subcaption{$r_p = 0.225$kpc}
%\label{ilr}
\end{subfigure}
\caption{3D structure of non-resonant wakes shown for various $r_p$ values, with the format the same as in figure \ref{fig_res_moll}. \emph{Left panel}: $\rho_{1,\rm nr}$ in the plane of the perturber in its rest frame (same as figure~\ref{fig_nr_wake2d_rps}). \emph{Right panel}: Mollweide projections of $\rho_{1,\rm nr}$ on the sphere of radius $r_{\rm nr, max}$ on which the point of the maximum $\rho_{1,\rm nr}$ lies. Note that $r_{\rm nr, max} = 0.3$, 0.26, 0.24kpc for $r_p=$ 0.3, 0.26,  0.225kpc, respectively. 
The non-resonant wake overdensity maintains an extended and high-amplitude structure throughout the considered range of $r_p$.  The two overdensities that sandwich the perturber at $r_p=0.3$kpc eventually merge together at smaller $r_p$, giving rise to a single overdensity. }
\label{fig_nr_moll}
\end{figure*} 

\begin{figure*}
\begin{framed} 
\begin{subfigure}{0.9\textwidth}
\centering
\includegraphics[width=0.8 \textwidth]{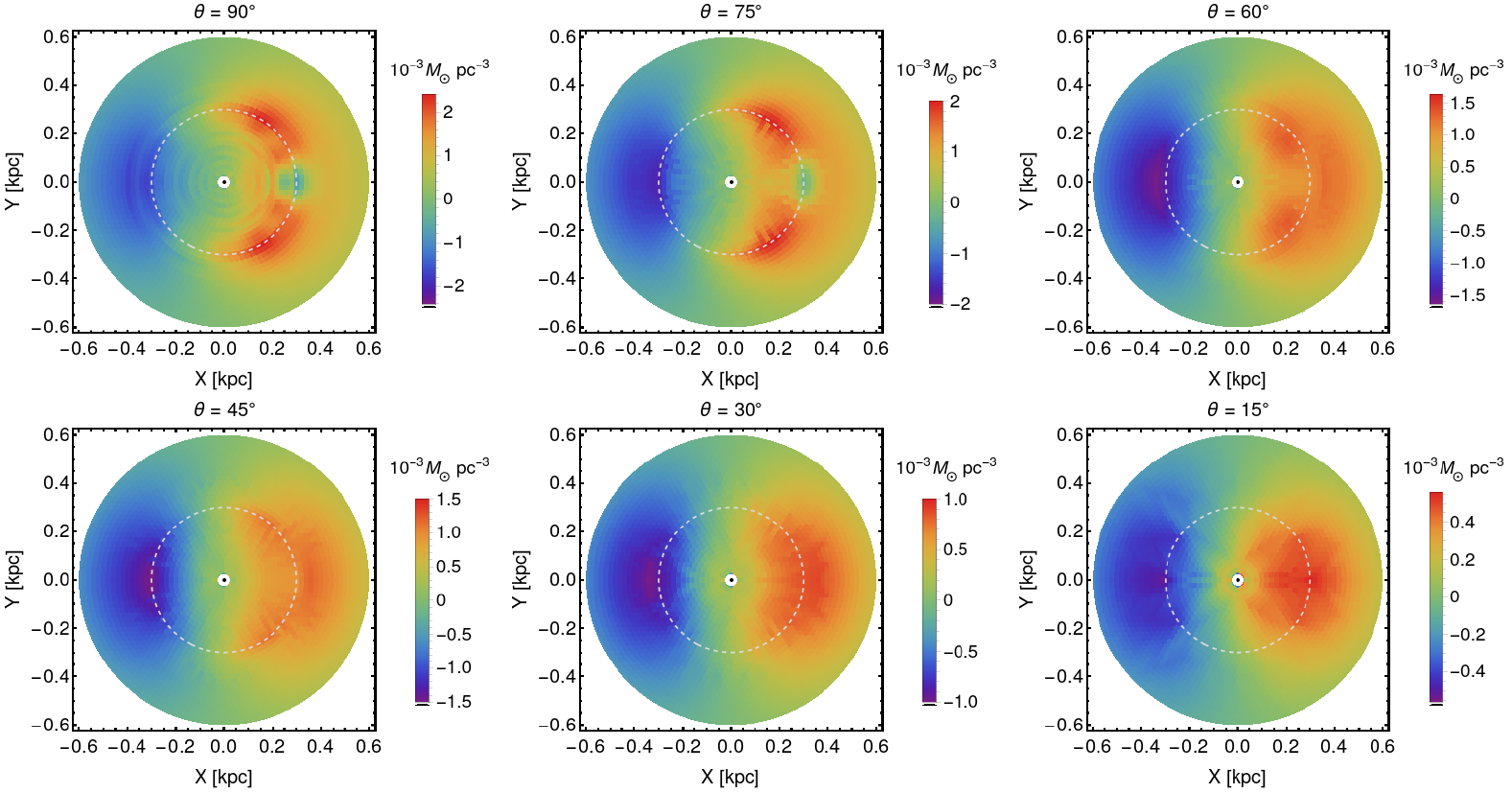}%,trim={5cm 9.5cm 1cm 6cm}
\subcaption{$r_p=0.3$kpc}
%\label{ilr}
\end{subfigure}
%\hline
\\
\begin{subfigure}{0.9\textwidth}
\centering
\includegraphics[width=0.8 \textwidth]{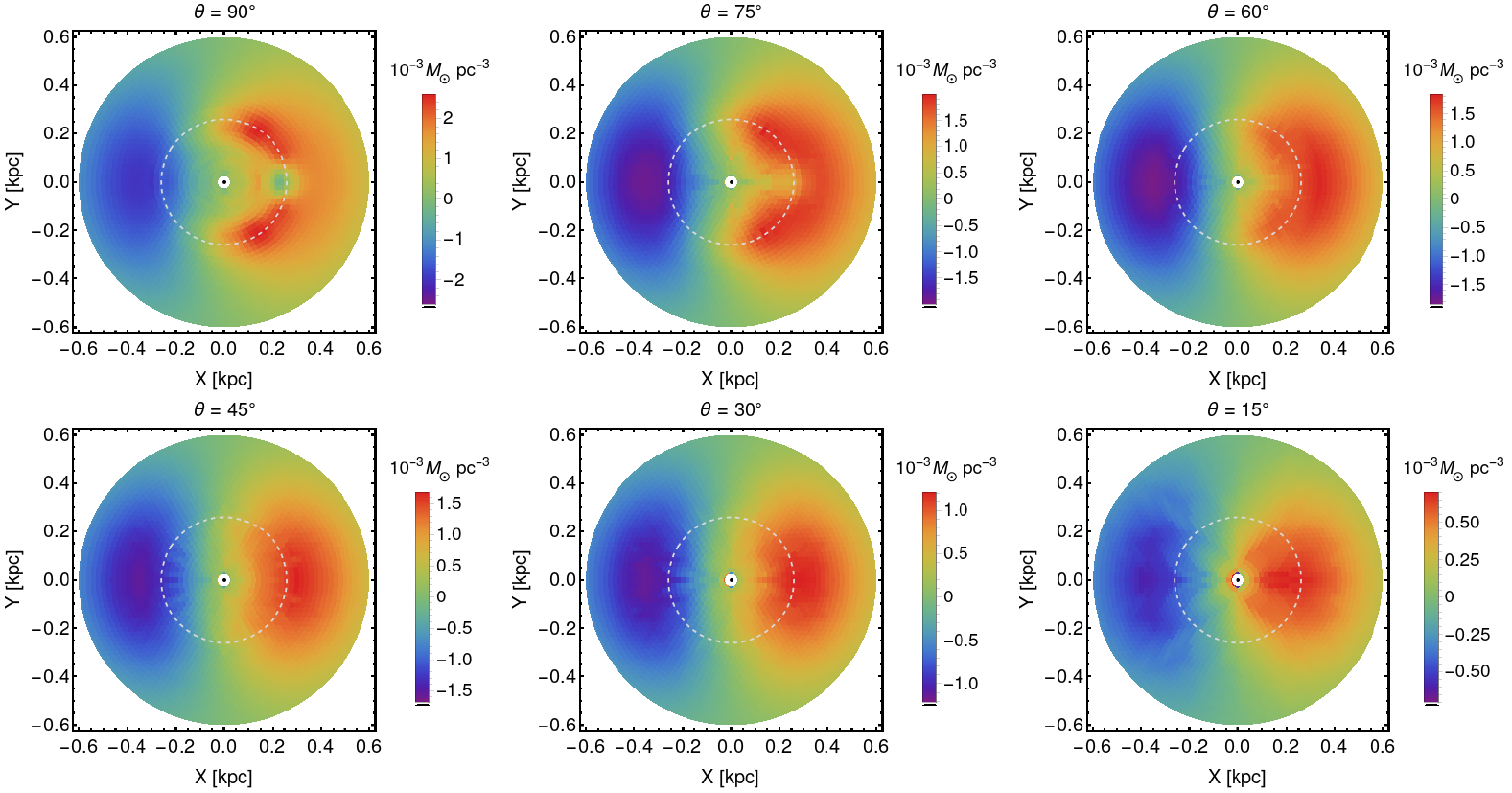}%,trim={5cm 9.5cm 1cm 6cm}
\subcaption{$r_p=0.26$kpc}
%\label{ilr}
\end{subfigure}
%\hline 
\\
\begin{subfigure}{0.9 \textwidth}
\centering
\includegraphics[width=0.8 \textwidth]{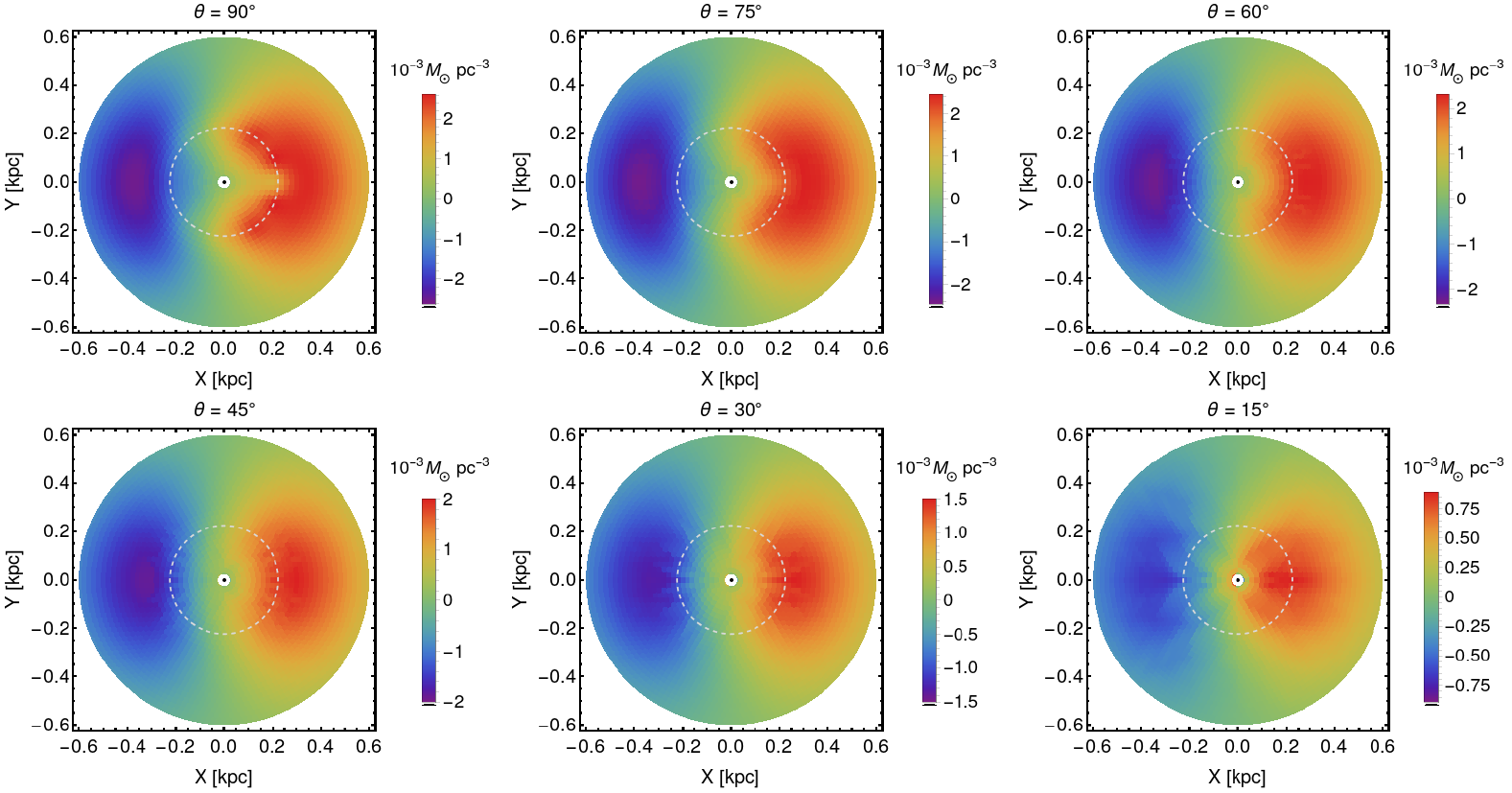}%,trim={5cm 9.5cm 1cm 6cm}
\subcaption{$r_p=0.225$kpc}
%\label{ilr}
\end{subfigure}
\end{framed}
\caption{
 Non-resonant wakes are shown in the perturber's rest frame in three sets for three $r_p$ values, with the same format as figure \ref{fig_res_cones}.   
 The qualitative structure of non-resonant wakes shows a dominant and extended overdensity around the perturber, where $x>0$, and an extended underdensity on the opposite side ($x<0$) of galactic center.  This structure remains roughly similar for all three $r_p$ and at all co-latitudes in the unfolded cones plotted here. 
 A partial exception is the small and weak underdense hole around perturber at large $r_p$ ($0.3,0.26$kpc) and low latitudes.    
}
\label{fig_nr_cones}
\end{figure*} 

\begin{figure*}
\centering
\includegraphics[width=1.0 \textwidth]{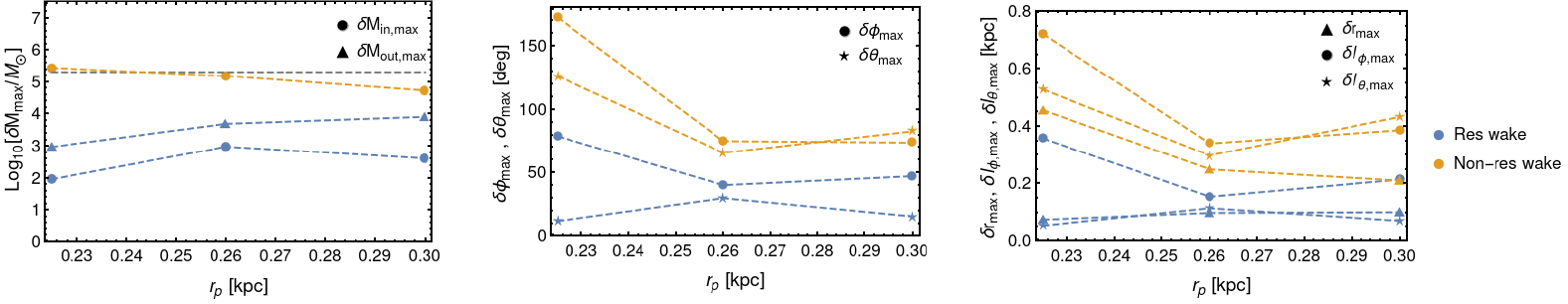}%,trim={5cm 9.5cm 1cm 6cm}
\caption{Properties of 3D wake overdensities (resonant in \emph{blue} and non-resonant in \emph{yellow}) as functions of the perturber semimajor axis $r_p$. \emph{Left panel}: the total mass $\delta M_{\rm in, max}$ and $\delta M_{\rm out, max}$, contained respectively within the inner solid and outer dashed black contours delineating wake overdensities in figure~\ref{fig_res_moll} or \ref{fig_nr_moll} (corresponding, respectively, to 1/2 and 1/10th of maximum $\rho_{1,\rm res}$ or $\rho_{1,\rm nr}$). The dashed gray horizontal line corresponds to the perturber's mass $M_p$. The mass $\delta M_{\rm in,max}$ contained in the inner non-resonant overdensity is comparable to $M_p$ (increasing only slightly with decreasing $r_p$) and remains greater than its resonant counterpart by a factor of $10^{2-3}$ for all $r_p$. \emph{Middle panel}: The angular extents of wake overdensities, $\delta \phi_{\rm max}$ (in azimuth) and $\delta \theta_{\rm max}$ (in latitude), about the point of maximum density perturbation ($\rho_{1,\rm nr}$ or $\rho_{1,\rm res}$).  \emph{Right panel}: spatial extents $\{ \delta r, \delta l_{\theta}, \delta l_{\phi} \}_{\rm max}$ of wake overdensity, defined about the point of maximum overdensity and considering the inner (solid black) contours as wake overdensity boundaries, as earlier. The resonant wake overdensity becomes more azimuthally extended with decreasing $r_p$, showing an increase in $\delta \phi_{\rm max}$ and $\delta l_{\phi,{\rm max}}$.  However, the resonant wake remains radially and vertically compact at all $r_p$.  The non-resonant wake overdensity is quite spatially extended in 3D for all $r_p$, and roughly doubles its radial extent $\delta r_{\rm max}$ as $r_p$ goes from $0.3$kpc to $0.225$kpc. It expands in other dimensions too.}
\label{fig_wake3dover_properties}
\end{figure*}

\begin{figure*}
\centering
\includegraphics[width=1.0 \textwidth]{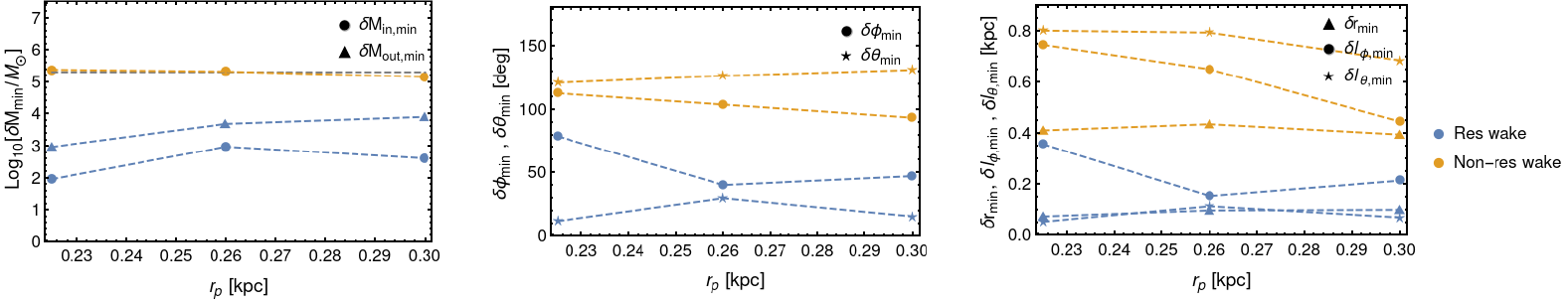}%,trim={5cm 9.5cm 1cm 6cm}
\caption{The properties of 3D wake underdensities (resonant in \emph{blue} and non-resonant in \emph{yellow}) as functions of $r_p$.  Figure format is the same as in figure \ref{fig_wake3dover_properties}, except now all masses and angular/spatial extents are defined about the points of the dominant {\it underdensity}. 
Non-resonant underdensities are quite extended, but do not show any strong trend with changing $r_p$. As with overdensities, the mass $\delta M_{\rm in}$ contained in non-resonant wake underdensities is comparable to $M_p$ and is 2-3 orders of magnitude higher than that in resonant wake underdensities.  Due to anti-symmetry, resonant underdensities have properties 
identical
to those of resonant overdensities, as previously described in figure~\ref{fig_wake3dover_properties}.  }
\label{fig_wake3dunder_properties}
\end{figure*}

Here we investigate the 3D structure of density wakes, decomposed into their resonant and non-resonant sub-parts.  
As before, we work in the perturber's rotating frame and choose three representative values of the perturber's orbital radius, $r_p = {0.3,0.26,0.225}$ kpc.  In 3D as in 2D, this narrow range in radii captures interesting morphological changes in the resonant wakes that occur close to the filtering radius $r_{\star}$. 

\emph{Resonant wakes:} The right panels of figure~\ref{fig_res_moll} illustrate the 3D structure of resonant wakes for various $r_p$ values. Specifically, this figure shows Mollweide projections of resonant wakes on a sphere passing through the points of maximum $\rho_{1 \rm res}$. Due to the anti-symmetry of the resonant wake, the points of minimum $\rho_{1, \rm res}$ also lie on these spheres. The main advantage of this representation is that we can also judge the vertical extent of various wake overdensities and underdensities, whose properties we only investigated within the perturber's orbital plane before.  

$\delta \theta$ measures the (angular) latitudinal extent of the wake as bounded by solid contours (i.e. contours corresponding to density perturbations that are 1/2 of the extremal $\rho_{1,\rm res}$), about the point of extremum $\rho_{1,\rm res}$ on this spherical surface. The corresponding measure of \emph{height} $\delta l_{\theta}$ of the wakes is the length of arc of constant $\{ r,\phi \}$ (for extremum point) bounding the solid contours. Note that the three lengthscales $\{\delta r, \delta l_{ \theta} , \delta l_{\phi}  \}$ defining the size of wake overdensities (underdensities) are all calculated about the maximum (minimum) points of density deformation. 

The resonant wakes are quite compact in the vertical direction, with $\delta l_{\theta} \sim 0.1$kpc. 
The three-dimensional extents of overdensities and underdensities for the three chosen $r_p$s are compared in figure~\ref{fig_wake3dover_properties}. Resonant wakes are especially compact in radial and vertical directions, with small $\delta r$ and $\delta l_{\theta}$, but they do have a relatively large azimuthal extent $\delta l_{\phi}$. This feature is more pronounced for the resonant wake at $r_p = 0.225$kpc, where $\delta l_\theta \sim \delta r \sim 1/4 \,\delta l_{\phi} \sim 0.1 $kpc. This also makes this wake look qualitatively different compared to those from larger $r_p$ values.  The low-$r_p$ wake has a more complicated structure, with (1) both smaller leading and relatively larger trailing dominant overdensities close to the plane, (2) a weak leading overdensity above and below the plane at higher latitudes $|\lambda|\gtrsim 20.4^{\circ} $, (3) a peak overdensity that lies away from the perturber at an azimuthal angle $\phi \simeq 66.3^{\circ}$ making the wake \emph{more global} in nature. This is in contrast to the simple structure of the wake at larger $r_p$, which is simply a compact, dominant overdensity following directly behind the perturber.  

Figure~\ref{fig_wake3dover_properties} also presents the mass contained in the resonant wakes. Masses $\delta M_{\rm in}$ and $\delta M_{\rm out}$ measure the mass contained within the inner (solid) and outer (dashed) contour surfaces, where the density deformation (here we refer to the resonant part $\rho_{1,\rm res}$) falls to 1/2 and 1/10th of the extremum value, respectively. For resonant wakes at all $r_p$, the total contained masses $\delta M_{\rm out}$ and $\delta M_{\rm in}$ are smaller than the perturber mass $M_p$ respectively by at least one and two orders of magnitude. For resonant wakes, both $\delta M_{\rm out}$ and $\delta M_{\rm in}$ decrease for smaller $r_p$, with $\delta M_{\rm out}\sim 10^{-2} M_p$ and $\delta M_{\rm in}\sim 10^{-3} M_p$ when $r_p=0.225$kpc. 

Figure~\ref{fig_res_cones} offers an alternative and more complete and view of 3D resonant wake structure at these three $r_p$ values.  This figure represents the 3D wakes on unfolded cones for various fixed colatitudes $\theta$. For smaller $\theta$ far away from the perturber's orbital plane, resonant wakes constitute a weak leading overdensity (and corresponding trailing underdensity) for all $r_p$ values.  This trend inverts close to the perturber's plane at $\theta = 90^{\circ}$. The resonant wake at $r_p = 0.225$ kpc has a visibly small radial extent at all $\theta$, when compared to its counterparts at larger $r_p$.

\emph{Non-resonant wakes:} Non-resonant wakes are much more massive and spatially extended than are the resonant wakes. Figure~\ref{fig_nr_moll} gives a 3D view of the structure of these wakes, by presenting their Mollweide projections on the spheres passing through the points of maximum $\rho_{1, \rm nr}$. In general, the minimum points corresponding to peak underdensity do not lie on these spheres, except for $r_p =0.3$kpc, in which case the two points of minima lie at longitude $\phi = 180^{\circ}$ and latitudes $\lambda \simeq \pm 24^{\circ}$, above and below negative $x$-axis. For all $r_p$ values, the maximum and minimum points for peak overdensity and underdensity lie in the orbital plane of perturber, excluding of course the minima for $r_p=0.3$kpc. As $r_p$ decreases, the small underdense region surrounding the perturber (at large $r_p$) steadily shrinks, giving each of these wakes a unique morphology. The non-resonant wakes are quite spacious, and the dashed contours corresponding to 1/10th of the extremum $\rho_{1\rm nr}$ extend beyond the spherical region of radius 0.6kpc in which this calculation was performed. As noticed in 2D analysis, two extended overdensities at $r_p=$0.3 kpc merge to form an even more extended single overdensity surrounding perturber at smaller $r_p$.  

As earlier, the two sets of three lengthscales $\{\delta r, \delta l_{\theta}, \delta l_{\phi}  \}$ measure the size of overdensities and underdensities (bounded by solid contours) around the respective points of maximum and minimum which they enclose; check figures~\ref{fig_wake3dover_properties} and \ref{fig_wake3dunder_properties}. Both the non-resonant overdensity and underdensity are quite extended in the three orthogonal directions, especially compared to their resonant counterparts. Their sizes are maximized for the smallest $r_p=0.225$ kpc, which is also clearly visible from its Mollweide projection of figure~\ref{fig_nr_moll}. % is even visibly the most extensive one. 
Apart from providing an alternative look at 3D structure, the unfolded conical surfaces (of fixed $\theta$) in figure~\ref{fig_nr_cones} re-emphasize these features of non-resonant wakes.         
 
The mass enclosed by both overdense and underdense non-resonant wakes (specifically $\delta M_{\rm in}$, the mass bounded by the solid contours) is roughly of the order of the perturber's mass $M_p$, as can be seen in figures~\ref{fig_wake3dover_properties} and \ref{fig_wake3dunder_properties}. This mass is 2-3 orders of magnitude higher compared to the mass contained in the resonant wakes. As $r_p$ decreases from $0.3$kpc to $0.225$kpc, the $\delta M_{\rm in}$ of the non-resonant wake witnesses a slight increase by a factor $\sim 2-5$, staying of the same order as $M_p$. We are unable to provide the mass $\delta M_{\rm out}$ contained within dashed contours for non-resonant wakes, because these structures extend outside the region in which this calculation is performed.

%\FloatBarrier

\section{Significance of Non-Corotating Torques and orbit evolution}
\label{sec_non_CR_torq}

\begin{figure*}
\centering
\includegraphics[width=0.99 \textwidth]{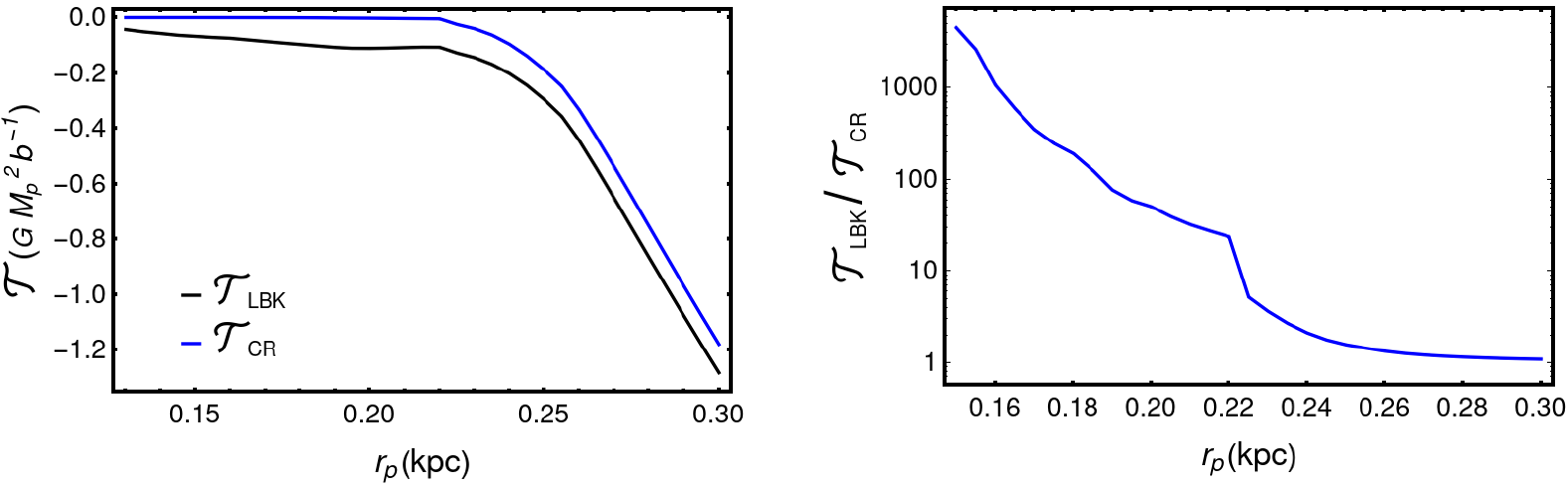}%,trim={5cm 9.5cm 1cm 6cm}
\caption{   \emph{LBK Torque and CR Torque}.  {\it Left Panel}: radial profiles for both the total LBK torque $\mathcal{T}_{\rm LBK}$ (in black) and the torque from corotating resonances only, $\mathcal{T}_{\rm CR}$ (in blue). $\mathcal{T}_{\rm LBK}-\mathcal{T}_{\rm CR} = \mathcal{T}_{\rm n CR}  \sim - 0.1 G M_p^2/b $ throughout the range of $r_p$. {\it Right panel}: the radial profile for the torque ratio $\mathcal{T}_{\rm LBK}/ \mathcal{T}_{\rm CR}$. Inside the filtering radius $r_{\star}$, $\mathcal{T}_{\rm CR}$ is highly suppressed and $\mathcal{T}_{\rm nCR}$ is the main contributor to $\mathcal{T}_{\rm LBK}$.      
}
\label{fig_torq_prof}
\end{figure*}

\begin{figure*}
\centering
\begin{subfigure}{0.48\textwidth}
\centering
\includegraphics[width=0.99 \textwidth]{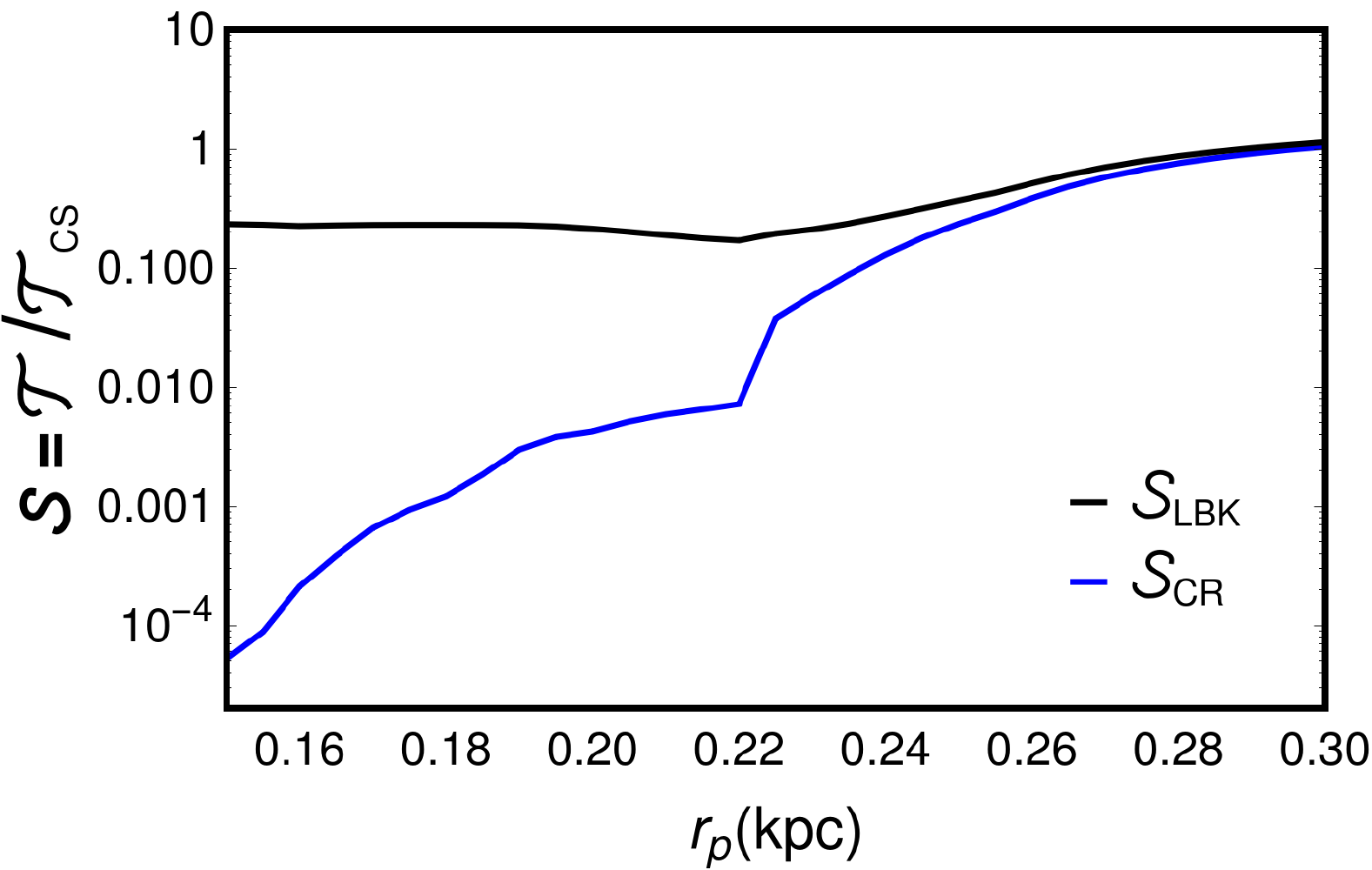}%,trim={5cm 9.5cm 1cm 6cm}
%\subcaption{$r_p=0.3$kpc}
%\label{ilr}
\end{subfigure}
\hfill
\begin{subfigure}{0.48\textwidth}
\centering
\includegraphics[width=0.99 \textwidth]{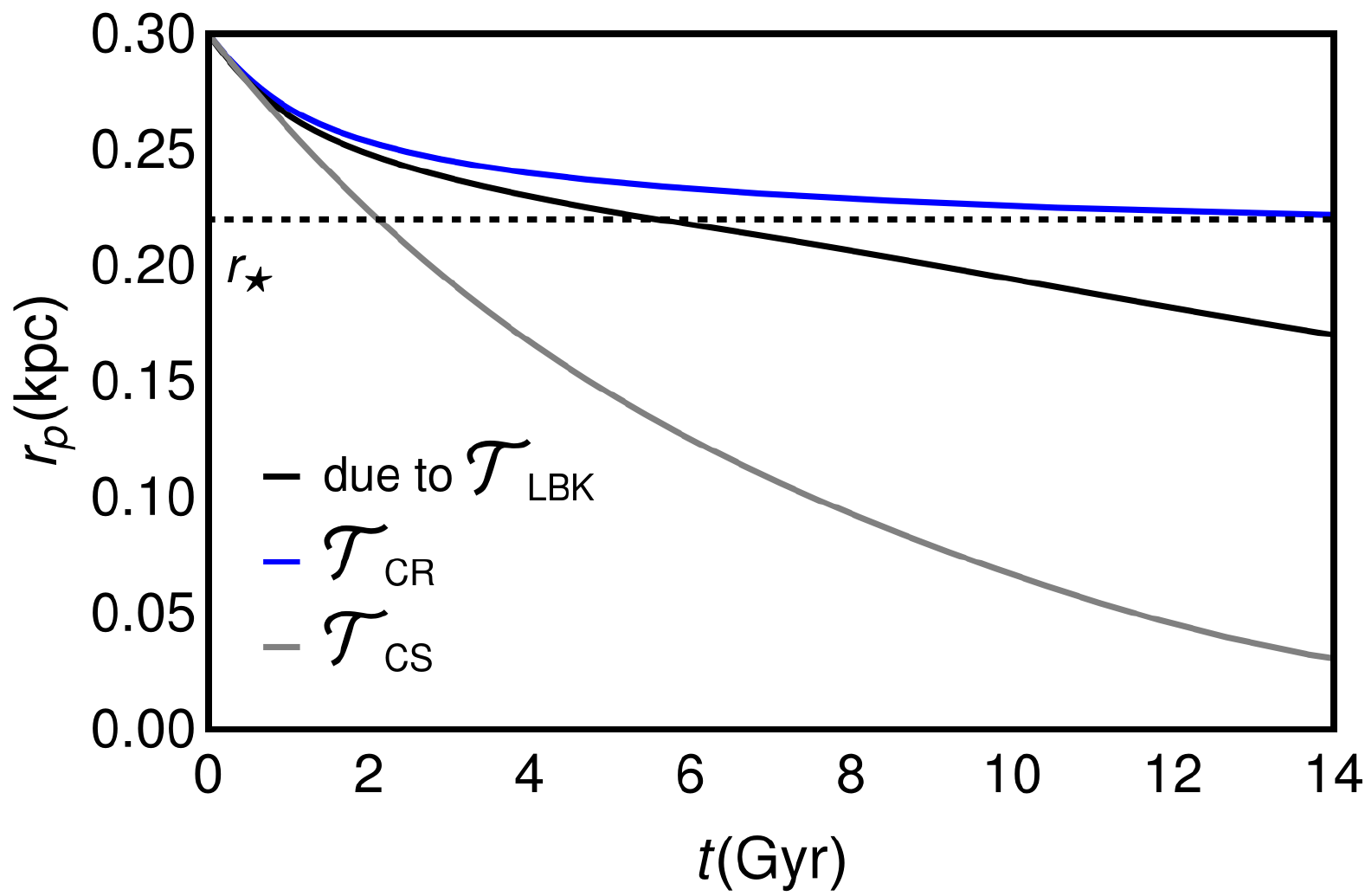}%,trim={5cm 9.5cm 1cm 6cm}
%\subcaption{$r_p=0.3$kpc}
%\label{ilr}
\end{subfigure}
\caption{ \emph{Suppression Factor \& Orbit evolution}. {\it Left panel}: the radial profile of suppression factors $\mathcal{S}_{\rm LBK}$ and $\mathcal{S}_{\rm CR}$, measuring the fraction of the Chandrasekhar torque that is produced in our global calculations. Inside $r_{\star}$, $\mathcal{S}_{\rm LBK} \sim 1/5$, while the corotating suppression factor $\mathcal{S}_{\rm CR}<10^{-2}$ steeply falls with decreasing $r_p$. {\it Right panel}: time evolution of the orbital radius $r_p$ of a perturber due to the full LBK torque $\mathcal{T}_{\rm LBK}$ (in black), the corotating LBK torque $\mathcal{T}_{\rm CR}$ (in blue) and the Chandrasekhar torque $\mathcal{T}_{\rm CS}$ (in gray). The Chandrasekhar formula ($\mathcal{T}_{\rm CS}$) predicts that the perturber would roughly reach the galactic center in a Hubble time. On the contrary, it stalls almost precisely at $r_{\star}$ due to $\mathcal{T}_{\rm CR}$. Considering the most realistic calculation $\mathcal{T}_{\rm LBK}$, with the inclusion of non-CR resonances, we find an intermediate trajectory: the perturber keeps inspiraling (even inside $r_{\star}$) but at a far slower rate than the Chandrasekhar theory predicts. }
\label{fig_S_orb_evo}
\end{figure*}

In this section, we investigate the contributions of different orbital resonances to the LBK torque acting on the perturber orbiting near the galaxy core. In TW theory, a linearly deformed galaxy with linear deformation $F_1(\bfI,\bfw)$, as in equation~(\ref{fin_F1}), exerts a DF torque $\mathcal{T}_{\rm LBK}$ on perturber, given by the LBK formula (see \S~2.4 of KS18):
\begin{subequations}
\begin{align}
& \mathcal{T}_{\rm LBK} \;=\; \sum_{n,\ell = -\infty}^{\infty}\;
\sum_{m \,=\, 1}^{\infty}\;\; \mathcal{T}_{n \ell m}\;,\qquad\quad\mbox{where}
\label{LBK_torq}\\[1em] 
&\mathcal{T}_{n \ell m} = 
16\pi^4\,m^2\,\Omega_p \!\! \int \rmd\bfI\;\frac{\rmd F_0}{\rmd E} \,
\delta\!\left(\bfl \centerdot \bfOmg \right)\,\vert\Phitilda_{\bfl}(\bfI)\vert^2 \, . 
\label{torq-comp}
\end{align}
\end{subequations}  
As is evident from the presence of the Dirac delta $\delta$-function inside the action-space integral for a torque component $\mathcal{T}_{n \ell m}$, only stellar orbits in resonance with perturber (i.e. satisfying the resonant condition $\bfl \centerdot \bfOmg = n \Omega_w + \ell \Omega_g - m \Omega_p = 0$) can contribute to the LBK torque. In KS18, the torques $\mathcal{T}_{\rm CR} = \sum_{\ell m} \mathcal{T}_{m \ell m}$ were computed only for corotating (CR) resonances with $n =m$ and hence, the corresponding resonance condition becomes $m (\Omega_w - \Omega_p)= - \ell \Omega_g$. Inside the galaxy core (i.e. $r_{\star} \lesssim r_p \lesssim b$), these CR resonant orbits are expected to be the main contributor to the LBK torque for low-order resonances, with small $|m|$ and $|\ell|$. This is due to orbital frequency hierarchies inside the galactic core, where $\Omega_w \sim \Omega_p \sim \Omega_b \gg \Omega_g$.  

But as the perturber reaches the inner core, close to the filtering radius $r_{\star}$ (where $\Omega_p(r_\star) = \Omega_b$), $\Omega_p$ becomes greater than the stellar orbital frequency $\Omega_w$ for most of the stars composing the galactic mass distribution. 
Hence, as the perturber sinks to smaller orbital radii near $r_{\star}$, CR resonances are expected to weaken and become fewer in number (see figures~4 and 5 of KS18) and the CR torque $\mathcal{T}_{\rm CR}$ diminishes accordingly (see %left panel of
figure~\ref{fig_torq_prof}). This naturally raises the question: \emph{do the non-CR resonances, which are subdominant at large radii, take over and contribute significantly to the total LBK torque $\mathcal{T}_{\rm LBK}$ at small $r_p$, once CR resonances are depleted?}  

We evaluate the torque components $\mathcal{T}_{n \ell m}$ using equation~(\ref{torq-comp}) for each integer of magnitude ${|n|,|\ell|,|m|} \leq 10$ for all odd or all even combinations (given property {\bf P2} of $\Phitilda_{\bfl}$ of appendix~\ref{app_Phi_nlm_properties}). Both CR and non-CR components are depicted in figure~\ref{fig_torq_comp}, with detailed descriptions of different resonant orbits given in appendix~\ref{sec_lbk_tor_comp}. Near $r_{\star}$, a small decrease in $r_p$ causes the number and strength of CR torque components to fall steeply, while non-CR components remain reasonably abundant with moderate strengths throughout the range of $r_p \in[0.18-0.3]$kpc we consider. The total non-CR torque\footnote{Non-CR torque components are non-vanishing only for $n = m +2 $ within the considered bounds of the action $I \leq I_{\rm max} = \epsilon I_b$ inside the galactic core.} $\mathcal{T}_{\rm n CR} = \sum_{n = m+2} \mathcal{T}_{n \ell m} \simeq - 0.1 G M_p^2/b$  for all $r_p$, while $\mathcal{T}_{\rm CR}$ falls by more than 3 orders of magnitude; see table~\ref{tab_tor_comp}.

The radial profiles for the LBK torque $\mathcal{T}_{\rm LBK}$ and its corotating part $\mathcal{T}_{\rm CR}$ are presented in figure~\ref{fig_torq_prof} (left panel). The magnitude of $\mathcal{T}_{\rm LBK}$ is on average about $\sim 0.1 G M_p^2/b$ larger than $\mathcal{T}_{\rm CR}$, a difference due to $|\mathcal{T}_{\rm nCR}|$. Also in figure~\ref{fig_torq_prof} (right panel) we present the torque ratio $\mathcal{T}_{\rm LBK}/\mathcal{T}_{\rm CR}$, which measures the factor by which total LBK torques $\mathcal{T}_{\rm LBK}$ are amplified when accounting for non-CR resonances. This amplification factor highlights the importance of non-CR resonances at small values of $r_p$. For large $r_p$ close to $r_p = 0.3$kpc, the torque ratio rises very slowly as $r_p$ decreases. But near $r_\star$ there is a sharp jump in the torque ratio, to a factor $\approx 10$, and it continues rising for smaller $r_p < r_\star$, up to $\sim 10^{3-4}$ (due to extremely small $\mathcal{T}_{\rm CR}$ at the smallest radii we consider).   

Finally, we compare both $\mathcal{T}_{\rm LBK}$ and $\mathcal{T}_{\rm CR}$ with the classic frictional torque $\mathcal{T}_{\rm CS}$ from Chandrasekhar's local formula. For $\mathcal{T}_{\rm CS}$, we make calculational choices corresponding to solid blue curve of figure~1(a) in KS18, i.e. selecting the smallest possible magnitude of the Chandrasekhar torque\footnote{These assumptions are: (1) the maximum impact parameter (in the Chandrasekhar Coulomb logarithm) is the orbital radius $r_p$ of perturber, and (2) the velocity distribution is obtained from the ergodic distribution function $F_0(E)$, equation~(\ref{df-iso}), appropriate for an unperturbed galaxy.}.  Figure~\ref{fig_S_orb_evo} (left panel) shows the profiles of suppression factors, $\mathcal{S}_{\rm LBK}= \mathcal{T}_{\rm LBK}/\mathcal{T}_{\rm CS}$ and $\mathcal{S}_{\rm CR}= \mathcal{T}_{\rm CR}/\mathcal{T}_{\rm CS}$. In the outer regions of the core with $r_p \approx 0.3$kpc, these torques are all comparable in strength i.e. $\mathcal{T}_{\rm LBK } \sim \mathcal{T}_{\rm CR} \sim \mathcal{T}_{\rm CS}$. Both the factors  $\mathcal{S}_{\rm LBK}$ and $\mathcal{S}_{\rm CR}$ gradually decrease as $r_p$ decreases, though the rate of decline for $\mathcal{S}_{\rm CR}$ is much larger. The trend continues until $r_\star$, where $\mathcal{T}_{\rm LBK} \sim \mathcal{T}_{\rm CS}/6$ and $\mathcal{T}_{\rm CR} \sim \mathcal{T}_{\rm CS}/100$. Within $r_p \leq r_\star$ the two suppression factors behave in a vividly distinct manner.  After a sudden fall at $r_\star$, $\mathcal{S}_{\rm CR}$ keeps decreasing for smaller $r_p$ to the extremely small magnitude of $\sim 10^{-4}$ at $r_p = 0.15$kpc. On the contrary, $\mathcal{S}_{\rm LBK}$ hits its global minimum at $r_\star$ and starts rising slightly for smaller $r_p$ with $\mathcal{S}_{\rm LBK} \sim 1/5$ as $r\to0$. 
Ultimately, the suppression of LBK torques inside the filtering radius (relative to a local calculation) is limited to a factor $\approx 5$ level by the persistence of non-CR resonances, so we expect the orbital radius $r_p$ of the perturber to decay faster than in the case of $\mathcal{T}_{\rm CR}$.

The orbital evolution of the perturber is evaluated using the general equation $M_p {\rmd (\Omega_p r_p^2 ) }/{\rmd t} = \mathcal{T}(r_p)$, for each of the three types of torques $\mathcal{T} = \mathcal{T}_{\rm CS}, \mathcal{T}_{\rm CR}, \mathcal{T}_{\rm LBK}$ of interest. It is plotted in figure~\ref{fig_S_orb_evo} (right panel). For evolution driven by the total LBK torque $\mathcal{T}_{\rm LBK}$, the perturber, upon crossing $r_\star$ at $\sim 5$Gyr, sinks slowly down to $r_p \sim 0.17$kpc at $13$Gyr. This is quite  different from both the orbital decay due to (1) $\mathcal{T}_{\rm CS}$, where torque magnitudes are high enough to
submerge the perturber deep into the galactic center with $r_p\sim 30$pc at 13 Gyr, and (2) $\mathcal{T}_{\rm CR}$, where torque magnitudes are so suppressed that the perturber's orbit stalls at $r_\star$ (as in KS18). This calculation, which is novel in its inclusion of non-CR resonances, implies a qualitative revision to the usual picture of \emph{stalling}.  In this picture, while the orbital decay of the perturber slows markedly inside the filtering radius $r_{\star}$, it does not come to a firm halt there.  In order to check the sensitivity of this result on the resonance order, we compare the strengths of $\mathcal{T}_{\rm LBK}$ derived up to different resonance orders and the resulting orbits of the perturber in appendix~\ref{sec_hi_res}, and conclude that the inclusion of the higher order resonances should not affect the qualitative nature of the perturber's orbit evolution due to $\mathcal{T}_{\rm LBK}$. We discuss below the implications of this result.

\subsection{Implications for Orbit Stalling}
\label{sec_stalling_impli}
 
The evolution of a perturber's orbit in a cored potential due to the total LBK torque $\mathcal{T}_{\rm LBK}$ (including both corotating {\it and} non-corotating resonances) turns out to be quite different from the evolution observed in the most $N$-body simulations, which witness nearly \emph{perfect} orbit stalling.  The $N$-body results are similar to the LBK results if non-CR resonances are excluded and one only computes $\mathcal{T}_{\rm CR}$, but the inclusion of non-CR resonances qualitatively alters the nature of stalling in the LBK picture. Here we discuss different possible causes of this discrepancy.  Before proceeding, we note that while this discrepancy is quite intriguing theoretically, LBK approach, due to extremely slow inspiral inside $r_{\star}$, can still explain in principle the presence of globulars well away from the galactic center in Fornax dSph. Additionally, as mentioned earlier the recent simulation study by \citet{Meadows+20} reports only a partial stalling featured by extremely slow inspiral inside inner core.

\begin{enumerate}
\item \emph{Resonance overlaps}: High eccentricity orbits (i.e. of pericenter $\lesssim r_p$ but relatively large semimajor axis) are the primary contributors to $\mathcal{T}_{\rm n CR}$; see figures~\ref{fig_res_lines} and \ref{fig_torq_st_ecc}. Such high-$e$ resonances tend to overlap with each other \footnote{As an example, this is well-appreciated in context of planetary dynamics; see chapter 8 of \citealt{MD99}.}, and therefore are expected to produce chaotic evolution of the stars residing in them. The stars that populate these overlapping high-e resonances may therefore be exhausted by chaotic diffusion to other parts of phase space, a process not captured by the linear perturbation theory employed here. This may reduce the non-CR contribution to total torque in a real galaxy core (or in simulations thereof), leading to stalling akin to what is seen in $N$-body simulations. We note that the possibility of chaotic diffusion during resonance overlap will have a much smaller effect on the stationary ($n = \ell = m$) CR resonance that dominates the torque at large $r_p \simeq 0.3$kpc. For large perturber orbits, this special CR resonance supplies $\sim 70\%$ of the total $\mathcal{T}_{\rm L BK}$ (see table~\ref{tab_tor_comp}) and most of its LBK torque is drawn from stars on low-$e$ orbits (see figure \ref{fig_torq_st_ecc}) that are less prone to overlap.

\item \emph{Neglect of wake self-gravity}: \citet{Weinberg1989} found that the orbital decay timescale is enhanced by a factor of 2-3 for a perturber orbiting in a $n=3$ polytrope galaxy, when one considers the effects of self-gravity. Hence, it is possible that the addition of wake self-gravity to the current study might slow the orbital decay of the perturber, 
although we note that comparable results (a factor of few slowdown) would not bring the LBK torque calculations here into agreement with the near-total stalling seen in $N$-body simulations.

\item \emph{Limitations of secular approximation}: In their generalized perturbative approach going beyond the secular approximation, \citet{Banik2021} have found that inside a critical radius in galactic core, the torques are anti-frictional as a result of ``memory effects'' (path-dependent torques due to past orbital evolution of the perturber).  Although non-CR Fourier modes are not taken into account in their non-secular theory, it is nonetheless possible that these anti-frictional torques might overwhelm $\mathcal{T}_{\rm n CR}$ and lead to stalling. 

\item \emph{Limitations of simulations}: Aside from physical (e.g. chaotic diffusion) reasons that stars may escape from high-$e$ resonances, it is also possible that these non-CR resonances are smeared out due to the inexact treatment of $N$-body gravity used in simulations that find evidence for core stalling \citep{Read2006,Inoue2011,DuttaChowdhury2019}.  Specifically, the use of tree codes rather than direct $N$-body integration, though necessary for computational efficiency, may prevent the accurate resolution of resonances driven by relatively distant orbits from the perturber.    

\end{enumerate}

Given the above list of possibilities, we cannot conclusively deduce the real nature of core stalling on the basis of present study alone. A more general investigation is needed in the future to resolve whether core stalling creates a hard lower limit on the radius $r_p$ to which a perturber can inspiral in a cored potential (as is seen in most $N$-body simulations and suggested by the non-secular ``memory torques'' of \citealt{Banik2021}), or whether the novel non-corotating resonances we have investigated will allow perturber to continue its inspiral, albeit at a greatly reduced pace.  For practical purposes, there will sometimes not be an astrophysically meaningful difference between these two possibilities.

\section{Orbit Stalling with an Additional Perturber}
\label{sec_LK_dynamics}

Regardless of whether orbit stalling is total (as is generally seen in numerical simulations) or partial (as in this work), the inspiral of a massive perturber will slow down as it moves closer to the center of a galaxy with a cored potential.  This suggests the possibility that many such perturbers will pile up, which is arguably realized by the large population of globular clusters in Fornax.  These perturbers will exert secular torques on each other, which may cause their orbits to evolve in the space of angular momentum.

In this section, we study the secular dynamics of two perturbers orbiting the galaxy core. An outer massive perturber, P2, is assumed to follow a fixed circular orbit.  An inner perturber, P1, is treated as a test particle. Perturbations due to P2 on the orbit of P1 are considered at the quadrupolar level.  We average the Hamiltonian over the mean anomalies of the two perturbers in order to study the long-term (secular) evolution of the orbit of P1. This analysis is similar to well-known Lidov-Kozai dynamics \citep{Lidov1962,Kozai1962,Merritt2013}, except that the usual central massive point particle is replaced by the non-Keplerian potential a cored spherical galaxy. Note that, for simplicity, we do not take into account the perturbing potential from the wake of P2.\footnote{Moreover, azimuthal averaging of $\rho_1$ would result in an axisymmetric, annular density distribution with a peak close to the perturber's orbital radius; see e.g. figure~\ref{fig_nr_wake2d_rps} for the structure of non-resonant wake. The resonant wake does not contribute to secular torques due to its antisymmetry under the transformation $\phi' \rightarrow - \phi'$). In the secular picture, the (non-resonant) wake would contribute as a smooth mass torus around the otherwise sharp circular ring of mass P2, adding an additional quadrupole moment to the problem.  For the sake of simplicity, we do not consider this additional effect here.} In addition, we neglect the dynamical friction acting on P1 and P2 due to their own wakes. This assumption remains valid if %globulars P1 and P2 orbit close to their respective filtering radii; we specialize to this case in \S~\ref{sub_sec_2stalled_perturbers}. 
either (i) globulars P1 and P2 have stalled, or (ii) the dynamical friction inspiral timescale is much longer (as expected close to and within $r_{\star}$) than the secular timescale, which is defined later in this section as $\tau_{\rm LK}$.

\begin{figure*}
\centering
\includegraphics[width=0.99 \textwidth]{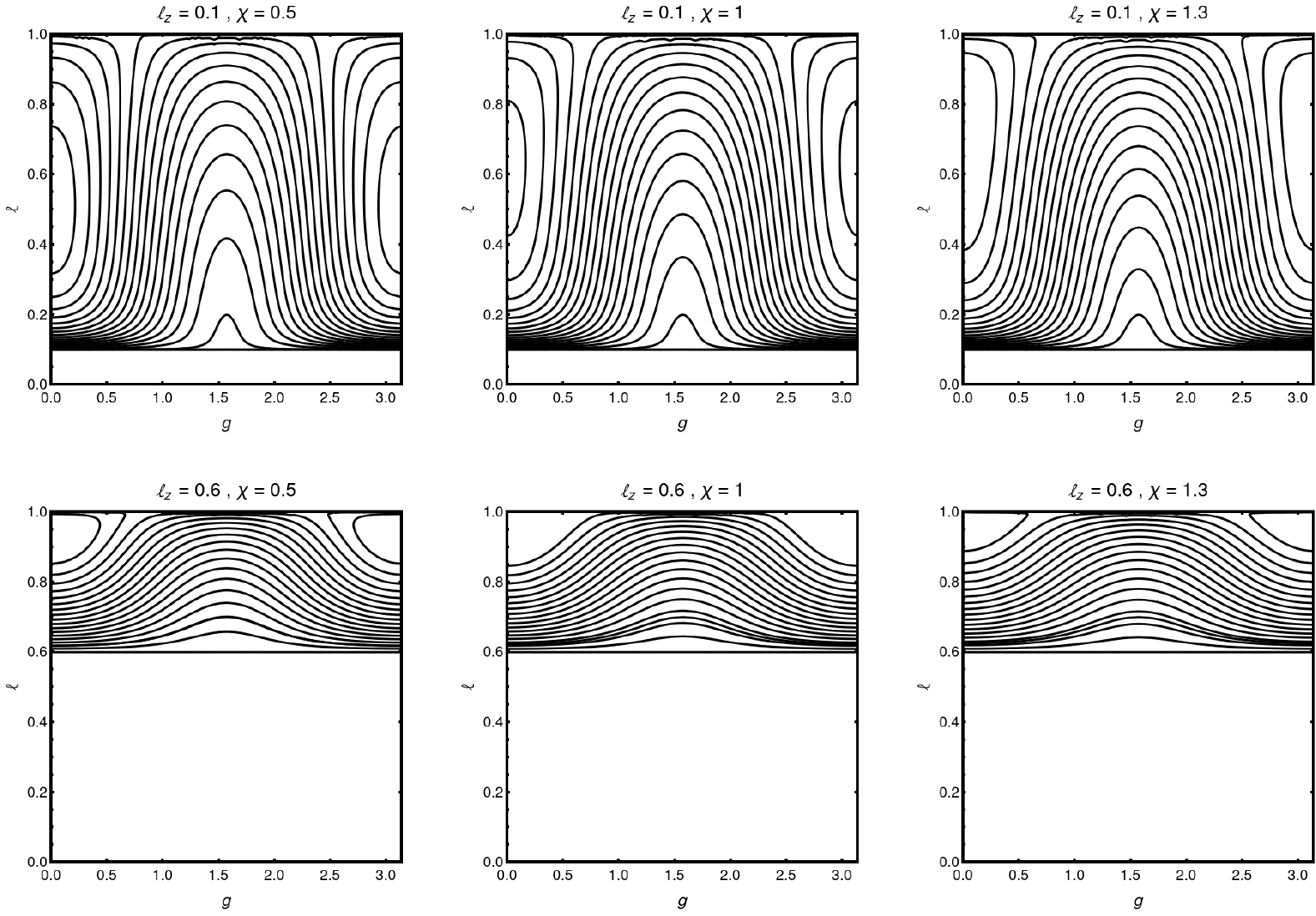}%,trim={5cm 9.5cm 1cm 6cm}
\caption{ Isocontours of the normalized secular Hamiltonian $H /(\alpha^2 /8)$ for a test particle orbiting an isochrone core in presence of an outer perturber P2. In all cases the isocontours are shown as functions of argument of periapse $g$ and dimensionless angular momentum $\ell$.  Upper and lower panels correspond to dimensionless $z$-components of angular momentum $\ell_z = 0.1$ and 0.6 respectively.  The three columns correspond to three different values of the parameter $\chi = 0.5, \, 1, \, 1.3$.        
}
\label{fig_LK_Hcont}
\end{figure*}
\begin{figure}
\centering
\includegraphics[width=0.45 \textwidth]{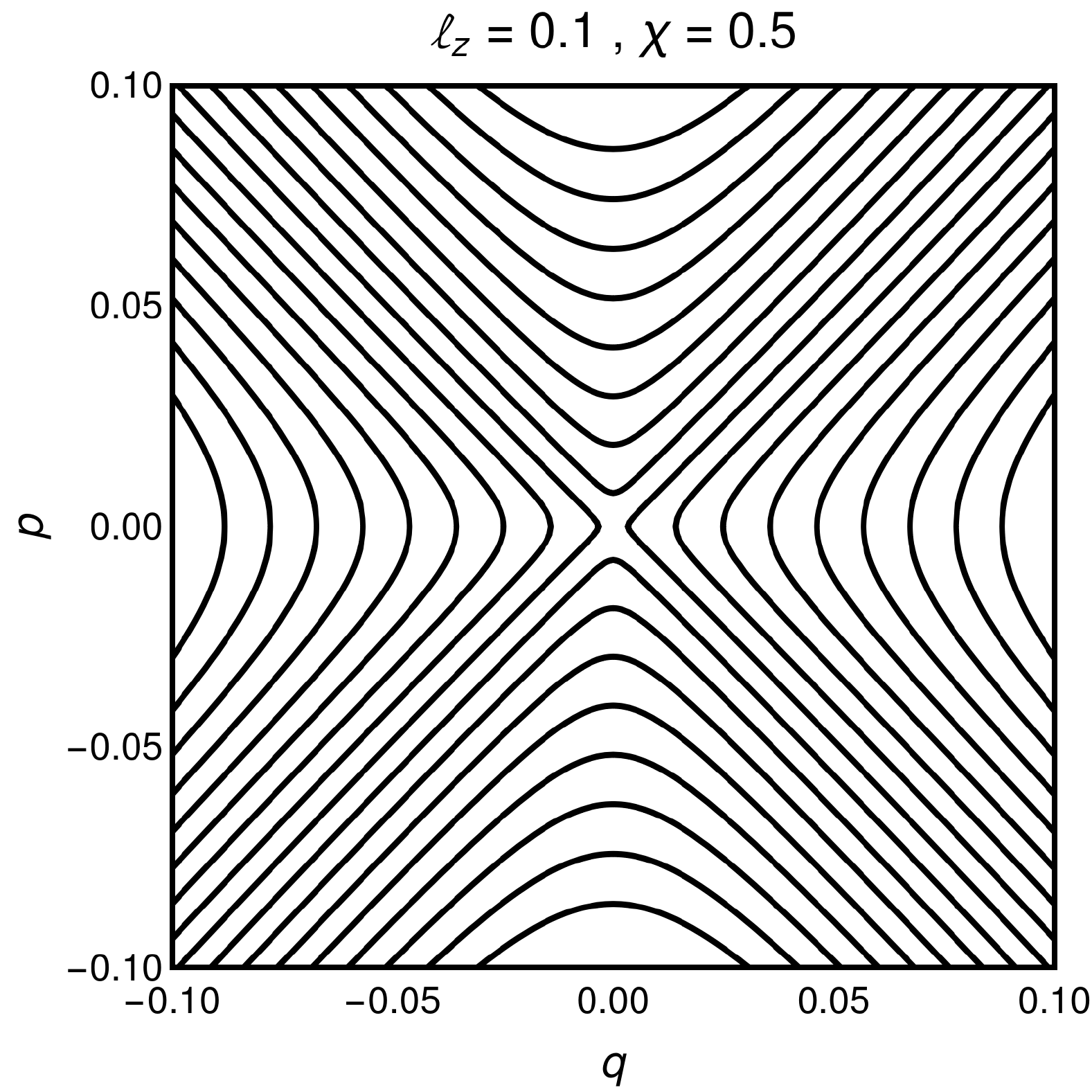}%,trim={5cm 9.5cm 1cm 6cm}
\caption{Isocontours of the normalized secular Hamiltonian $H_s /(\alpha^2 /8)$ in the $q,p$-plane, in the vicinity of $\{p=0, q=0 \}$ or $\ell =1$, which is clearly a hyperbolic fixed point.
}
\label{fig_LK_Hl1_cont}
\end{figure}

We consider an outer massive perturber (P2) of mass $M_{p2}$ following a circular orbit of radius $r_2$ with orbital frequency $\Omega_{p2}$. The coordinates of P2 are:
\beq
x_2 = r_2 \cos{w_2} \quad ; \quad y_2 = r_2 \sin{w_2} \quad ; \quad w_2 = \Omega_{p2} t 
\label{P2_orbit}
\eeq
where $w_2$ is the mean anomaly (also true anomaly for this case). The inner perturber P1 is assumed to be a test particle orbiting the galaxy core under the combined influence of the gravitational potential of the galaxy and that of P2. The instantaneous spatial coordinates $\{x,y,z \}$ of P1 are given by equation~(\ref{gen_coord_expl}) in terms of actions and angles $\{ \bfw,\bfI \}$.  

The orbital evolution of P1, with phase space coordinates $\{ {\bf r, v}\}$ ($\bfr$ and {\bf v} being the position vector and velocity in physical space), is governed by the specific Hamiltonian:
\beq
H_p = \frac{v^2}{2} + \Phi_0(r) - \frac{G M_{p2}}{\left| \bfr - \bfr_2  \right|} + G M_{p2} \frac{\bfr \centerdot \bfr_2 }{ r_2^3  } \, . 
\label{gen_H}
\eeq
The first two terms correspond to unperturbed dynamics in an isochrone galaxy and are equal to specific energy $E(I,L)$. The third term (III) is due to the \emph{direct} Newtonian pull of P2; while the fourth (IV) \emph{indirect} term arises from the choice of coordinate system whose origin is located at the galaxy center. Term III$+$IV forms the perturbing potential and for $r_2 > r$, it can be expanded as a series in the small parameter $(r/r_2)$:
\beq 
\begin{split}
{\rm III + IV} &= - \frac{G M_{p2}}{r_2} \sum_{\ell=1}^{\infty} P_{\ell}(\cos{\widetilde{\phi}}) \left( \frac{ r }{ r_2 }  \right)^{\ell} + G M_{p2} \frac{r}{r_2^2} \cos{\widetilde{\phi}} \\[1ex] 
&= - \frac{G M_{p2}}{r_2} \sum_{\ell =2}^{\infty} P_{\ell}(\cos{\widetilde{\phi}}) \left( \frac{ r }{ r_2 }  \right)^{\ell}
\end{split}
\eeq 
where $\widetilde{\phi}$ is the angle between $\bfr$ and $\bfr_2$. Keeping only the first $\ell =2$ term, we have a quadrupole-order perturbing potential: 
\beq 
\begin{split}
\widetilde{\Phi}_{p2} &=  - \frac{G M_{p2}}{2 r_2^3} r^2 \left( 2 \cos^2{\widetilde{\phi}} -1 \right) \\[1ex]
&= - \frac{ G M_{p2} }{2} \left[ \frac{ 3 ( x x_2 + y y_2 )^2  }{ r_2^5  } - \frac{r^2}{ r_2^3  }    \right]
\end{split}
\label{quad_Phi_p2}
\eeq 
We rewrite the quadrupolar Hamiltonian $E(I,L) + \widetilde{\Phi}_{p2}$ in terms of AA variables, and then average it over the mean anomalies of two perturbers ($w$ and $w_2$), resulting in the following {\it secular} Hamiltonian: 
\beq 
\widetilde{H} = E(I,L) + <\! \!\widetilde{\Phi}_{p2}\! \! >  \quad;  \quad < \!\!\widetilde{\Phi}_{p2}\!\! >  = \frac{1}{(2 \pi)^2} \oint \!\!\! \oint \rmd w \,\rmd w_2 \, \widetilde{\Phi}_{p2} \,. 
\eeq 
Employing equations~(\ref{gen_coord_expl}) and (\ref{P2_orbit}), it is straightforward to express the secular perturbing potential $<\!\!\widetilde{\Phi}_{p2}\!\!>$ in an explicit form: 
\beq 
<\!\!\widetilde{\Phi}_{p2}\!\! >  = - \frac{3 G M_{p2}}{8 r_2^3} \left( \frac{ I}{\Omega_b} \right) \left[ 1+ \cos^2{i} - e \sin^2{i}  \cos{(2 g)}  \right] \, . 
\eeq 
As a result of orbital phase averaging, $I$ and hence mean size of an orbit is conserved. In addition, $L_z = L \cos{i}$ is conserved due to axisymmetry. Note that $e = \sqrt{1- L^2/I^2}$. The reduced 4D phase space $\{ L,L_z,g,h \}$ can be transformed to dimensionless variables $\{ \ell,\ell_z , g,h  \}$, where $\ell = L/I$ and $\ell_z = L_z /I = \ell \cos{i}$. We introduce a dimensionless time $\tau = \Omega_b \, t$ and the new specific Hamiltonian becomes:
\beq 
H = \widetilde{H}/(I \Omega_b)  = E_0(\ell) + \Phi_{2p}(\ell,\ell_z,g) \,
\label{H_sec_LK}
\eeq 
where $E_0(\ell) = E(I,L)/(I \Omega_b)$ and $\Phi_{p2}(\ell,\ell_z,g) = <\!\!\widetilde{\Phi}_{p2}\!\! >/(I \Omega_b)$. We introduce two dimensionless parameters: $\alpha = \sqrt{I/\Omega_b}/b$, the average orbital size of P1, and $\alpha_2 =  r_2/b$, the radius of the circular orbit of P2 (both in units of the core radius $b$). We can then write the new energy:
\beq 
%\begin{split}
E_0(\ell) =  \frac{-32}{  \displaystyle{ \alpha^6 \left[ 1 + \frac{4}{\alpha^2} \sqrt{ 1 + \frac{ \alpha^4 \ell^2  }{16  }  } \right]^2 } }  \simeq  \frac{ \alpha^2 \ell^2 }{ 8  } \,. %1 - \frac{ 2 }{ \alpha^2 } +
%\end{split}
\eeq 
The approximate form on the right is valid for $\alpha \ll 1$ (and ignores the constant term irrelevant for secular dynamics), which is approximately valid for the present scenario, with P1 having stalled (or nearly stalled) deep inside the galactic core. In the remainder of this section, we use this approximate form. 

Written explicitly, the secular perturbing potential is
\beq  
\Phi_{p2} = -\frac{3}{2} \frac{M_{p2}}{M \alpha_2^3} \left[ 1 + \frac{\ell_z^2}{\ell^2} - \sqrt{1-\ell^2} \left( 1 - \frac{\ell_z^2}{\ell^2} \right) \cos{(2 g)}   \right] \, .
\eeq 
Note that the strength of this perturbing potential is governed by a single combination $\xi = M_{p2}/(M \alpha_2^3)$ of most of the problem's parameters. The Hamiltonian equations of motion in the $\{\ell,g\}$-phase plane are given below. 
\begin{equation} 
\begin{split}
& \dot{\ell} = - \frac{\p H }{\p g} = 3 \xi \sqrt{1-\ell^2} \left( 1 - \frac{\ell_z^2}{\ell^2} \right) \sin{(2 g)} \\[1em]
& \dot{g} = \frac{\p H}{\p \ell} =  \frac{\alpha^2}{4} \ell - \frac{3}{2} \xi \bigg[ -2 \frac{\ell_z^2}{\ell^3} \left( 1 + e \cos{(2 g)}  \right) \\
& \hspace{2cm} + \frac{\ell}{\sqrt{1-\ell^2}}  \left( 1 - \frac{\ell_z^2}{\ell^2} \right) \cos{(2 g)} \bigg]
\end{split}  
\end{equation}
From these equations, it is apparent that LK dynamics occurs on a timescale $\tau_{\rm LK} \sim \xi^{-1} \Omega_b^{-1}$. Inside the galaxy core, the orbital timescale $\tau_w \sim \Omega_b^{-1} \simeq 0.02$Gyr for the choice of isochrone parameters in this study. If we assume further that the outer perturber is inspiralling close to 
its filtering radius, then $\alpha_2 \simeq  (M_{p2}/M_c)^{1/5}$ (see also the beginning of  \S~\ref{sub_sec_2stalled_perturbers}), where $M_c = 0.12 M$ is the mass enclosed within isochrone core radius $b$. This implies a simpler form $\tau_{\rm LK} \sim 3 (M/M_{p2})^{2/5} \Omega_b^{-1}$. For a range of GC masses $10^{5-6}\Msun$, and previously used galaxy properties, this timescale $\tau_{\rm LK} \sim 1-3$Gyr. However, we note that the scenario where both perturbers are stalled or inspiralling   near their filtering radii will strain the earlier assumption that $r/r_2$ can be expanded out as a small parameter.  Secular interactions between two perturbers will in any event begin earlier, when P2 is at substantially larger radii.

\bigskip

\noindent
{\bf Fixed Points} 

\medskip 

From the isocontours of $H$ (figure~\ref{fig_LK_Hcont}), we see that there are two elliptic fixed points (at $g= 0 ,\pi$). The presence of these fixed points makes the circular orbit $\ell =1$ unstable. Clearly  $\dot{\ell} = 0$ for these fixed points, with $g= 0 ,\pi$. In order for $\dot{g} = 0$ at these fixed points, we have: 
\beq 
\chi =\frac{\alpha^2}{6 \xi} = \frac{1}{e_1} \left[ 1 - \frac{\ell_z^2}{(1-e_1)^2} \right]  
\eeq 
where we have introduced a new parameter $\chi$ for convenience.  We have also made use of $e_1 = \sqrt{1-\ell_1^2}$ for fixed points $(\ell = \ell_1 , g = 0)$  and $(\ell = \ell_1 , g = \pi)$. This leads to following cubic equation in $e_1$:
\beq 
\mathcal{P}(e_1) = \chi e_1^3 - (1 + 2 \chi ) e_1^2 + (2 + \chi) e_1 - (1- \ell_z^2) = 0 \,.
\eeq 
Note that $\mathcal{P}(0) = \ell_z^2 - 1 \leq 0$ and $\mathcal{P}(1) = \ell_z^2 \geq 0$. This implies that there is at least one real value of $e_1 \in [0,1]$ for all $\chi$ and $\ell_z$. Hence the elliptic fixed points at $g = 0, \pi$ always exist. 

Now, we explicitly explore the dynamics close to $\ell =1$. To avoid a coordinate degeneracy, we choose the new canonically conjugate coordinates $\{q,\,p\}$, defined as: 
\beq
p = \sqrt{2(1-\ell)} \cos{g} \quad  ; \quad
q = - \sqrt{2(1-\ell)} \sin{g} \, ,
\eeq
such that $\{p =0,\, q=0\}$ for $\ell = 1$. Hence the secular Hamiltonian of equation~(\ref{H_sec_LK}) attains the following approximate form in vicinity of the point $\{ p=0, q=0 \}$ (up to leading orders in $\{|p|,|q|\} \ll 1$):
\beq
H_s = \frac{\alpha^2}{8} (1 - p^2 - q^2) - \frac{3 \xi}{2} \bigg[ 1 + \ell_z^2 (1+p^2 +q^2) - (1-\ell_z^2) \frac{(p^2 - q^2)}{\sqrt{p^2 + q^2}}   \bigg] \, . 
\label{H_sec_LK_l_1}
\eeq 
Figure~\ref{fig_LK_Hl1_cont} shows the isocontours of $H_s$ in the $q, p $-plane for a representative set of $\chi$ and $\ell_z$ values. The nature of the isocontours does not change with other parameter values. This is due to the dominance of the leading order term $(p^2-q^2)/\sqrt{p^2+q^2}$ in $H_s$, which ensures that the point $\{ p=0, q =0 \}$, i.e. $\ell = 1$ is a hyperbolic fixed point for all $\chi$ and $\ell_z$ (except for $\ell_z =\pm 1$ corresponding to a coplanar circular orbit of P1). So an initially circular orbit with a non-zero initial inclination $i_0 = \cos^{-1}{\ell_z}$ undergoes LK oscillations\footnote{ This secular dynamics is notably distinct from the usual Lidov-Kozai dynamics around a central (Keplerian) point mass, where $\ell = 1$ is a hyperbolic point only for initial mutual orbital inclination 
$39.2^{\circ} \le i_0 \le 141.8^{\circ}$.}
(with the conserved quantity $\ell_z = \sqrt{1-e^2} \cos{i}$), attaining a maximum eccentricity $e_m$. We evaluate an explicit expression for $e_m$ below. 

\bigskip

\noindent
{\bf Maximum Excited Eccentricity }

\medskip

We follow the separatrix passing through $(\ell = 1, g = \pi/2)$, which corresponds to following $H$-isocontour:
\beq 
H_{\rm sep} = H(\ell = 1, g = \frac{\pi}{2}) = \frac{\alpha^2}{8} - \frac{3}{2} \xi (1+\ell_z^2) \, .
\eeq 
Since the separatrix also passes through $(\ell = \ell_m, g = 0)$ corresponding to the maximum eccentricity $e_m = \sqrt{1-\ell_m^2}$, we have: 
\beq 
\begin{split}
H_{\rm sep} & = H(\ell = \ell_m , g = 0) =  \frac{\alpha^2}{8} \ell_m^2 - \frac{3}{2} \xi \bigg[  1 + \frac{\ell_z^2}{\ell_m^2}\\
& \hspace{3cm} - \sqrt{1-\ell_m^2} \left( 1 - \frac{\ell_z^2}{\ell_m^2} \right) \bigg] \,.
\end{split}
\eeq 
Equating above expressions, we get the following cubic equation in $e_m$:
\beq 
\frac{\chi}{2} e_m^3 - e_m^2 - \left(\frac{\chi}{2}+\ell_z^2 \right) e_m + 1 - \ell_z^2 = 0
\eeq 
$e_m = -1$ is one of the roots and remnant quadratic equation is: 
\beq 
\mathcal{P}_2(e_m) = \frac{\chi}{2} e_m^2 - \left( 1 + \frac{\chi}{2} \right) e_m + 1 -\ell_z^2 \,= \,0 .
\eeq 
This gives following unique physical value for the maximum excited eccentricity $e_m$ (the other root being always greater than unity): 
\beq 
e_m = \frac{  2 + \chi - \sqrt{ ( \chi -2 )^2 + 8 \chi \ell_z^2 } }{ 2 \chi}
\eeq 
$e_m$ is a monotonically decreasing function of $\chi = {\alpha^2}/{ (6 \xi) } = { \alpha^2 \alpha_2^3 M}/{(6 M_{p2})}  $.
Effectively, $\chi$ is proportional to the ratio of strengths between the secular part of the unperturbed Hamiltonian and the perturbing potential. Hence $e_m$ increases with increasing strength of the perturbing potential ($\propto \xi$), and decreases as the apsidal precession of the unperturbed orbit ($\propto \alpha^2$) increases. 

From this formula we see that $e_m$ decreases with increasing $|\ell_z| = |\cos{i_0}|$; the maximum eccentricities are excited for initially polar circular orbits with $\ell_z =0$ ($i_0 = 90^{\circ}$). For $\chi \leq 2$ this maximum value is unity; while for $\chi > 2$ the maximum value is $e_m = 2/\chi$ which is smaller than unity.

So far we have only evaluated the maximum eccentricity achievable for circular initial conditions.  However, for eccentric initial conditions, P1 can achieve maximum eccentricities that are either lower (if librating around the fixed points) or higher (if circulating) than $e_m$.

\begin{figure}
\centering
\includegraphics[width=0.45 \textwidth]{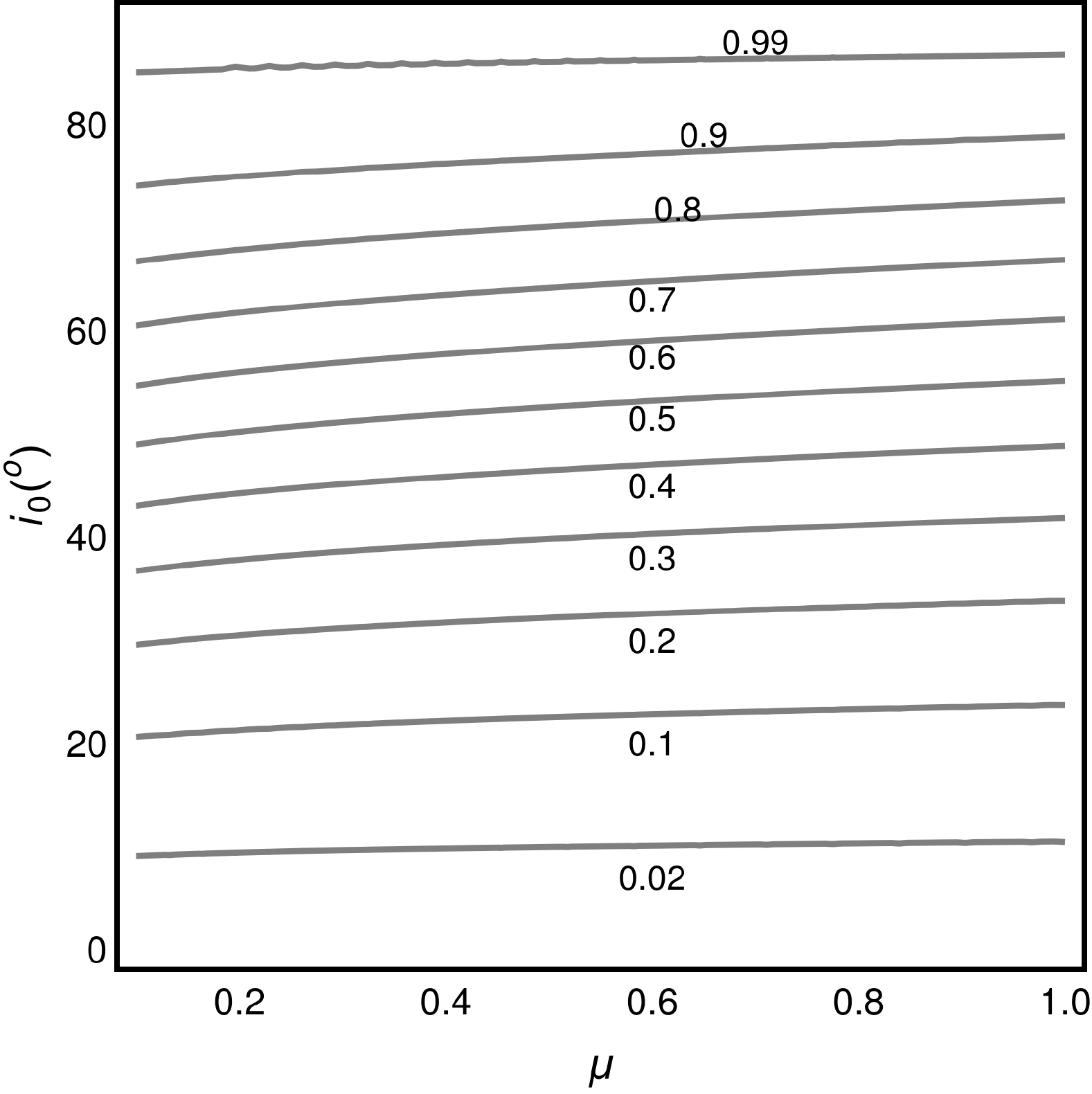}%,trim={5cm 9.5cm 1cm 6cm}
\caption{Isocontours of maximal eccentricity $e_m$  of P1 due to LK oscillations in $\mu,i_0$-plane for $f=1$. The average orbital size $\alpha$ of P1 and orbital radius $\alpha_2$ of P2 correspond to their respective filtering radii.     
}
\label{fig_LK_emax}
\end{figure}

\subsection{Consequences for Stalling}
\label{sub_sec_2stalled_perturbers}
We now consider a scenario where the inner perturber P1 orbits within its filtering radius $r_\star$, such that further orbital decay is significantly slow (owing to the residual contribution of non-CR resonances, as discussed in \S~\ref{sec_non_CR_torq}). For simplicity, here we assume that P1 orbits the galactic center on a circle of fixed radius $\sim f r_\star$ with factor $f \lesssim 1$. From KS18, an isochrone galaxy has $r_\star \simeq b \,(M_p/M_c)^{1/5} $, where $M_p$ is mass of perturber, and $M_c$ is the galaxy mass enclosed within the galaxy core radius $b$. Hence, the average (relative) orbital size of P1 defined above is just $\alpha = f (M_p/M_c)^{1/5}$. The more massive outer perturber P2 spirals in due to dynamical friction until it \emph{settles} near its own filtering radius $r_{\star2} \simeq b \,(M_{p2}/M_c)^{1/5}$, inside which its rate of orbital decay slows down significantly. Here we study the implications of Lidov Kozai dynamics with P2 orbiting on a fixed circle of radius $r_{\star2}$; this implies its relative orbital size $\alpha_2 = (M_{p2}/M_c)^{1/5}$. {It is to be noted that the ratio of orbital sizes of the two perturbers $\alpha/\alpha_2 = f (M_p/M_{p2})^{1/5}$ need not be a very small number given that globulars generally spend just two orders of magnitude in masses $10^{4-6} \Msun$, and the smallest value achieved is $\alpha/\alpha_2 \sim 0.4 f$. So, this study, based on ignoring the higher order terms in the Hamiltonian (see equations~(\ref{gen_H})-(\ref{quad_Phi_p2})), is more relevant for: (i) globulars lying on the two extremes of the mass range, and (ii) the case where P1 has already inspiralled inside $r_\star$ (making $f < 1$), while the outer perturber P2 has just reached close to its filtering radius $r_{\star2}$.  }

Substituting the above values of $\alpha$ and $\alpha_2$, the parameter $\chi$ becomes:
\beq 
\chi = 1.397 f^2 \mu^{2/5} ; \quad
{\mbox{where  }}  \mu = \frac{M_p}{M_{p2}} < 1 \, .
\label{chi_sp}
\eeq 
Here we used the fact that $M_c = 0.12 M$ for isochrone galaxy. Hence the maximum excited eccentricity $e_m$ for the orbit of P1 becomes function of the three parameters -- mass ratio of perturbers $\mu$, their initial mutual orbital inclination $i_0$ {and the factor $f$ that measures the extent to which P1 has inspiralled inside its filtering radius.}\footnote{It is interesting to note that $f$ and $\mu$ occur in the function $e_m$ only as combination $f \mu^{1/5}$. } $e_m$ is higher for higher mass $M_{p2}$ of P2, inspite of the fact that its orbital size $\alpha_2$ also increases with $M_{p2}$. In addition, $e_m$ is higher for lower mass $M_p$ of P1 and smaller values of $f$, because its average orbital size $\alpha$ decreases which diminishes the effect of secular apsidal precession inside galactic core. As expected, $e_m$ is high for  higher initial inclinations $i_0$. Figure~\ref{fig_LK_emax} shows these variations of $e_m$ explicitly. Note that for parameter ranges of interest $e_m$ is a stronger function of $i_0$. 

\begin{figure}
\centering
\includegraphics[width=0.5 \textwidth]{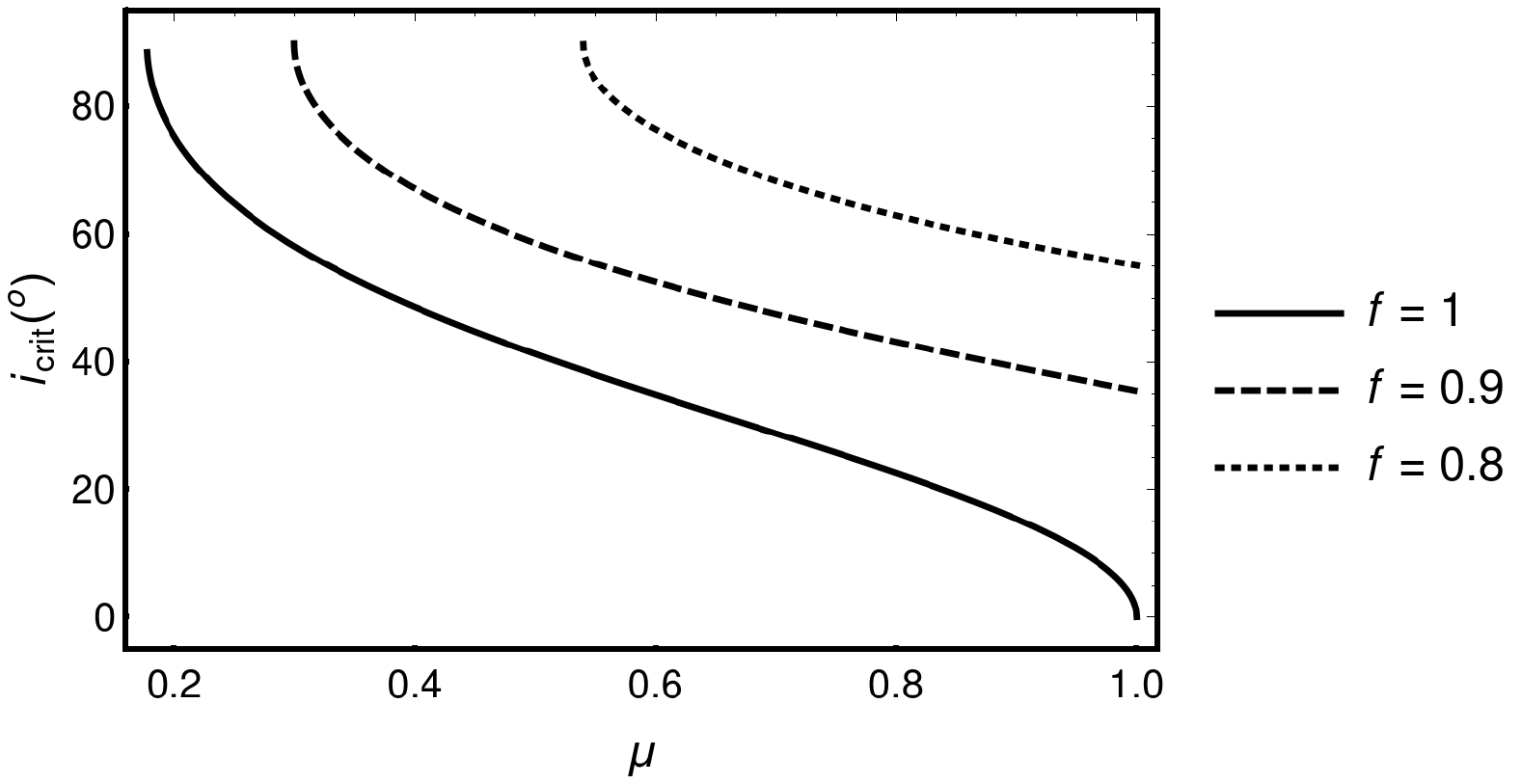}%,trim={5cm 9.5cm 1cm 6cm}
\caption{Critical initial inclination $i_{\rm crit}(\mu,f)$ of an initially circular orbit of P1 above which instability appears, i.e. maximally eccentric orbit of P1 during LK oscillations intersects the outer circular stalled orbit of P2. The instability is possible for $\mu \geq \{ 0.177, 0.299, 0.54  \}  $ for the three tentative values $f = \{ 1, 0.9 ,0.8\}$.    
}
\label{fig_LK_i_crit}
\end{figure}

Lidov Kozai oscillations of orbital eccentricity of P1 may result in close encounters with massive perturber P2 which would destabilize its state of stalling. Such an instability would creep in if the semi-major axis $\alpha \sqrt{1+e_m}$ of P1 just exceeds the orbital radius $\alpha_2$ of outer perturber. The corresponding instability condition is: 
\beq 
f \mu^{1/5} \sqrt{1+e_m(f \mu^{1/5},i_0)} \geq 1 \, . 
\eeq 
 Higher values of both $i_0$ and $f \mu^{1/5}$ tend to lead to instability. Note that the instability happens only for $f \mu^{1/5} \geq 1/\sqrt{2}$ (or $\mu \geq 0.177/f^5$).
  The above instability criterion gives a critical initial mutual inclination $i_{\rm crit}(\mu , f)$:
\beq 
i_{\rm crit}(\mu , f) \!=\! \cos^{-1}{\!\!\! \sqrt{ 1.397 f^2 \mu^{2/5} - 0.0955 - 0.3015 f^{-2} \mu^{-2/5}  } }
\eeq 
such that if $i_0 \geq i_{\rm crit}$, the orbit of P1 would destabilize. $i_{\rm crit}$ is a decreasing function of $\mu$ and $f$ (shown in the figure~\ref{fig_LK_i_crit}). Hence larger $\mu$ and $f$ values increases the odds for instability, despite maximum excited $e_m$ being smaller. Higher values of mass ratio $\mu$ may correspond to either higher $M_p$ and hence larger orbital size $\alpha$ of inner perturber P1, or lower mass $M_{p2}$ and hence smaller orbital size $\alpha_2$ of outer perturber P2. This implies that for larger $\mu$, initial orbits of P1 and P2 are closer and LK oscillations of even modest amplitudes can lead to orbital intersection. 

For parameter values considered in this paper, $M_p = 2  \times 10^{5} \Msun$, there can exist a finite range of $i_0$ leading to instability for external perturber with mass $M_{p2} \lesssim f^5 10^6 \Msun$. 
For a tentative value of $M_{p2} = 4 \times 10^5 \Msun$, instability occurs for even modest values of $i_0 \geq i_{\rm crit} = \{ 41.24^{\circ} , 58.63^{\circ} \} $ (for $f=\{1, 0.9\}$). This analysis shows that for a dwarf galaxy hosting a system of multiple globular clusters, there is a finite probability of destruction of the state of orbit stalling due to their mutual orbital interactions, {though it is sensitive to the status of inspiral of globulars within their filtering radii.} Also, for case of a less massive outer perturber, orbital intersection is hard to avoid due to its smaller filtering radius. In addition, it is more likely to have a lighter incoming outer perturber for dynamical friction timescale (from Chandrasekhar's formula, which is a good approximation outside galaxy core) is shorter for more massive perturber, which would hence reach its filtering radius faster. This favours the odds of finding the globulars on excited eccentricity orbits, rather than just peacefully {undergoing a slow circular inspiral close to their filtering radii}.

{This simple secular study cannot take into account close\footnote{ {The close interactions are almost inevitable for globulars with similar masses (mass ratio $\mu \lesssim$ 1), such that their stalled orbital radii $\alpha_2 \sim \alpha$.     }   } and/or resonant interactions of the two perturbers. Nevertheless, $N$-body simulations \citep{Inoue+09,DuttaChowdhury2019} of a cored galaxy containing a multi-globular system indicates recurrent disturbance of the state of stalling due to close encounters which in fact restores \emph{stalling} by driving globulars away from galaxy center on more energetic and eccentric orbits. In the current study, we have not considered resonant interactions which might be important for GCs orbiting inside the galactic core. We consider this as a major caveat of our analysis and do not expect it to hold for interactions of GCs of comparable masses which would \emph{stall} close-by making resonant interactions quite important. As an example, in their numerical simulations of five GCs of equal masses (extreme scenario) orbiting a galaxy core, \citet{Goerdt2006} found them to stall at same radial distance  without undergoing destabilizing interactions for a number of orbits. We speculate that the resonant interactions might be at work to stabilize such a state of stalling. }

\section{Discussion and Conclusions}
\label{sec_conclusion}

In this paper, we have 
solved the linearized collisionless Boltzmann equation to deduce the physical structure of density wakes produced by an infalling massive perturber in a cored galactic potential. This calculation was undertaken to better understand the nature of ``core stalling'' evidenced in both $N$-body simulations and observations {of globulars orbiting dwarf galaxies}.  In our methodology, we have paralleled and extended the pioneering work of \citet{Weinberg1986}, who first calculated the physical geometry of density wakes.% 

Our primary conclusions are as follows.
\begin{enumerate}
    \item The physical structure of density wakes in cored potentials is in agreement with the past work \citep{TremaineWeinberg84,Weinberg1986}. In particular, non-resonant wakes are symmetric, and resonant wakes anti-symmetric, on leading and trailing sides of perturber in its rest frame. This provides a simple visual explanation for why the dynamical friction torque is produced only by stars on resonant orbits in the LBK approach.
    
    \item As the perturber's orbital radius $r_p$ approaches the filtering radius $r_{\star}$, the resonant wake becomes weak and assumes a more intricate and extended structure lying mostly close to, but outside its orbit. %  (figure~\ref{fig_res_wake2d_rps}).
    This is in contrast to the compact and strong overdensity trailing behind the perturber, inside its orbit, for larger $r_p$. This  
    morphological transition can be understood in terms of the dominant resonant interactions between the perturber and field stars. For a perturber orbiting at large $r_p$, corotating resonant stars on orbits of smaller semimajor axis can %(figure~\ref{fig_res_lines}) 
    torque the perturber effectively. As the perturber sinks in close to or inside $r_{\star}$, however, these CR resonances are depleted and only non-CR resonant stars on eccentric orbits (of larger semimajor axes) contribute strongly to the LBK torque $\mathcal{T}_{\rm L BK}$. Likewise, as $r_p$ approaches $\sim r_{\star}$, the overall wake changes its morphology from a trailing overdensity (with a dominant resonant part) to a roughly symmetric overdensity (corresponding to the non-resonant part) shrouding the perturber. This explains the great decrease in DF torque as the perturber reaches $r_{\star}$ in the inner galaxy core.

    \item The total mass enclosed in the overdense portion of the wake is comparable to the perturber's mass, and is dominated by the (torque-free) non-resonant component of the wake. The mass contained in the resonant wake's overdensity remains 2-3 orders of magnitude smaller throughout the range of $r_p$, and decreases with decreasing $r_p$ as the perturber's orbit shrinks.

    \item In contrast to our earlier work \citep{KaurSridhar18}, we find a more subtle role for ``resonance depletion'' in the core stalling phenomenon.  Once the perturber enters the inner core of the isochrone potential defined by $r_{\star}$, the LBK torque drops by roughly one order of magnitude, not by multiple orders of magnitude as was found by \citet{KaurSridhar18}.  The reason for this difference is that in this paper, we have also considered higher-order, {\it non-corotating} (non-CR) resonances.
    The broader suite of resonances considered in the construction of density wakes in this work leads to a significant slowing of dynamical friction inspirals inside constant-density cores, but not a complete stalling.
    
    \item We also identify special CR orbits (in $\{m,m,m\}$ resonance with the perturber) which contribute heavily to the DF torque when it is farther out from filtering radius $r_{\star}$. These are the \emph{stationary} CR orbits whose libration frequency vanishes in the rest frame of perturber, such that their guiding center is at rest with respect to perturber; see figure~\ref{fig_res_orbits}. The corresponding real stellar orbits (responding to the full gravity of massive perturber, beyond our approach of linear perturbation theory), should be identified as horseshoe orbits, which have been earlier recognized as a major driver of DF \citep{QuinnGoodman1986,Fujii2009,Inoue2011,Banik2021b}.   
    
        \item We have furthermore considered the problem of interactions between multiple perturbers orbiting an isochrone core. In particular, we study the secular orbital evolution of one perturber under the influence of an external (and more massive) perturber on a fixed circular orbit.  In general, the inner orbit undergoes eccentricity/inclination oscillations akin to the classic Lidov-Kozai effect (but which begin for any non-zero mutual inclination, not just the restricted range in the classic Lidov-Kozai mechanism).  These oscillations can lead to orbital intersection with outer perturber, potentially destabilizing a stalled orbit due to close encounters. 
\end{enumerate}

Of these conclusions, (iv) is the most surprising, and merits further discussion.  Our work shows that while the loss of low-order CR resonances plays an important role in the core stalling phenomenon, {\it this resonance depletion cannot produce complete stalling}, given the residual torque contribution from non-CR resonances. % this surprising result.
In other words, {\it we predict that core ``stalling'' may only be a partial phenomenon, with DF inspiral continuing at a reduced rate due to previously neglected, high-order resonances.}  This result is in disagreement with most past $N$-body simulations of DF in cored potentials \citep[][and others all find complete core stalling]{Read2006, Goerdt2006, Inoue2011, Cole2012}, although we note that very recent $N$-body work finds incomplete stalling akin to our predictions \citep{Meadows+20}.  

We have discussed possible explanations for this discrepancy in greater detail in \S~\ref{sec_stalling_impli}, but to summarize, several possibilities present themselves. These include limitations of our analytic treatment, which was based on secular and linear perturbation theory, and by construction neglects self-gravity of wake and non-secular effects. 
The inclusion of wake self-gravity may slow down inspiral rates further \citep{Weinberg1989}, and by relaxing the secular approximation, anti-frictional ``memory torques'' in the inner galactic core \citep{Banik2021} arise which may be able to balance the frictional, non-CR torques first identified in this work. Another possibility to consider is that the high orbital eccentricities associated with non-CR resonances make them vulnerable to resonance overlap. Stars initially in overlapping non-CR resonances may leave through chaotic diffusion, rendering them unable to exert long-lived torques on perturbers in a real or simulated galaxy. Finally, it is possible that the origin of the discrepancy lies on the side of  existing $N$-body simulations: a sufficiently approximate treatment of $N$-body gravity can smear out the weak and relatively distant non-CR resonances. This interesting dynamical puzzle is worthy of careful exploration in future.   

%\bigskip
\section*{Acknowledgements}

KK is supported by the Foreign
Postdoctoral Fellowship Program of the Israel Academy of Sciences and Humanities. Both KK and NCS gratefully acknowledge support from the Israel Science Foundation (Individual Research Grant 2565/19). Data underlying the findings of this study will be shared by the authors upon reasonable request.

\bibliographystyle{mnras}
\bibliography{main} % if your bibtex file is called example.bib

\appendix 

\onecolumn

\section{Unperturbed 3D Orbits in isochrone Core}
\label{app_3d_orb}

Unperturbed stellar orbits in the isochrone core are two-dimensional rosettes, described by the following relations in the orbital plane:
\beq
\begin{split}
& r^2 = \frac{ I }{ \Omega_b } \left[ 1 - e \cos{(2 w)}    \right] \quad \\
& \psi = g + \begin{cases} \arctan{\left(  \sqrt{ \frac{ 1 + e  }{ 1 - e  } } \tan{w}  \right) }  \hspace{1.2cm} ; \; w \in [0 , \pi] \\
\pi + \arctan{\left(  \sqrt{ \frac{ 1 + e  }{ 1 - e  } } \tan{w}  \right) }
 \quad ; \; w \in (\pi, 2\pi ) \end{cases}
 \label{core_orbits}
\end{split}
\eeq
where $w$ and $g$ advance uniformly in time, with orbital frequencies $\Omega_w(I)$ and $\Omega_g(L)$, respectively. Here $\psi$ represents the true phase of star in its orbital plane, measured from the ascending node. 
Spatial coordinates $\{ x',y' \}$ in the orbital plane are:
\beq
 \begin{split}
 x' = r \cos{(\psi - g)} = \sqrt{\frac{ I  }{\Omega_b  } (1-e) } \cos{w} \; \\
 y' = r \sin{(\psi -g)} = \sqrt{\frac{ I  }{\Omega_b  } (1+e) } \sin{w} \; .
\label{planar_coord}
\end{split}
\eeq
Here we have taken periapses (or minor axis) to be aligned with the $x'$ axis.

Spatial coordinates $\{ x,y,z \}$ in a general reference frame are related to $\{x',y' \}$ by an Eulerian angular transformation, composed of these three angular transformations: 
\begin{itemize}
\item[(a).] Clockwise rotation about $z'$ axis by $g$ (the argument of periapsis measured with respect to ascending node)
\item[(b).] Clockwise rotation about $x''$ axis by $i$ (the inclination angle of the stellar orbit with respect to $x,y$-plane)
\item[(c).] Clockwise rotation about $z$ axis by $h$ (the longitude of ascending node with respect to $x$-axis).
\end{itemize}
This leads to the following rotational matrix transformation:   
 \begin{equation*}
{\left( \begin{array}{c} x \\ \\ y \\ \\ z \end{array} \right) } 
 = 
{\left( \begin{array}{ccc} 
   C_{g} C_{h} - C_{i} S_{h} S_{g} & \quad -S_{g} C_{h} - C_{i} S_{h} C_{g} & 
   \quad S_{i} S_{h}  \\ \\
   C_{g} S_{h} + C_{i} C_{h} S_{g} & \quad -S_{g} S_{h} + C_{i} C_{h} C_{g} & 
   \quad -S_{i} C_{h} \\ \\
   S_{i} S_{g}               & \quad S_{i} C_{g}            & \quad C_{i}
	\end{array} \right)} {\left( \begin{array}{c} 
   x' \\  \\ y' \\  \\ 0 \end{array} 
   \right)}
\label{gen_coord_matrix}
\end{equation*}
 where $S \equiv \sin$ and $C \equiv \cos$ of the angles given as subscripts. Hence we can explicitly express  $\{x,y,z \}$ as follows in terms of AA variables:
 \beq
 \begin{split}
& x = \sqrt{\frac{ I }{ \Omega_b  } } \left[ \sqrt{1-e}\, C_w ( C_{g} C_{h} - C_{i} S_{h} S_{g} ) - \sqrt{1+e} \, S_w (  S_{g} C_{h} + C_{i} S_{h} C_{g})   \right] \\[1em]
& y = \sqrt{\frac{ I }{ \Omega_b  } } \left[ \sqrt{1-e}\, C_w (C_{g} S_{h} + C_{i} C_{h} S_{g}) + \sqrt{1+e}\, S_w (  -S_{g} S_{h} + C_{i} C_{h} C_{g} ) \right] \\[1em]
& z =  \sqrt{\frac{ I }{ \Omega_b  } } S_i \left[ \sqrt{1-e}\, C_w S_g + \sqrt{1+e}\, S_w C_g    \right] \,.
 \end{split}
 \label{gen_coord_expl}
  \eeq

Apart from the above explicit relations, it is useful, for evaluation of the density deformation ($\rho_1$) integral, to express the coordinates $\{ x,y,z \}$ in an alternative form including an implicit dependence on $\psi$. The relevant coordinate transformation includes angular transformations (b) and (c) as presented above. This gives the following rotational transformation:  
 \beq
\left( \begin{array}{c}
x \\[1ex] y \\[1ex] z
\end{array}
 \right)  \; = \; 
\left( \begin{array}{ccc}
C_h & - S_h C_i & S_h S_i \\[1ex]
S_h & C_h C_i & -C_h S_i \\[1ex]
0 & S_i & C_i
\end{array}
\right)
\left( 
\begin{array}{c}
r \cos \psi \\[1ex]
r \sin \psi \\[1ex]
0
\end{array}
\right)
\eeq     
leading to these useful expressions for spherical polar angles $\{ \theta = \arccos{(z/r)}, \phi = \arctan{(y/x)}  \}$: 
\beq
 \tan \phi = \frac{ \tan h + \tan \psi \cos i  }{ 1 - \tan h \tan \psi  \cos i  }   \; , \hspace{1cm}
\cos \theta = \sin \psi \sin i \, . \label{polar_ang}
\eeq

\section{Fourier Coefficients of Perturbing Potential}
\label{app_Phi_nlm_properties}

In the chosen rotating reference frame, the perturber is stationary on the $x$-axis at a distance $r_p$ from the galactic center. %From the equation~(\ref{Phi_p}), it 
Its potential has the following explicit form in physical coordinates: 
\begin{equation}
 \Phi_{\rm p} = - \frac{ G M_p }{ \sqrt{ a^2 + r_p^2 - 2 r_p x + r^2  }  } + \frac{ G M_p r_p x}{ ( a^2 + r_p^2 )^{3/2} }
\label{Phi_p}
\end{equation}
where $r$ and $x$ can be explicitly expressed in AAs using equations~(\ref{core_orbits}) and (\ref{gen_coord_expl}).
$\Phi_p$, being only a function of $r$ and $x$ for this choice of reference frame, is symmetric under any physical rotation about $x$-axis. Hence, it can readily be seen that $\Phi_p$ remains invariant under the following transformations in angles $\bfw$: 
\begin{itemize}
\item[\bf T0.] $w  \rightarrow 2 \pi- w$,  $g \rightarrow 2 \pi - g$,  $h \rightarrow 2 \pi -h$ 
(or equivalently $\{x,y,z \} \rightarrow \{ x,-y,-z \}$ for given set of actions $\bfI$)

\item[\bf T1.] $w \rightarrow w \pm \pi$ , $g \rightarrow g \pm \pi$ (the identity transformation $\{x,y,z \} \rightarrow \{ x,y,z \}$ in physical coordinates for given $\bfI$)

\item[\bf T2.] $g \rightarrow g \pm \pi$ , $h \rightarrow h \pm \pi$ (or $\{x,y,z \} \rightarrow \{ x,y,-z \}$ for given $\bfI$)

\item[\bf T3.] $h \rightarrow h \pm \pi$ , $w \rightarrow w \pm \pi$ (or $\{x,y,z \} \rightarrow \{ x,y,-z \}$ for given $\bfI$; this is equivalent to {\bf T2} in physical coordinates). The angular transformation is a combination of {\bf T1} and {\bf T2}. 

\item[\bf T4.] $i \rightarrow \pi - i$, $h \rightarrow 2 \pi - h$ (or $\{x,y,z \} \rightarrow \{ x,-y,z \}$ for given $\{I,L\}$; this is a combination of {\bf T0} and {\bf T2} (or {\bf T3}) in physical coordinates. )

\end{itemize} 
 
As a result of the above symmetry properties of $\Phi_p$, its Fourier coefficients
\beq
\Phitilda_{\bfl} = \frac{1}{(2 \pi)^3} \oint \rmd^3 \bfw \, \Phi_p \exp{[- \rmi \bfl \centerdot \bfw]} 
\label{Phi_til}
\eeq
satisfy the following properties:
\begin{itemize}
\item[\bf P1.] $\Phitilda_{\bfl}(\bfI)$ is real, with its imaginary part $\im [\Phitilda_{\bfl} ] = 0$. This follows from invariance of $\Phi_p$ under transformation {\bf T0}. This also implies $ \Phitilda_{-\bfl} = \Phitilda_{\bfl} $. 

\item[\bf P2.] $ \Phitilda_{\bfl} \neq 0$ only for integer sets $\bfl = \{n,\ell,m \}$ composed of 
either three entirely even or three entirely odd integers.  $\Phitilda_{\bfl}$ vanishes for all other general $\bfl$.  This results from the invariance of $\Phi_p$ under transformations {\bf T1}, {\bf T2} and {\bf T3}, which respectively imply that $n \pm \ell$, $\ell \pm m$ and $n \pm m$ have to be even integers for non-zero $\Phitilda_{\bfl}$. 

\item[\bf P3.] $\Phitilda_{n,\ell,-m}(I,L,L_z) = \Phitilda_{n,\ell,m}(I,L,-L_z) $ as a result of transformation symmetry {\bf T4} of $\Phi_p$. 

\end{itemize}

\section{Angles $\bfw'$ in terms of $\{ \bfr' , \bfI \}$}
\label{app_8_cases}

\begin{table}
%\begin{wraptable}{r}{9cm}
\centering
%\tiny
\footnotesize
% \captionsetup{font=footnotesize}
\begin{tabular}{c c c c c}%{ |s|p{2cm}| }
  \hline
  
  Case & $w'$ & $\psi'$  & $g'$ & $h'$ \\
  
  \hline
  
A1. &  $w_1'$ & $ \psi_1'$ & $ \psi_1' - \chi_1' $ & $\phi' - \zeta_1'$ \\ 
 
A2. & $w_1'$ &  $\pi - \psi_1'$ & $\pi - \psi_1' - \chi_1'$ & $\phi' + \zeta_1'-\pi$ \\ 

A3. & $\pi - w_1'$ & $\psi_1'$ & $\psi_1' + \chi_1'-\pi$ & $\phi' - \zeta_1'$ \\ 

A4. & $\pi - w_1'$ & $\pi - \psi_1'$ & $  - \psi_1' + \chi_1'  $ & 
$ \phi' + \zeta_1' -\pi $ \\

A5. & $\pi + w_1'$ & $\psi_1'$ & $\psi_1' -\chi_1'-\pi $ & $\phi' -\zeta_1' $
 \\ 
  
A6. & $\pi + w_1'$ & $\pi - \psi_1'$ & $ - \psi_1'- \chi_1' $ & $\phi' + \zeta_1'-\pi$ \\

A7. & $2 \pi - w_1'$ & $ \psi_1' $ & $\psi_1' + \chi_1'-2\pi$ & $ \phi' -\zeta_1' $ \\ 

A8. & $2 \pi - w_1'$ & $\pi - \psi_1'$ & $ - \psi_1' + \chi_1' -\pi$ & $ \phi' + \zeta_1'-\pi $ \\

\hline
 
\end{tabular}
 
\caption{ Eight cases of multi-valued functions $(w',g',h')$ to solve $\rho_1'$ integral.  } 
\label{tbl:wgh_cases}
%\end{wraptable}
\end{table}

Here we devise a scheme to find expressions for angular functions $\{w',g',h'\}$ in terms of physical coordinates $\bfr'$ for given actions $\bfI$ to simplify the integral of equation~(\ref{Il_integral_initial}).  
As evident from equation~\ref{core_orbits}a, for a given $r'$, we have four values of $w' = w_1', \pi - w_1', \pi + w_1' , 2 \pi - w_1'$, where:
\beq
w_1' = \frac{1}{2} \cos^{-1} \left[ \frac{1}{e} \left( 1 - \frac{r'^2 \Omega_b}{I} \right)  \right]
\label{w1'}
\eeq
lies in $[0,\pi / 2]$. Also, using equation~\ref{polar_ang}, we have two effective values of $\psi' = \psi_1', \pi - \psi_1'$ for a given $\theta'$, where:
\beq
\psi_1' = \sin^{-1}\left( \frac{ \cos{\theta'} }{ \sin{i} }  \right) \,.
\label{psi1'}
\eeq
Note that it is sufficient to consider just these two cases for $\psi'$ because we are not dealing with fractions of this angle. 

For given values of $w'$ and $\psi'$, we have unique values of $g'$ (from equation~\ref{core_orbits}b) and $ h'$ (from equation~\ref{polar_ang}):
\beq
\begin{split}
& g' = \psi' - \chi' \quad ; 
\quad \chi' = \begin{cases}   \arctan{\left( \sqrt{ \frac{ 1 + e }{1 - e  }   } \tan{w'} \right)} \qquad \quad \mbox{for} \;w' \in [0,\pi)
 \\
 \pi + \arctan{\left( \sqrt{ \frac{ 1 + e }{1 - e  }   } \tan{w'} \right)} \quad \mbox{for} \; w' \in [\pi,2 \pi) \end{cases} 
 \\[1 ex]
& h' = \phi' - \zeta' \quad ; \quad
 \zeta' = \begin{cases} \arctan{\left( \cos{i} \tan{\psi'}  \right)} \qquad \quad \mbox{for} \; \psi' \in [0, \pi) \\
 \pi + \arctan{\left( \cos{i} \tan{\psi'}  \right)} \quad \mbox{for}  \; \psi' \in [\pi,2\pi)  
 \end{cases} 
 . 
\end{split}
\label{g'_h'}
\eeq

This leads us to the 8 combinations of $(w',g',h')$ (see Table~\ref{tbl:wgh_cases}), that we need to take into account while computing $\rho_1$ integral. Note that $\chi_1' = \chi'(w_1')$ and $\zeta_1' = \zeta'(\psi_1')$. It is interesting to see that, there are four pairs of physically identical combinations \{A1,A5\}, \{A2,A6\}, \{A3,A7\}, \{A4,A8\}. Orbits inside galaxy core, being centered ellipses, have two periapses and corresponding periapsidal angles separated by $\pi$. Hence, $\{w,g\}$ and $\{\pi+w,g-\pi\}$ correspond to the same point on an orbit. So, we need to consider only 4 combinations {\bf A1} to {\bf A4} to evaluate $\rho_1$ integral. 

\section{Numerical Methods to evaluate $\boldsymbol{\rho_1}$}
\label{sec_num_methods}

We first constructed tables for Fourier coefficients $\Phitilda_{\bfl}$ of perturbing potential in $\{ \alpha = r_p^{-1} \sqrt{I/\Omega_b} , \, e = \sqrt{1-L^2/I^2}, \,  \cos{i} = L_z/L \}$ space for all-even and all-odd integer combinations $\bfl$ with each of integer magnitude $\{|n|,|\ell|,|m|\} \leq 10$. Using equation~(\ref{Phi_p}) in (\ref{Phi_til}) and the fact that $\Phitilda_{\bfl}$ is real, we have the following explicit form of the integral: 
\beq
\frac{\Phitilda_{\bfl}}{(G M_p / r_p)} =  \frac{1}{(2\pi)^3} \oint \rmd^3\bfw \left[  - \frac{1 }{ \sqrt{ \xi+ 1 - 2 \tilde{x} + \tilde{r}^2  }  }  + \frac{\tilde{x}}{ ( 1 + \xi  )^{3/2} }   \right]  \cos{(\bfl \centerdot \bfw)}
\eeq 
where $\tilde{r}=r/r_p$ and $\tilde{x}=x/r_p$ (see equations~\ref{core_orbits} and \ref{gen_coord_expl}) are functions of only $\{ \alpha, e, \cos{i} \}$ and angles $\bfw$. Here $\xi = {(a/r_p)}^2$ is the softening parameter, whose value is chosen as $10^{-3}$ for this calculation. These integrals are solved in \texttt{Mathematica} up to a precision of $10^{-2}$. To construct these tables, we choose 3234 points uniformly distributed in 3D parameter space with the boundaries: $ 0.1 \leq \alpha \leq 4$, $ 0 \leq e \leq 1 $ and $ -1 \leq \cos{i} \leq 1$. When $\Phitilda_{\bfl}$ is called during the calculation of the $\rho_1$ integral (or both its sub-parts $\rho_{1,\rm nr}$ and $\rho_{1,\rm res}$), its value is interpolated linearly using these tables.  

Resonant $\rho_{1,\rm res}$ and non-resonant part $\rho_{1,\rm nr}$ of density wakes are evaluated by numerically solving the integrals of equation~(\ref{rho1_fin}) in action space. To study the 2D wake structure in perturber's orbital plane, we evaluate these integrals at 1400 uniformly distributed points with $ r \in [0.05,0.6]$kpc in the two leading quadrants with $y \geq 0$. For 3D wakes, we choose 2913 uniformly distributed points (for the same radial range) in two leading octants with $y \geq 0$ and $z \geq 0$. Owing to symmetry properties, the wake structure can be fully deduced in all directions.       

The non-resonant part of density deformation $\rho_{1,\rm nr}$ (the $\cos$ term in expression~\ref{rho1_fin}) is a 3D improper integral due to singularity at $\bfl \centerdot \bfOmg =0$ in $\{I,L \}$-plane. But fortunately it can be deduced by evaluating Cauchy principal value. Instead we choose an equivalent method which is numerically more convenient. We manually soften the singularity by making this replacement in the integral:
\beq 
\frac{1}{\bfl \centerdot \bfOmg} \rightarrow \frac{\sigma \, \Omega_b^{-1}}{10^{-3} + \Omega_b^{-1} |\bfl \centerdot \bfOmg  |} 
\eeq 
where $\sigma = \pm 1$ represents the sign of $\bfl \centerdot \bfOmg$. All other functions in the integrand of $\rho_{1,\rm nr}$ are analytic (except $\Phitilda_{\bfl}$) and well defined. To evaluate $\rho_{1,\rm nr}$ at a given point $\bfr'$, we solve a 3D integral over actions $\bfI$, all Fourier modes in $\bfl$ being summed up inside the integrand. 
 
 The resonant part of density deformation $\rho_{1,\rm res}$ (the $\sin$ term in the expression~\ref{rho1_fin}) at a given $\bfr'$, is effectively a sum (over resonances defined by respective Fourier modes $\bfl$) of 2D integrals on respective resonant planes defined by $\bfl \centerdot \bfOmg =0$. All the numerical integrals for $\rho_{1,\rm res}$ and $\rho_{1,\rm nr}$ are solved up to a precision of $10^{-2}$ in \texttt{Mathematica}. 
  
\section{LBK Torques}  
\label{sec_lbk_tor_comp}

\begin{figure}
\centering
\begin{subfigure}{0.49\textwidth}
\centering
\includegraphics[width=1. \textwidth]{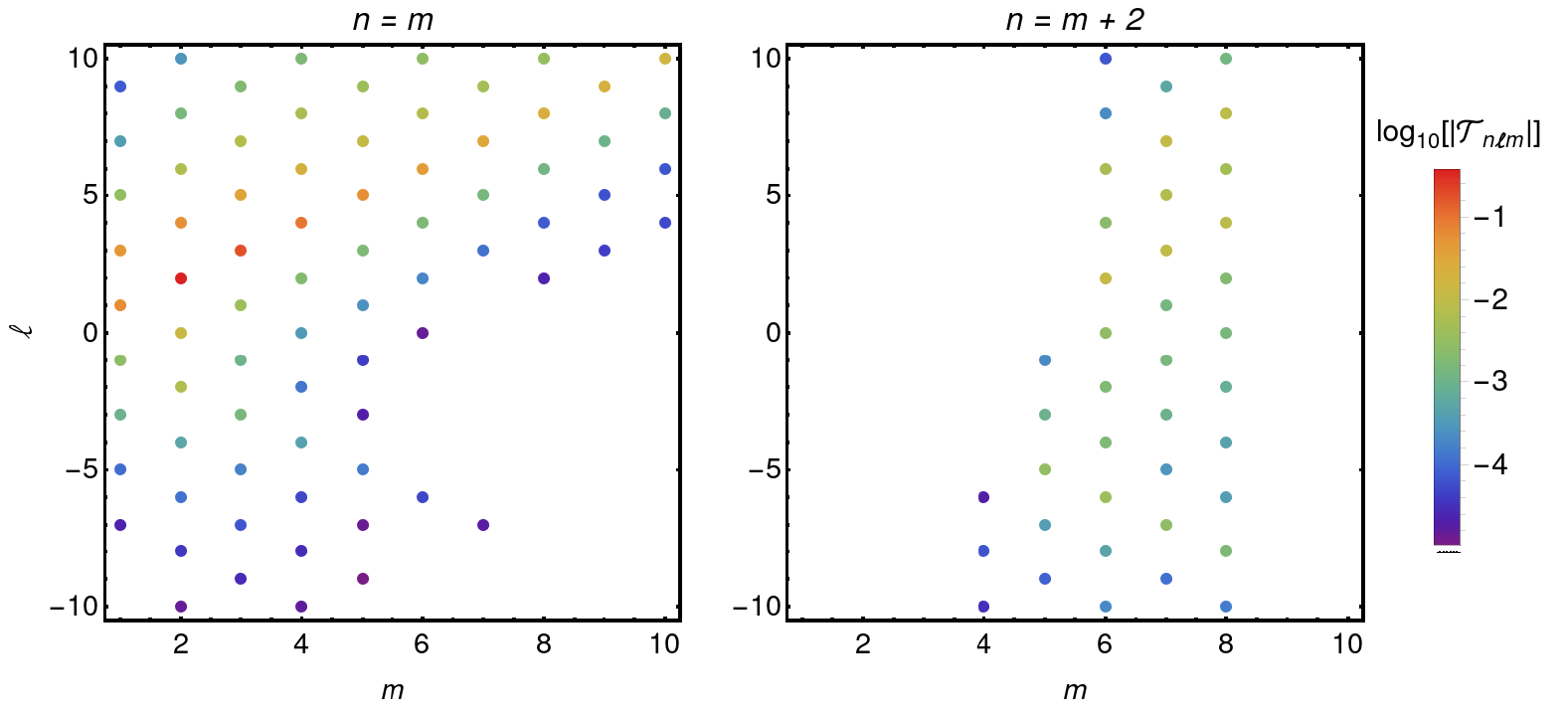}%,trim={5cm 9.5cm 1cm 6cm}
\subcaption{$r_p = 0.3$kpc}
%\label{ilr}
\end{subfigure}
\begin{subfigure}{0.49\textwidth}
\centering
\includegraphics[width=1. \textwidth]{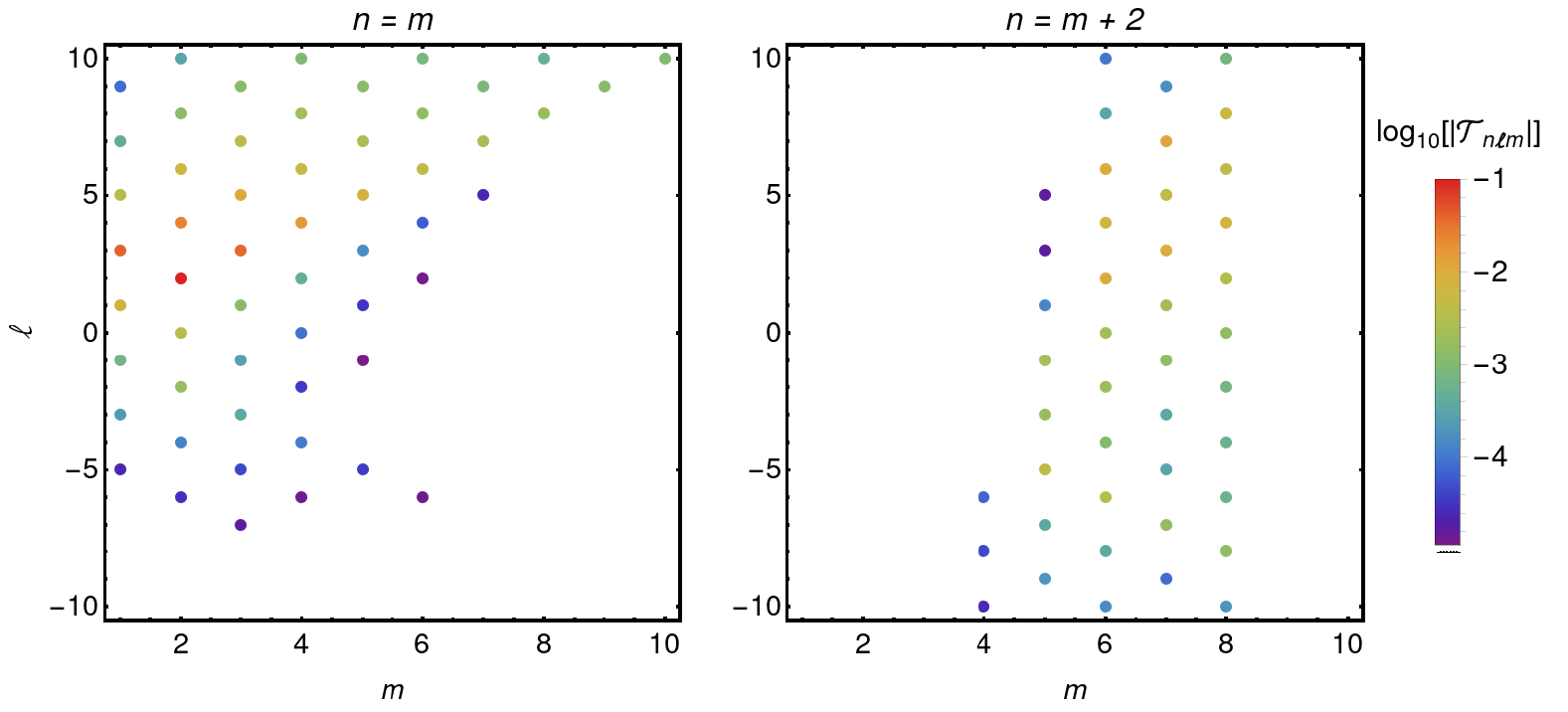}%,trim={5cm 9.5cm 1cm 6cm}
\subcaption{$r_p = 0.26$kpc}
%\label{ilr}
\end{subfigure}
\\[1em]
\begin{subfigure}{0.49\textwidth}
\centering
\includegraphics[width=1. \textwidth]{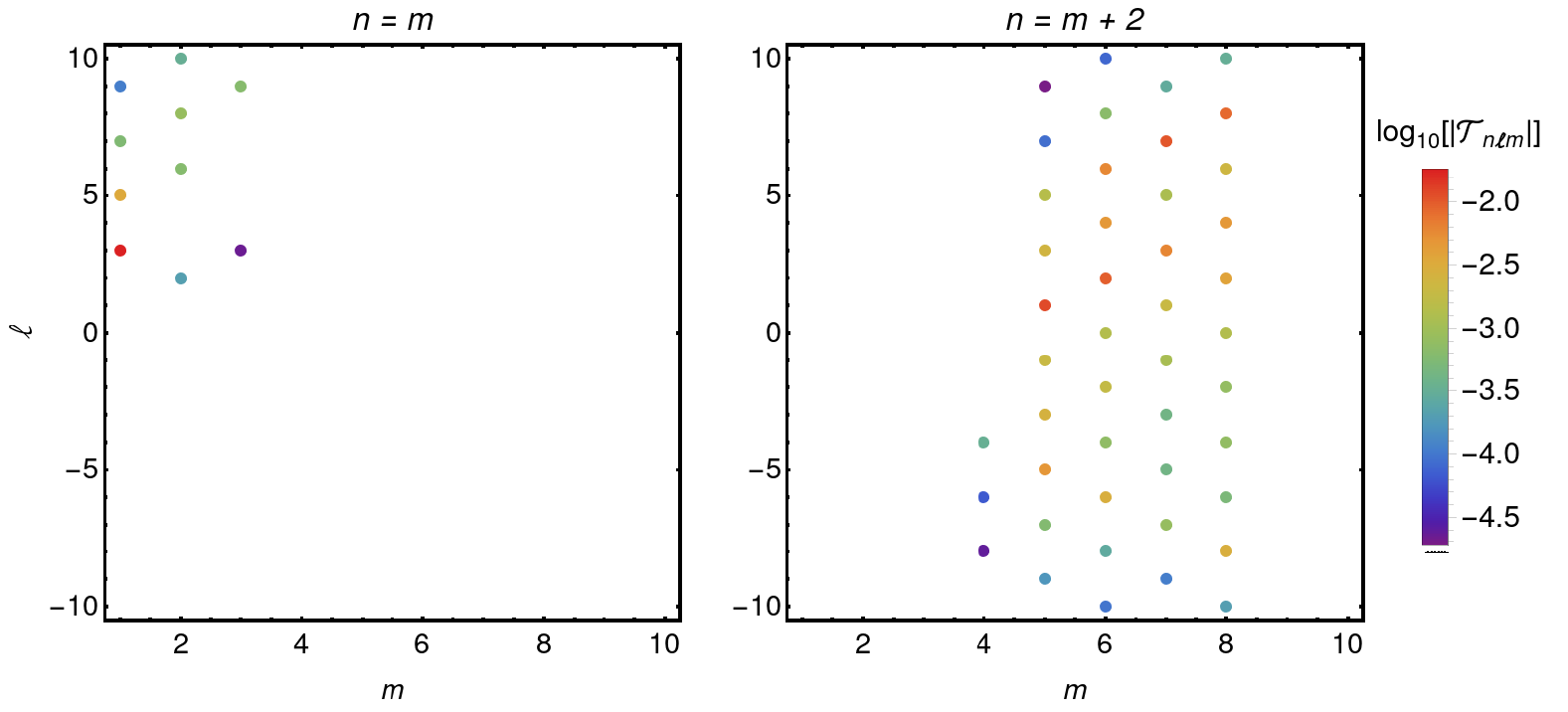}%,trim={5cm 9.5cm 1cm 6cm}
\subcaption{$r_p = 0.225$kpc}
%\label{ilr}
\end{subfigure}
\begin{subfigure}{0.49\textwidth}
\centering
\includegraphics[width=1. \textwidth]{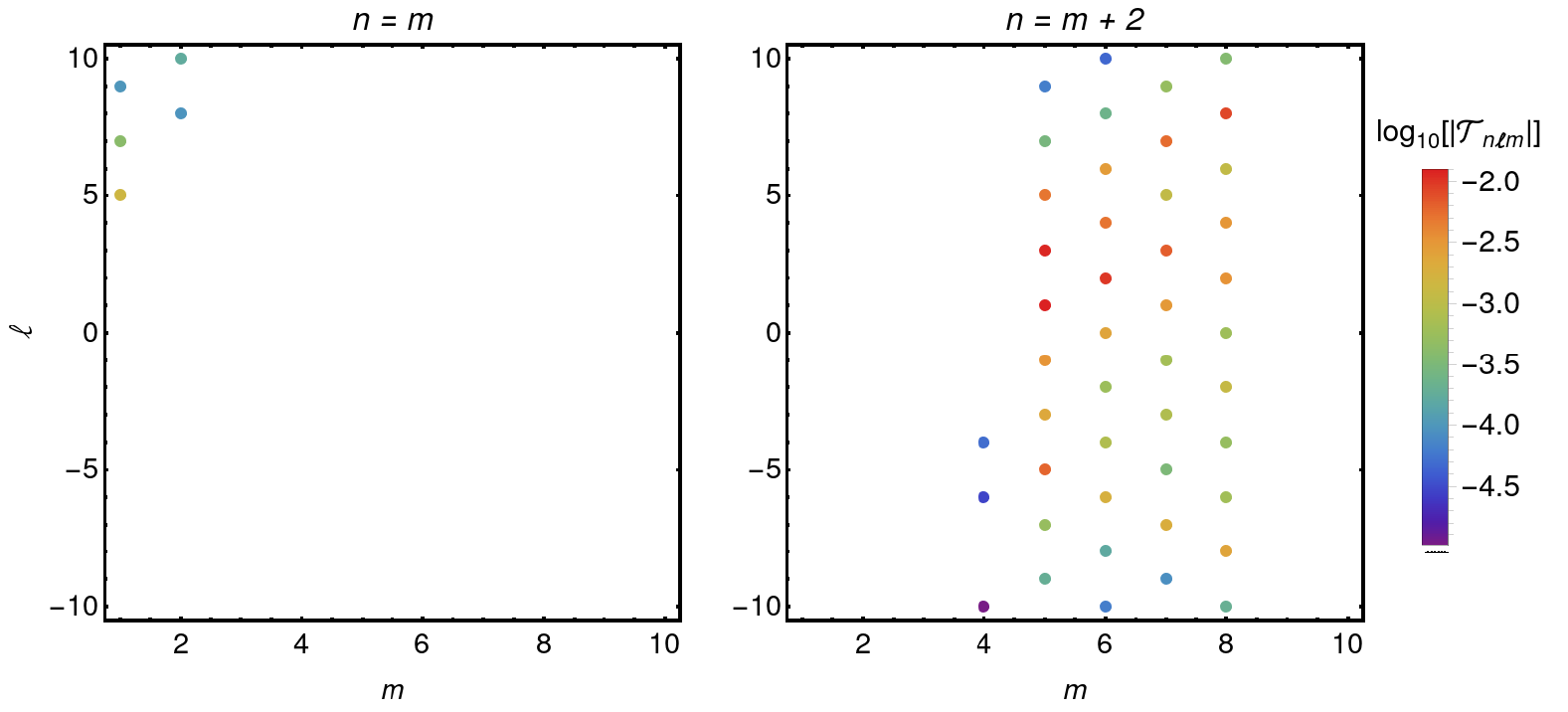}%,trim={5cm 9.5cm 1cm 6cm}
\subcaption{$r_p = 0.2$kpc}
%\label{ilr}
\end{subfigure}
\\[1em]
\begin{subfigure}{0.49\textwidth}
\centering
\includegraphics[width=1. \textwidth]{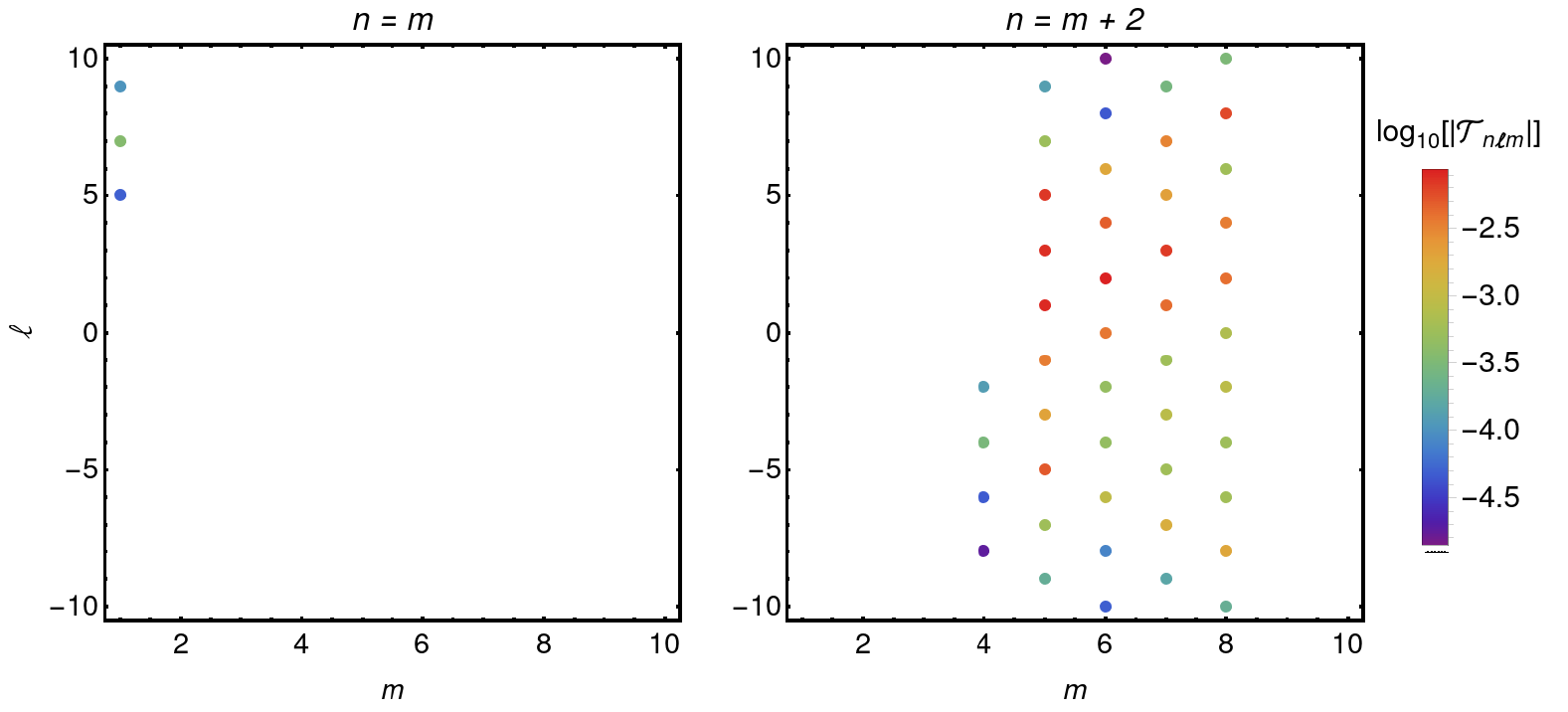}%,trim={5cm 9.5cm 1cm 6cm}
\subcaption{$r_p = 0.18$kpc}
%\label{ilr}
\end{subfigure}
\caption{  Torque components $\mathcal{T}_{n \ell m}$ ($G M_p^2 b^{-1}$) for CR (with $n = m$, the left figure of each panel) and non-CR (with $n = m+2$, the right figure) resonances are shown in $m \ell$-plane for various $r_p$.   }
\label{fig_torq_comp}
\end{figure} 

\begin{figure}
	\centering
	\includegraphics[width=0.65\textwidth]{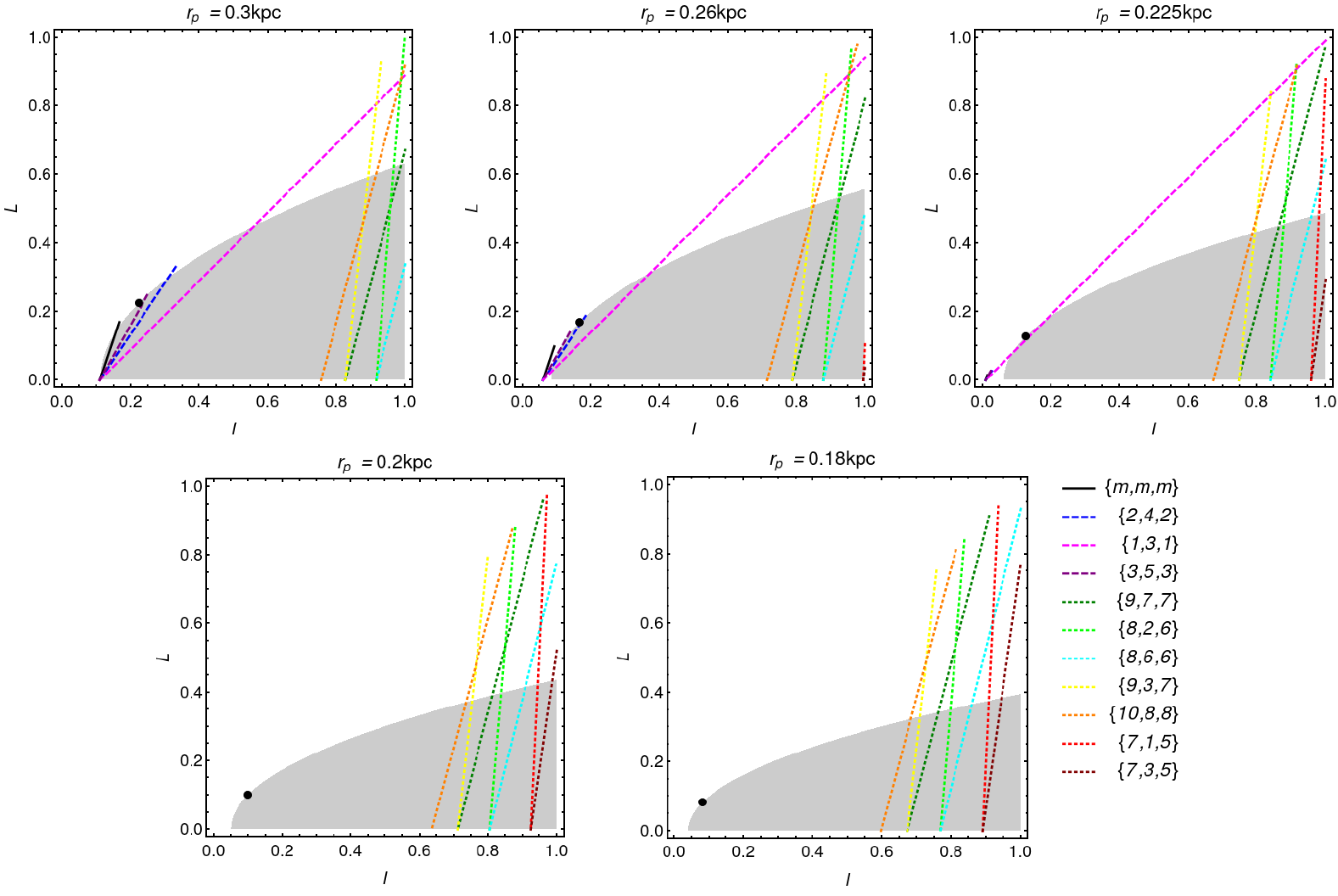}
	\caption{  Dominant resonant lines are shown in the $I,L$-plane for various $r_p$. Both $I$ and $L$ are in units of $I_{\rm max} = \epsilon I_b$. Black dot denotes the orbit of perturber. Shaded region corresponds to the orbits with $ r_{\rm peri} \leq r_p \leq r_{\rm apo}$. }
	\label{fig_res_lines}
\end{figure}

\begin{figure}
	\centering
	\includegraphics[width=0.9\textwidth]{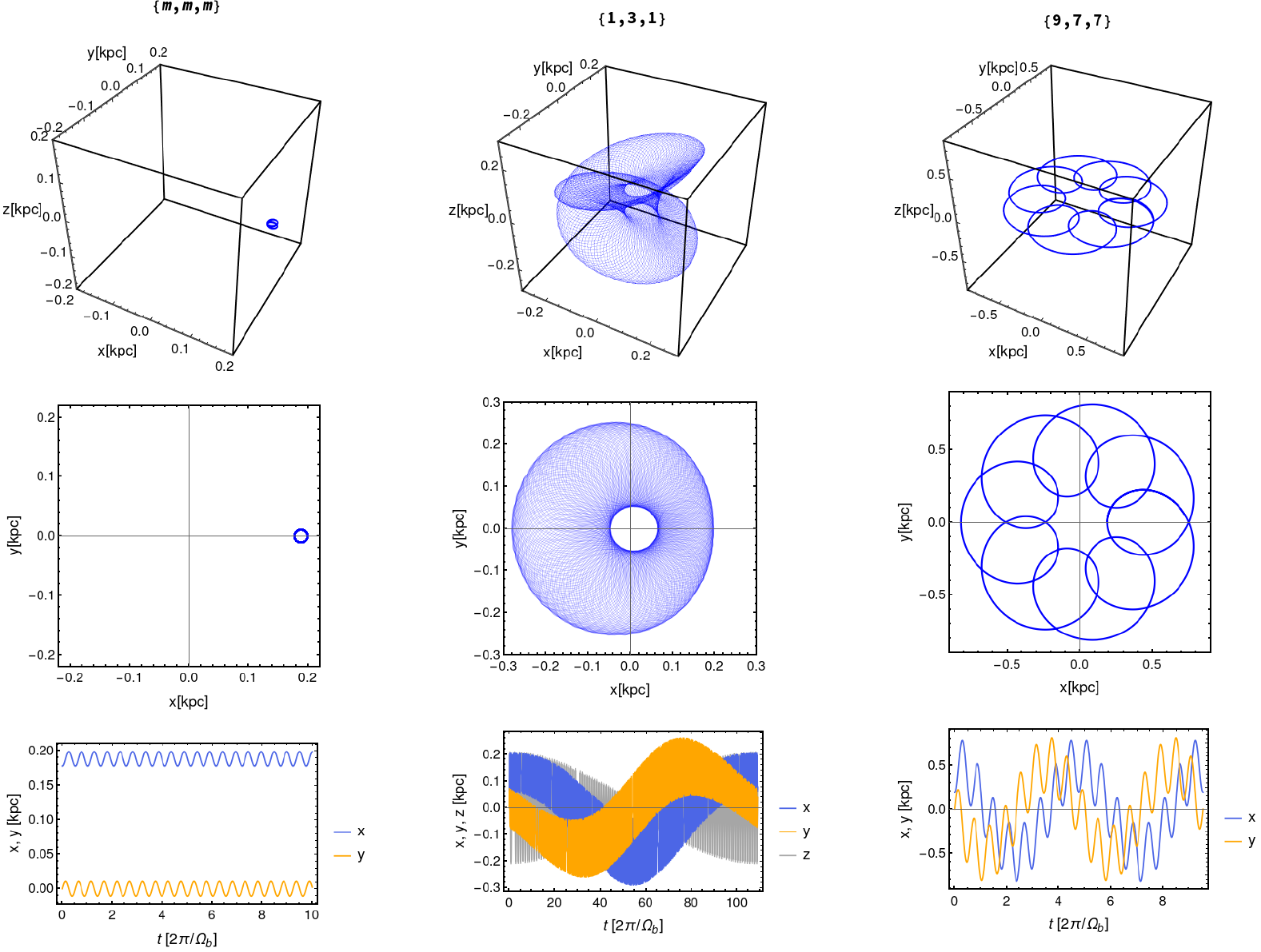}
	\caption{Three resonant orbits are shown in the rest frame of perturber for  $r_p = 0.26$kpc; [\emph{left panel}] $\{m,m,m \}$ is the stationary CR resonance, [\emph{middle panel}] $\{1,3,1  \}$ is the dominant CR resonance of general form and [\emph{right panel}] $\{ 9,7,7 \}$ is one of the dominant non-CR resonances. 3D orbits are shown on the top; while the left and right orbits are just confined to the $x,y$-plane. Orbital projections in the $x,y$-plane are presented in the middle. Temporal variation of spatial coordinates is depicted at the bottom. Choice of actions is such that the contribution to the LBK torque is maximum.     }
	\label{fig_res_orbits}
\end{figure}

\begin{figure}
	\centering
	\includegraphics[width=0.8\textwidth]{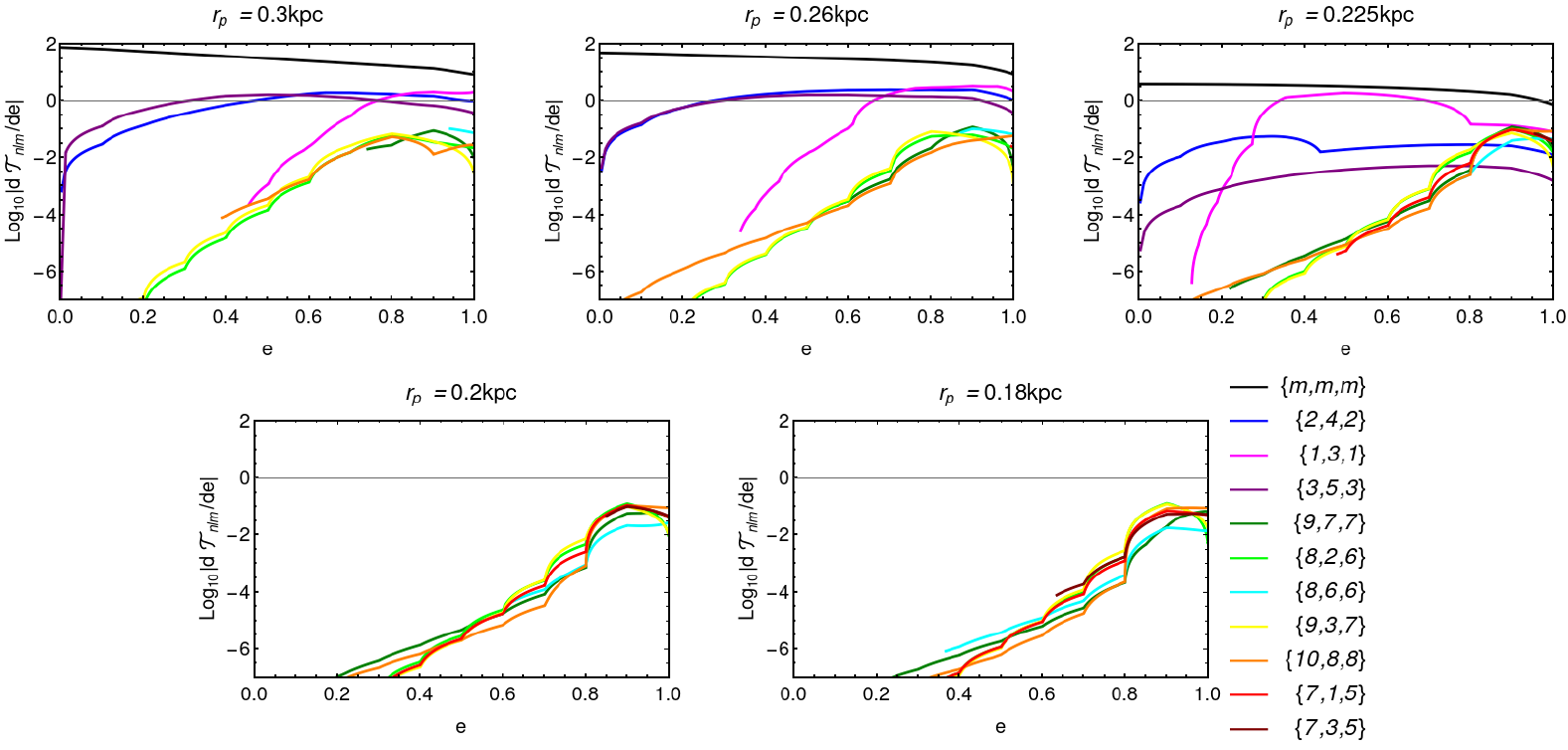}
	\caption{Measure of torque strengths for dominant resonances as function of eccentricity $e$ of resonant stellar orbits. Non-corotating resonances have highest contributions from high $e$ orbits.  }
	\label{fig_torq_st_ecc}
	\end{figure}

We can broadly divide resonances into two categories -- CR (with $n = m$) and non-CR (with $n \neq m$), as in KS18. For computing $\mathcal{T}_{\rm LBK}$, here we have considered resonances till order 10 (such that each of integer magnitudes $\{|n| ,|\ell|, |m|\}\leq 10$). Figure~\ref{fig_torq_comp} showcases these torque components $\mathcal{T}_{n\ell m}$, evaluated using equation~(\ref{torq-comp}), for various $r_p$s in $m \ell$-plane; left figure of each panel corresponds to CR torques with $n=m$ and right figure corresponds to a subset of non-CR torques with $n=m+2$. We find that the other possible forms of non-CR torques (with $n \neq m+2$) vanish for the chosen limits on resonance order and bounds of action space.\footnote{Non-CR resonant orbits of the form $\{ n, \ell, m \}$ with $n \geq m+3$ are larger orbits with $I > I_{\rm max}$ and hence do not contribute to the evaluated torque. Nonetheless we expect these resonant orbits to interact weakly with perturber, leading to smaller torque contributions. } Values of the total torque $\mathcal{T}_{\rm LBK}$, CR torque $\mathcal{T}_{\rm CR} = \sum_{m,\ell} \mathcal{T}_{m \ell m}$ and non-CR torque $\mathcal{T}_{\rm nCR} = \sum_{m,\ell}\mathcal{T}_{(m+2)\ell m}$ are quoted for various $r_p$s in table~\ref{tab_tor_comp}.  
\begin{table}
  \begin{center}
   \begin{tabular}{ c c c c c } 
    \hline
      $r_p ({\rm kpc})$ & $\mathcal{T}_{\rm LBK}$ & $\mathcal{T}_{\rm CR}$ & $\mathcal{T}_{\rm nCR}$ & $\mathcal{T}_{\rm CR,st}$ \\
      \hline
      0.3 & -1.274 & -1.171 (92\%)  &  -0.103 (8\%) & -0.903 (71\%) \\
      0.26 & -0.443 & -0.332 (75\%) & -0.111 (25\%)  & -0.203 (46\%)\\
      0.225 & -0.128 & -0.025 (19\%) & -0.103 (81\%) & -0.0002 (0.18\%) \\
      0.2 & -0.111 & -0.002 (2\%) & -0.108 (98\%) & 0 \\ 
      0.18  & -0.097 & -0.0005 (0.5\%) &  -0.096 (99.5\%) & 0  \\
      \hline
    \end{tabular}
     \caption{  Variation of $\mathcal{T}_{\rm LBK}$ and its CR and non-CR parts (denoted as $\mathcal{T}_{\rm CR}$ and $\mathcal{T}_{\rm nCR}$ respectively) with $r_p$. $\mathcal{T}_{\rm CR,st}$ for stationary CR resonances ($n=\ell =m$) is also quoted. All torques are in the units $G M_p^2/b$. Percent contribution of $\mathcal{T}_{\rm CR}$, $\mathcal{T}_{\rm nCR}$ and $\mathcal{T}_{\rm CR,st}$ to total torque $\mathcal{T}_{\rm LBK}$ is shown in parentheses. }
    \label{tab_tor_comp}
  \end{center}
\end{table}

It is interesting to note that, with decreasing $r_p$, {\bf(a).} $|\mathcal{T}_{\rm LBK}|$ decreases monotonically, with its magnitude dropping roughly by a factor of 10 from $r_p=0.3$kpc to $r_\star$. This fall gets increasingly shallow inside $r_\star$, and $|\mathcal{T}_{\rm LBK}|$ does not change much at an order of magnitude level. {\bf (b).} number of CR resonances and resulting CR torque magnitude $|\mathcal{T}_{\rm CR}|$ fall sharply. $|\mathcal{T}_{\rm CR}|$ decreases by a factor of 50 from $r_p=0.3$kpc to $r_\star$. It diminishes at a much faster pace inside $r_\star$ falling by more than 3 orders of magnitude at $r_p=0.18$kpc (when compared with its counterpart at $r_p=0.3$kpc).   
{\bf (c).} number of non-CR resonances and resulting non-CR torque magnitude $|\mathcal{T}_{\rm nCR}|$ do not change significantly throughout the range of $r_p$ considered ( $|\mathcal{T}_{\rm nCR}| \sim 0.1 G M_p^2/b $). But, the percent contribution of $\mathcal{T}_{\rm nCR}$ to the total torque $\mathcal{T}_{\rm LBK}$ increases significantly from just $8\%$ at $r_p=0.3$kpc to $81\%$ at $r_\star$ and a whopping $99.5\%$ at $r_p=0.18$kpc. 

At larger $r_p = 0.3 \, \mbox{and} \, 0.26$kpc, a special CR resonance with $n = m = \ell$ (with corresponding resonance condition $\Omega_w + \Omega_g - \Omega_p =0$) contribute significantly, 71\% and 46\% respectively, to $\mathcal{T}_{\rm LBK}$. Planar orbits with vanishing libration frequency $\Omega_s = \Omega_w + \Omega_g - \Omega_p $ are the major contributors to the torque $\mathcal{T}_{\rm CR,st}= \sum_{m} \mathcal{T}_{m m m} $ arising from this \emph{stationary} CR resonance. Note that the orbits that mainly contribute to the general CR resonances with $n = m \neq \ell$ have a slow libration frequency $\Omega_s = (m-\ell)/m \, \Omega_g$. On the contrary, non-CR resonant orbits have a libration frequency $\Omega_s$ comparable in magnitude to the fast dynamical frequency $\Omega_w$. Figure~\ref{fig_res_orbits} showcases the
example resonant orbits falling under these three categories, while the perturber orbits at radial distance $r_p=0.26$kpc.   

We choose a subset of dominant resonances that stand out in the figure~\ref{fig_torq_comp} and plot the corresponding resonant lines in $I L$-plane in figure~\ref{fig_res_lines}. Solid black line shows the stationary CR resonance $n = m = \ell $, and dashed lines correspond to general CR resonances. This set of resonant orbits generally have average sizes less than or comparable to $r_p$. As $r_p$ decreases ($\Omega_p$ increases), these resonant orbits become smaller and smaller (so as to have a $\Omega_w$ comparable to $\Omega_p$), and get vanishingly small as  $r_p \rightarrow r_\star^{+}$. This trend is straightforward to understand because $\Omega_p$ increases more sharply compared to $\Omega_w$, which saturates at the core frequency $\Omega_b$. This explains the steeply falling magnitudes of $\mathcal{T}_{\rm CR}$ with decreasing $r_p$. Non-CR resonant orbits (shown by the colored dotted lines) generally correspond to orbits with average size larger than $r_p$. For these dominant resonances, we also check the contributions $|\rmd \mathcal{T}_{n \ell m}/ \rmd e|$ to the torque coming from a unit range of orbital eccentricities around $e$ (along the resonant line in $I,L$-plane); see figure~\ref{fig_torq_st_ecc}. While contribution to CR torques is evenly distributed in $e$ space, the highest contributions to non-CR torques come from high $e$ orbits. The larger non-CR stellar orbits with high $e$ can approach perturber closely which leads to more effective torquing. We speculate that this trend does not appears for smaller CR orbits, because it is counterbalanced by strong net torquing from low $e$ orbits with small libration amplitudes.

\section{Inclusion of High-order Resonances}
\label{sec_hi_res}

\begin{figure}
	\centering
	\includegraphics[width=1\textwidth]{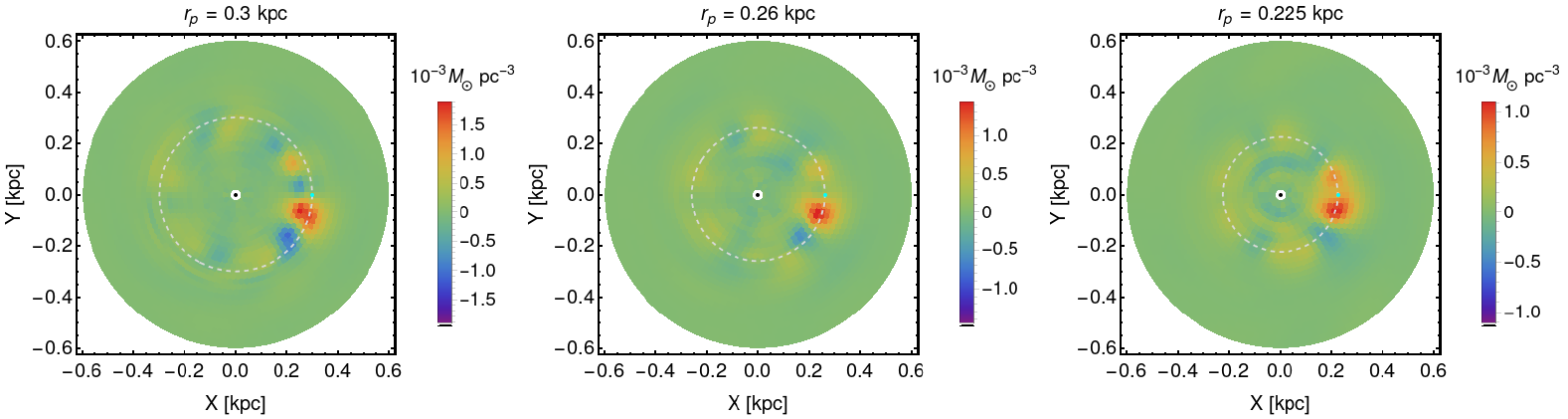}
	\caption{Higher order Fourier mode contributions to the overall wake (in units of $10^{-3}\Msun {\rm pc}^{-3}$) for different values of $r_p$, shown in the perturber's rotating rest frame and orbital plane. These perturbed density profiles include only those integer sets $\bfl$ where at least one integer has magnitude $> 5$ (for computational purposes all integers are still $\leq 10$). For comparison, figure \ref{fig_wake2d_rps} shows the overall wake with the full set of Fourier modes (again with $\{|n|, |\ell|, |m|\} \leq 10)$.}
	\label{fig_wake_2d_hi}
\end{figure}
\begin{figure}
\centering
\includegraphics[width=0.9\textwidth]{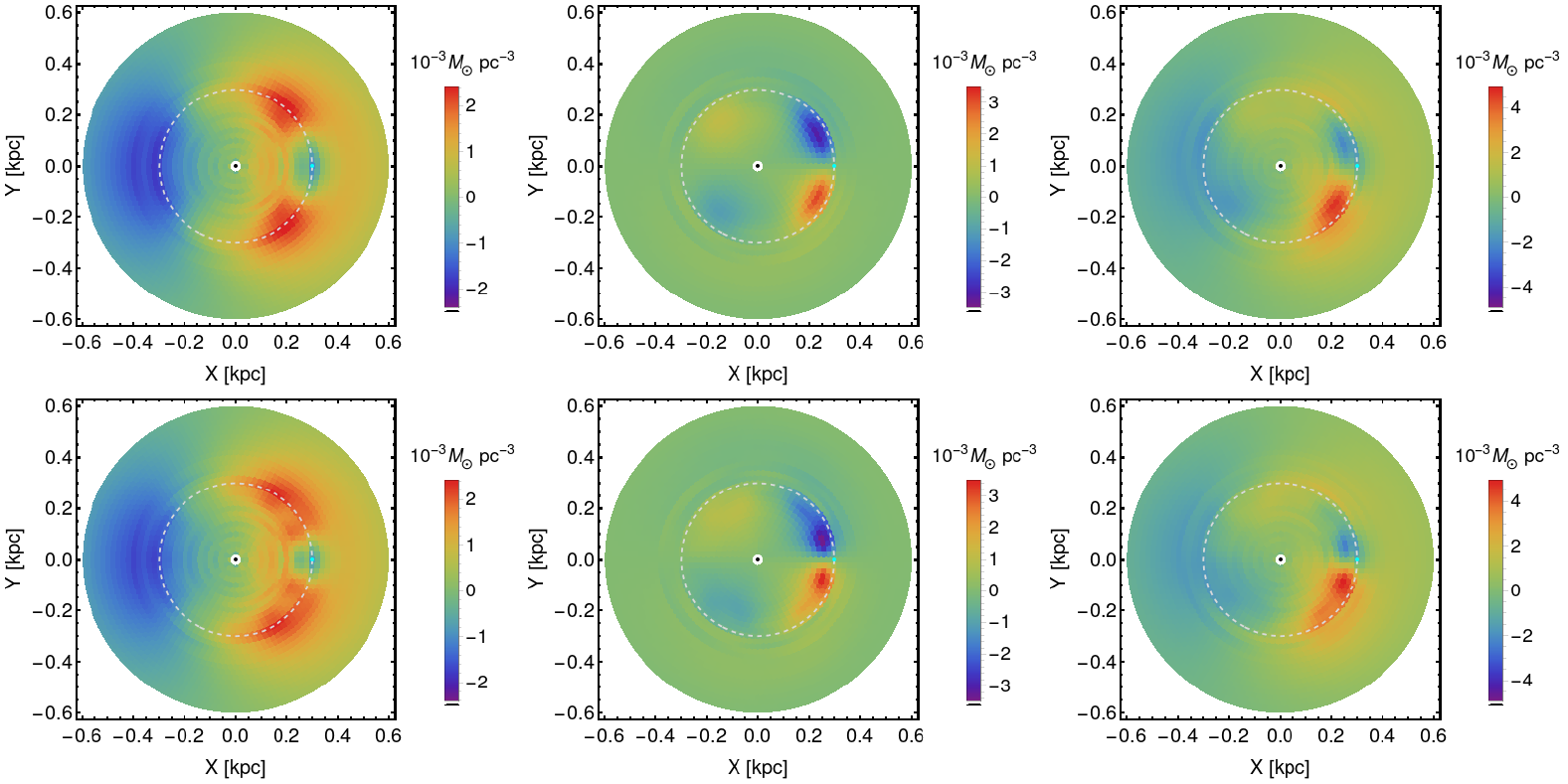}%,trim={5cm 	
\caption{The non-resonant ({\it left}), resonant ({\it middle})  and overall density deformations ({\it right}), in units of $10^{-3}\Msun {\rm pc}^{-3}$, for $r_p = 0.3{\rm kpc}$ in the perturber's rest frame. Upper panels include the integer set $\bfl$ of Fourier modes with magnitudes of individual integers $\leq 5$. Lower panels includes the Fourier integers up to magnitudes $\leq 10$.} 
\label{fig_wakes_fourier_comp}
\end{figure}

\subsection{Density Wakes}

In this sub-appendix we check the effects of adding higher order Fourier terms into the density deformation. We find that the higher order contribution intensifies the overdensity close to the perturber. Figure~\ref{fig_wake_2d_hi} showcases higher order contributions (by ``higher order,'' we mean that at least one Fourier integer $|n|$, $|\ell|$ or $|m|$ is greater than 5) to the total wakes in the orbital plane of the perturber; for comparison, check figure~\ref{fig_wake2d_rps}. This effect manifests itself for both resonant and non-resonant parts of the wake; figure~\ref{fig_wakes_fourier_comp} compares the lower order contribution (up to integer magnitudes $\leq 5$) %, shown in the upper panels,
to the \emph{total} wakes (up to integer magnitudes $\leq 10$) %, shown in lower panels,
for $r_p=0.3$kpc. Including higher order contributions leads to: (1) a shrinking of the underdensity around the perturber for the non-resonant wake, and (2) a strengthening of the trailing overdensity (and leading underdensity) close to the perturber that come from the resonant wake.

%\FloatBarrier

\subsection{LBK Torque Suppression Factor \& Orbital Evolution}

\begin{figure}
\begin{subfigure}{0.48\textwidth}
\centering
\includegraphics[width=1 \textwidth]{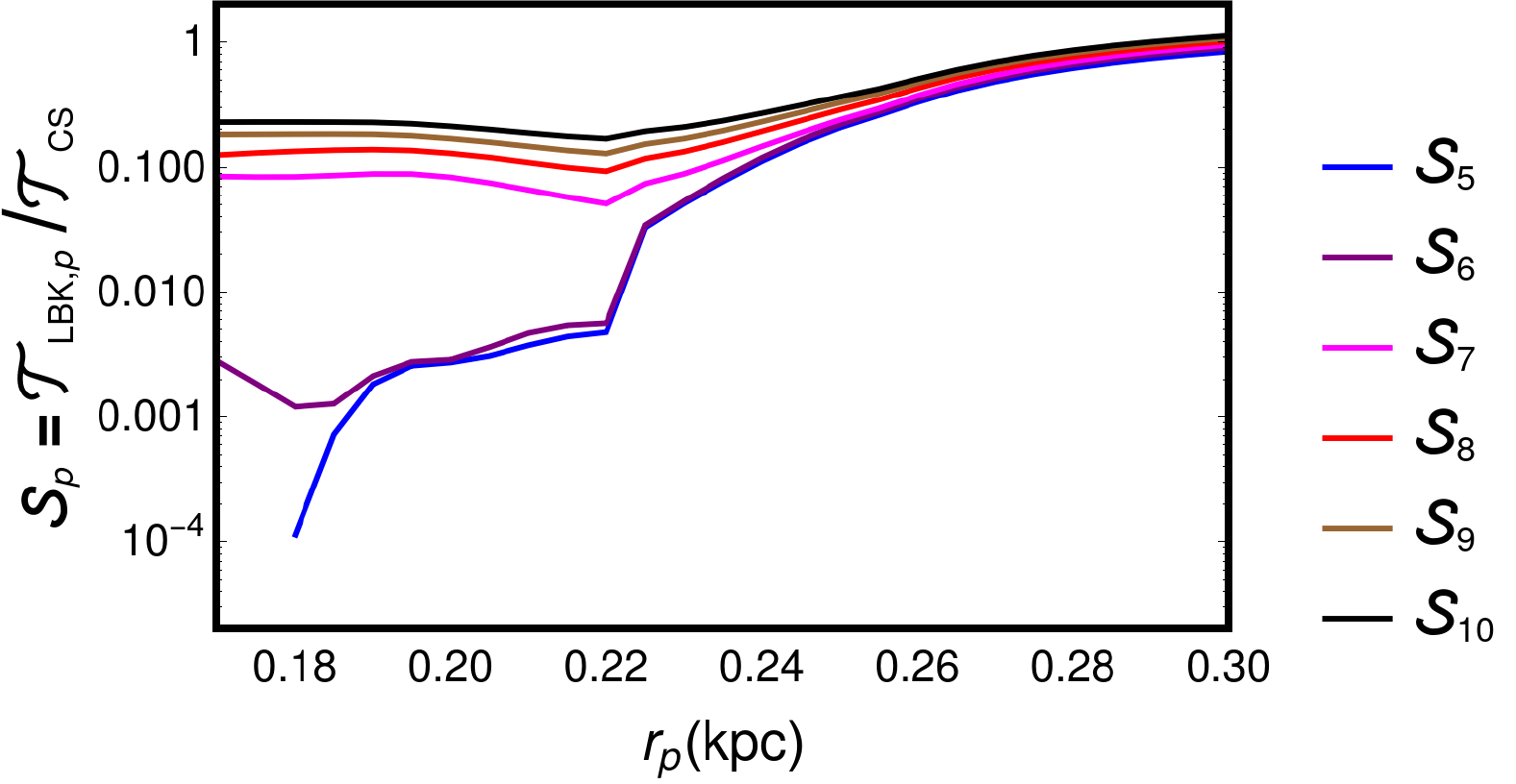}%,trim={5cm 9.5cm 1cm 6cm}
%\subcaption{$r_p=0.3$kpc}
%\label{ilr}
\end{subfigure}
\hfill
\begin{subfigure}{0.48\textwidth}
\centering
\includegraphics[width=1 \textwidth]{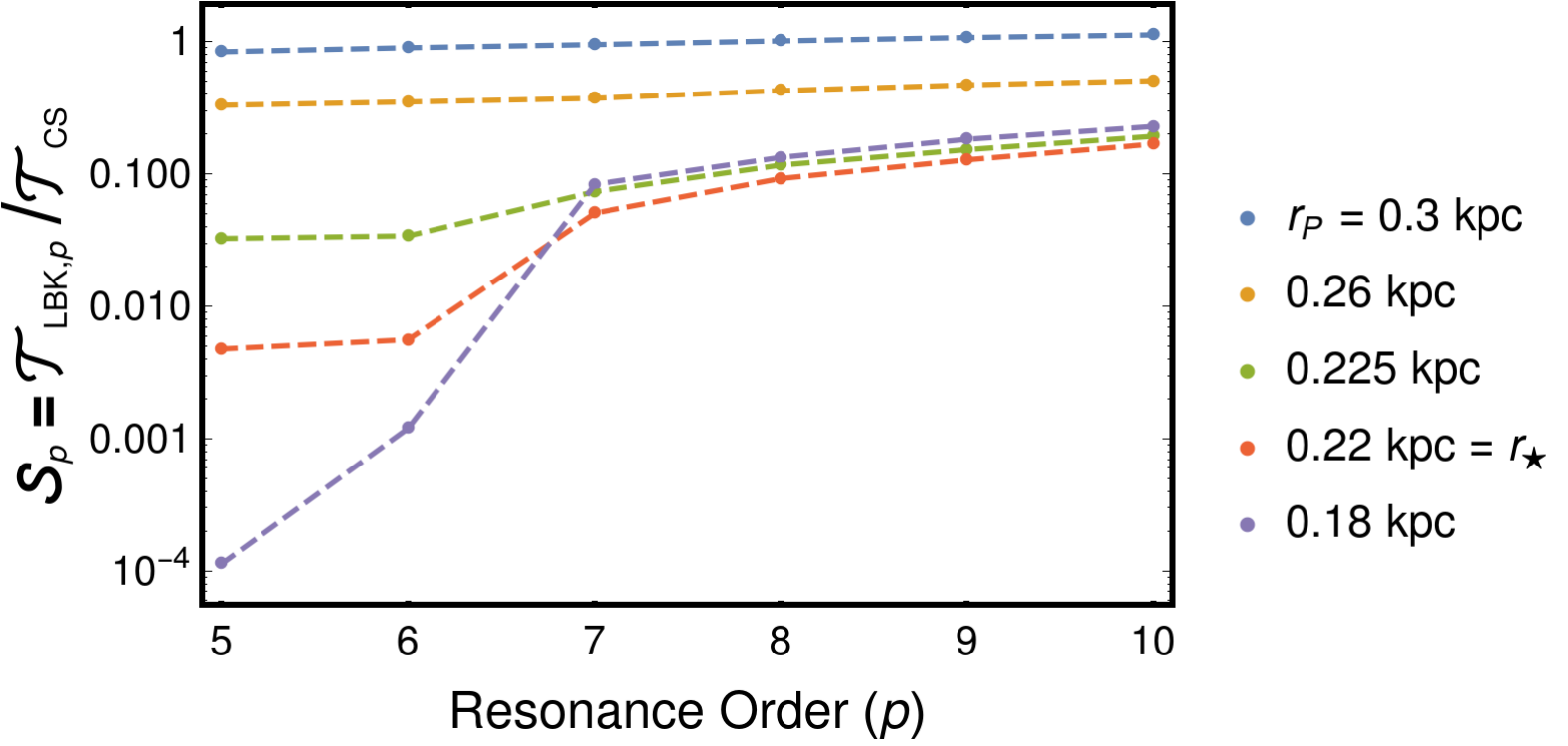}%,trim={5cm 9.5cm 1cm 6cm}
%\subcaption{$r_p=0.26$kpc}
%\label{ilr}
\end{subfigure}
\caption{
  Suppression factors $\mathcal{S}_p$ due to LBK torque for various resonances orders $p$. Left panel showcases the radial profile of $\mathcal{S}_p$ for $p=5-10$. Right panel presents $\mathcal{S}_p$ as a function of $p$ for a fixed perturber's orbital radius $r_p$. 
}
\label{fig_S_res_ord}
\end{figure} 
\begin{figure}
    \centering
    \begin{minipage}{0.46\textwidth}
        \centering
        \includegraphics[width=0.9\textwidth]{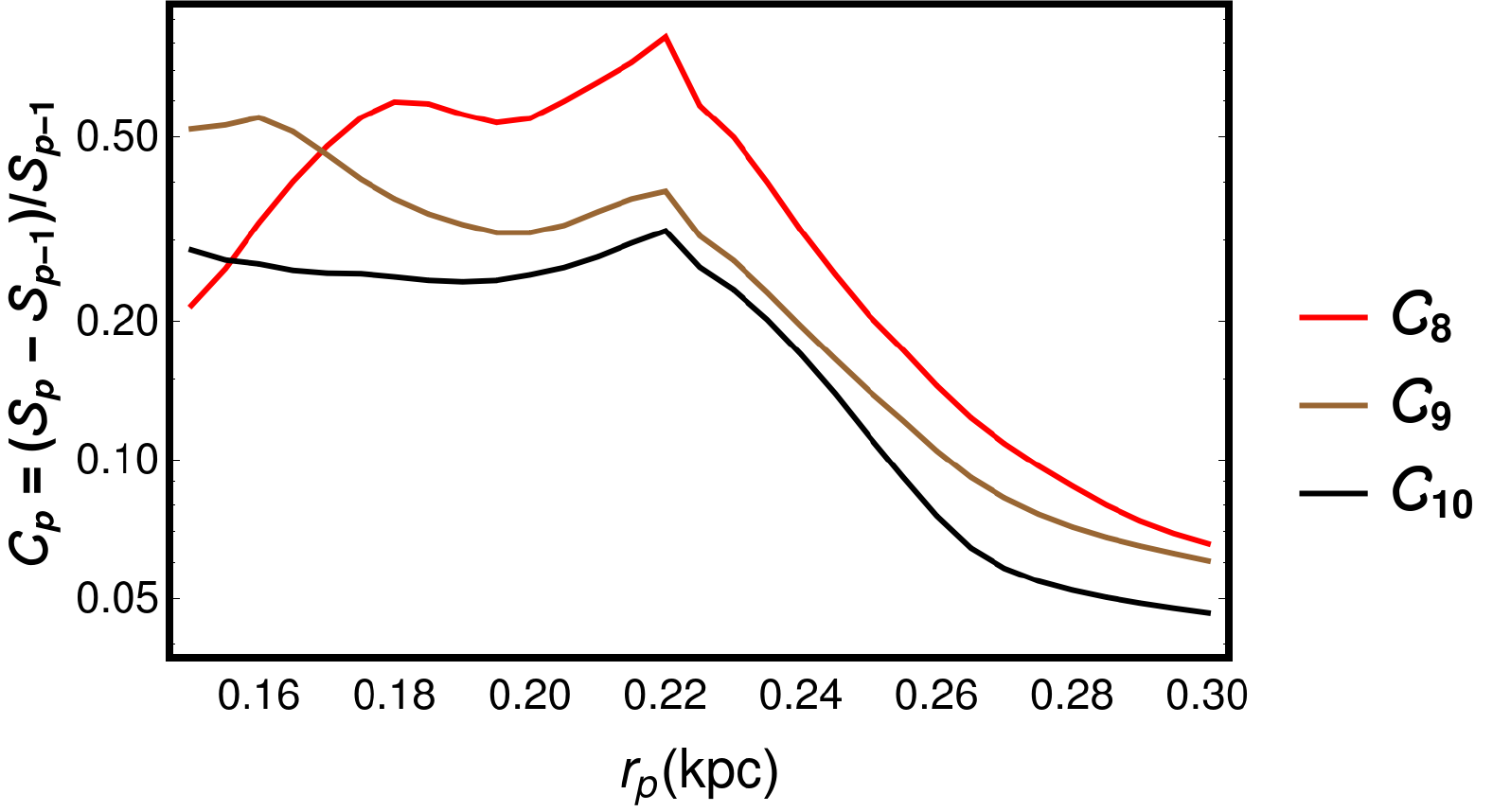}%,trim={5cm 	
\caption{  Radial profile of the fractional change $\mathcal{C}_p$ in LBK suppression factor for including $p^{\rm th}$ order term to $\mathcal{S}_{p-1}$.  }
\label{fig_S_conv}
    \end{minipage}\hfill
    \begin{minipage}{0.48\textwidth}
        \centering
        \includegraphics[width=1\textwidth]{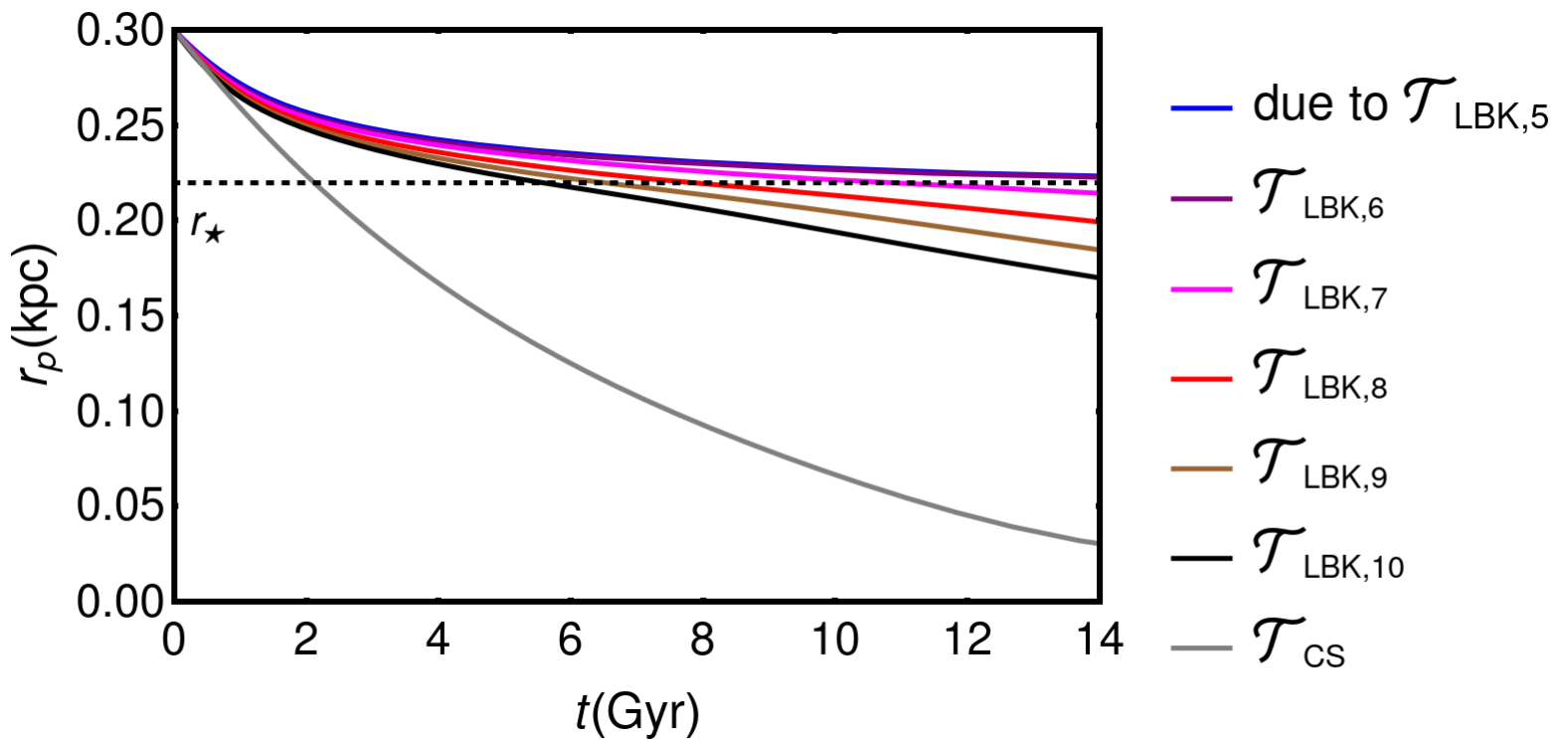}%,trim={5cm 	
\caption{  Time evolution of perturber's orbital radius due to LBK torques $\mathcal{T}_{{\rm LBK},p}$ of various orders (shown in various colors). Case for perturber's orbit evolving due to Chandrasekhar's dynamical friction torque is also shown (in gray) for comparison.  }
\label{fig_orbit_diff_ord}
    \end{minipage}
\end{figure}

Here we investigate the effect of adding higher order torque components into calculations of the torque suppression factor $\mathcal{S}_{\rm LBK}$, defined previously in \S\ref{sec_non_CR_torq}. We define the $p^{\rm th}$ order suppression factor $\mathcal{S}_p = \mathcal{T}_{{\rm LBK}, p}/\mathcal{T}_{\rm CS}$, that includes LBK torque components $\mathcal{T}_{n \ell m}$ such that each of integer magnitudes $\{|n|, |\ell|, |m| \} \leq p$. Figure~\ref{fig_S_res_ord} represents $\mathcal{S}_p$ as a function of resonance order $p$ and orbital radius of perturber $r_p$. 

For large $r_p \lesssim 0.3$kpc, all $S_p$s with $p \in \left[5,10\right]$ are of similar magnitude $\sim 1$; with $\mathcal{S}_{10} / \mathcal{S}_5 \simeq 1.3$ at $r_p = 0.3$kpc. Low order CR torques (especially from stationary CR resonances of the form $n = \ell = m$) dominate over higher order contributions from both CR and non-CR torques; figure~\ref{fig_torq_comp} gives a record of dominant torque components for various $r_p$.

For smaller $r_p \leq r_{\star}$, there is significant contribution of high order non-CR torques ($p \geq 7$) to $\mathcal{T}_{\rm LBK,10}$ and $\mathcal{S}_p$ ($p = 7 - 10$) are greater than $\mathcal{S}_{5}$ and $\mathcal{S}_{6}$ by one or two orders of magnitude. At these small $r_p$, (otherwise strong) stationary CR resonances of the form $n = \ell = m$ do not contribute and low order CR torques are weak. High order non-CR torques are significant contributors; check figure~\ref{fig_torq_comp}(d,e) for $r_p < r_\star$.    

In order to keep a better check on convergence, we define the fractional change $\mathcal{C}_p = ( \mathcal{S}_p - \mathcal{S}_{p-1} )/\mathcal{S}_{p-1}$ introduced in the suppression factor by including $\mathcal{S}_{p}$. Figure~\ref{fig_S_conv} gives radial profiles of $\mathcal{C}_p$ as a function of $r_p$ (only $p=8,9,10$ are plotted for clarity). It turns out that, the  fractional change at highest order i.e. $\mathcal{C}_{10}$ is less than $10\%$ for $r_p > 0.25$kpc and it lies in the range $\sim 30-40 \%$ for $r_p \leq r_\star$.  

Then we check the orbital evolution of perturber due to $\mathcal{T}_{{\rm LBK},p}$ corresponding to various resonance orders $p= 5-10$; see figure~\ref{fig_orbit_diff_ord}. It is apparent that the inclusion of higher order resonances tends to increase the LBK torque at smaller $r_p$s more significantly than at larger $r_p$s. Hence, we speculate that going beyond resonance order $p=10$ (in $\mathcal{T}_{{\rm LBK},p}$) would make the decay of orbital radius $r_p$ even sharper. Still, the qualitative nature of stalling, the slow fall of perturber even inside $r_\star$ (black orbit in the figure), is not expected to be effected due to inclusion of higher order resonances.

 \end{document}